\documentclass[aps,prd,nofootinbib,twocolumn,superscriptaddress]{revtex4-1}
\usepackage{amsmath}
\usepackage[scientific-notation=true]{siunitx}
\usepackage{graphicx}
\usepackage{placeins}
\usepackage{bm}
\usepackage[export]{adjustbox}
\usepackage{enumitem}

\usepackage[activate={true,nocompatibility},final,tracking=false,kerning=true,spacing=true,factor=1100,stretch=10,shrink=10]{microtype}

\usepackage{scalerel}
\usepackage{tikz}
\usetikzlibrary{svg.path}

\definecolor{orcidlogocol}{HTML}{A6CE39}
\tikzset{
	orcidlogo/.pic={
		\fill[orcidlogocol] svg{M256,128c0,70.7-57.3,128-128,128C57.3,256,0,198.7,0,128C0,57.3,57.3,0,128,0C198.7,0,256,57.3,256,128z};
		\fill[white] svg{M86.3,186.2H70.9V79.1h15.4v48.4V186.2z}
		svg{M108.9,79.1h41.6c39.6,0,57,28.3,57,53.6c0,27.5-21.5,53.6-56.8,53.6h-41.8V79.1z M124.3,172.4h24.5c34.9,0,42.9-26.5,42.9-39.7c0-21.5-13.7-39.7-43.7-39.7h-23.7V172.4z}
		svg{M88.7,56.8c0,5.5-4.5,10.1-10.1,10.1c-5.6,0-10.1-4.6-10.1-10.1c0-5.6,4.5-10.1,10.1-10.1C84.2,46.7,88.7,51.3,88.7,56.8z};
	}
}

\newcommand\orcidicon[1]{\href{https://orcid.org/#1}{\mbox{\scalerel*{
				\begin{tikzpicture}[yscale=-1,transform shape]
				\pic{orcidlogo};
				\end{tikzpicture}
			}{1}}}}

\usepackage{hyperref} 

\let\vec\bm

\newcommand{\approptoinn}[2]{\mathrel{\vcenter{
			\offinterlineskip\halign{\hfil$##$\cr
				#1\propto\cr\noalign{\kern2pt}#1\sim\cr\noalign{\kern-2pt}}}}}

\newcommand{\sigmav}{\ensuremath{\langle \sigma v \rangle}}
\newcommand{\diff}{\ensuremath{\mathrm{d}}}
\newcommand{\TRH}{\ensuremath{T_\mathrm{RH}}}
\newcommand{\Tdom}{\ensuremath{T_\mathrm{dom}}}
\newcommand{\xdom}{\ensuremath{\Tdom/\TRH}}
\newcommand{\kcut}{\ensuremath{k_\mathrm{cut}}}
\newcommand{\kRH}{\ensuremath{k_\mathrm{RH}}}

\newcommand{\xcut}{\ensuremath{\kcut/\kRH}}
\newcommand{\rmax}{\ensuremath{r_\mathrm{max}}}
\newcommand{\mmax}{\ensuremath{M_\mathrm{max}}}

\begin{document}
\title{Breaking a dark degeneracy: The gamma-ray signature of early matter domination}
\author{M. Sten Delos \orcidicon{0000-0003-3808-5321}}
\email{delos@unc.edu}
\affiliation{Department of Physics and Astronomy, University of North Carolina at Chapel Hill, Phillips Hall CB3255, Chapel Hill, North Carolina 27599, USA}
\author{Tim Linden \orcidicon{0000-0001-9888-0971}}
\email{linden.70@osu.edu}
\affiliation{Center for Cosmology and AstroParticle Physics (CCAPP), and \\ Department of Physics, The Ohio State University, Columbus, Ohio 43210, USA}
\author{Adrienne L. Erickcek \orcidicon{0000-0002-0901-3591}}
\email{erickcek@physics.unc.edu}
\affiliation{Department of Physics and Astronomy, University of North Carolina at Chapel Hill, Phillips Hall CB3255, Chapel Hill, North Carolina 27599, USA}

\begin{abstract}
The Universe's early thermal history is poorly constrained, and it is possible that it underwent a period of early matter domination driven by a heavy particle or an oscillating scalar field that decayed into radiation before the onset of Big Bang nucleosynthesis.  The entropy sourced by this particle's decay reduces the cross section required for thermal-relic dark matter to achieve the observed abundance.  This degeneracy between dark matter properties and the thermal history vastly widens the field of viable dark matter candidates, undermining efforts to constrain dark matter's identity.  Fortunately, an early matter-dominated era also amplifies density fluctuations at small scales and leads to early microhalo formation, boosting the dark matter annihilation rate and bringing smaller cross sections into the view of existing indirect-detection probes.  Employing several recently developed models of microhalo formation and evolution, we develop a procedure to derive indirect-detection constraints on dark matter annihilation in cosmologies with early matter domination.  This procedure properly accounts for the unique morphology of microhalo-dominated signals.  While constraints depend on dark matter's free-streaming scale, the microhalos make it possible to obtain upper bounds as small as $\langle\sigma v\rangle \lesssim 10^{-32}$~cm$^3$s$^{-1}$ using Fermi-LAT observations of the isotropic gamma-ray background and the Draco dwarf galaxy.
\end{abstract}

\pacs{}
\keywords{}

\maketitle

\section{Introduction}

The thermal history of the Universe prior to Big Bang nucleosynthesis (BBN) is largely unprobed. Light-element abundances \cite{kawasaki1999cosmological,kawasaki2000mev,hannestad2004lowest,ichikawa2005oscillation}, along with density variations inferred from the cosmic microwave background and galaxy surveys \cite{ichikawa2007constraint,de2008new}, demand only that the maximum temperature of the last radiation-dominated epoch be at least 3~MeV.  Our sole hint at earlier history is that to solve the horizon and flatness problems and explain the nearly scale-invariant spectrum of primordial density variations, the Universe is believed to have undergone a period of inflation prior to BBN \cite{guth1981inflationary,albrecht1982cosmology,linde1982new}.  However, the energy scale associated with inflation could be as high as $10^{16}$~GeV \cite{liddle1994inflationary,guo2011observational}, and we have no constraints on the Universe's evolution between inflation and BBN.

There is little reason to assume the Universe was radiation dominated from the end of inflation until BBN (see Ref.~\cite{Amin_2014} for a review of proposed dynamics).  Since the energy density of relativistic particles decreases more rapidly than that of nonrelativistic particles, any heavy field left over from the inflationary epoch would naturally come to dominate the energy density of the Universe, leading to an early matter-dominated era (EMDE); such a field is only required to decay into radiation before the onset of BBN.  Well motivated examples of such heavy fields include hidden-sector particles \cite{Pospelov_2008,Arkani_Hamed_2009,Hooper_2012,Abdullah_2014,Berlin_2014,Martin_2014,zhang2015long,Berlin_2016a,Berlin_2016b,Dror_2016,Tenkanen:2016jic,Dror_2018,Tenkanen:2019cik}, moduli fields in string theory \cite{coughlan1983cosmological,de1993model,banks1994cosmological,banks1995cosmological,banks1995modular,acharya2014bounds,Kane_2015,Giblin_2017}, and certain spectator fields invoked to generate primordial curvature variations during inflation \cite{mollerach1990isocurvature,linde1997non,lyth2002generating,moroi2001effects,*moroi2002erratum}.  After inflation ends, the inflaton itself can also behave as a pressureless fluid before its decay \cite{albrecht1982reheating,turner1983coherent,traschen1990particle,kofman1994reheating,kofman1997towards,dufaux2006preheating,allahverdi2010reheating,jedamzik2010collapse,easther2011delayed,musoke2019lighting}.

This gap in our understanding of the early Universe gravely impairs our capacity to constrain the properties of thermal-relic dark matter candidates.  If the dark matter froze out from the thermal plasma during the last radiation-dominated epoch, its annihilation cross section must be close to the canonical $\sigmav=3\times 10^{-26}$~$\si{cm^3s^{-1}}$ in order to produce the observed relic abundance, a value that astrophysical indirect-detection searches have begun to test \cite{ackermann2015searching,ahnen2016qkx,albert2017searching,hess2016}.  However, if dark matter froze out during or before an EMDE, its relic density would have been diluted by entropy produced by the decay of the species driving the EMDE.  In this scenario, a smaller cross section is required to effect the observed relic abundance, potentially making a broad new range of dark matter candidates viable \cite{kamionkowski1990thermal,giudice2001largest,gelmini2006neutralino,Kane_2016,Drees_2018,Drees:2018dsj,Bernal2019,Bernal:2018kcw,Cirelli:2018iax,allahverdi2019nonthermal,Evans:2019jcs}.

Fortunately, an EMDE also amplifies the range of dark matter cross sections accessible to indirect-detection searches.  Subhorizon density perturbations grow rapidly when pressureless fluids dominate the Universe.  Consequently, an EMDE can dramatically enhance small-scale density variations, resulting in the formation of a plethora of highly dense sub-Earth-mass dark matter microhalos long before dark matter halos would otherwise be expected to form \cite{erickcek2011reheating,barenboim2014structure,fan2014nonthermal,erickcek2015dark,erickcek2016bringing,blanco2019annihilation}.  These microhalos in turn boost the rate of dark matter annihilation for a given cross section.  The purpose of this work is to develop a procedure through which existing indirect-detection experiments can be applied to constrain thermal-relic dark matter candidates that freeze out during or before an EMDE.  We improve on previous efforts \cite{erickcek2015dark,erickcek2016bringing,blanco2019annihilation} by employing newly developed models of microhalo formation and evolution to characterize both the magnitude and the morphology of the microhalo-dominated annihilation signals that result from an EMDE.

The signal from annihilation within unresolved microhalos is morphologically similar to that of decaying dark matter; it follows the microhalo distribution, which is similar to the dark matter mass distribution.  However, microhalos in dense environments suffer gradual disruption due to tidal effects and encounters with other objects, so their annihilation signal is suppressed within these environments.  These effects are particularly important near the centers of host halos, and we account for them in our analysis by employing several recently developed models to characterize the microhalo population.  We use the results of Ref.~\cite{delos2019predicting} (hereafter Paper~I) to model the microhalos that result from EMDE scenarios.  This work predicts the population of halos and their density profiles given the (linear-theory) power spectrum of density variations.  Additionally, we use the results of Refs.~\cite{delos2019tidal,delos2019evolution} (hereafter Papers II and~III, respectively) to predict how these microhalos evolve within host halos.  Paper~II traces the dynamical evolution of subhalos due to tidal forces, while Paper~III treats the evolution of microhalos due to encounters with stars.

As a demonstration, we use Fermi-LAT data \cite{atwood2009large} to derive new constraints on thermal-relic dark matter candidates.  We first consider the isotropic gamma-ray background (IGRB), translating published limits on the dark matter lifetime derived therefrom \cite{liu2017constraints,blanco2019constraints} into bounds on dark matter annihilation within unresolved microhalos.  These constraints depend strongly on dark matter's free-streaming scale and its relation to horizon scales during the EMDE, but for reasonable sets of parameters, we obtain bounds as small as $\sigmav\lesssim 10^{-32}$~$\si{cm^3 s^{-1}}$ on dark matter's annihilation cross section.  We also consider gamma rays from the Draco dwarf spheroidal galaxy (dSph), employing Fermi-LAT data to derive limits on annihilation within microhalos inside Draco.  These limits take into account the unique signal morphology induced by disruptive tidal effects within galactic systems.  While Draco yields weaker limits than the IGRB on $\sigmav$, its signal morphology could potentially discriminate between microhalo-dominated emission and dark matter decay.

This work is organized as follows.  Section~\ref{sec:EMDE} reviews the impact of an EMDE on the dark matter abundance and density variations.  In Sec.~\ref{sec:halos}, we detail how the model in Paper~I is employed to predict the microhalo populations resulting from EMDE scenarios.  Section~\ref{sec:igrb} uses the IGRB to derive limits on dark matter cross sections, while Sec.~\ref{sec:draco} uses gamma rays from the Draco dwarf; in both cases, the suppression of microhalo annihilation rates due to tidal effects is considered in detail.  Section~\ref{sec:conc} presents our conclusions.  We include further technical details in an array of appendices.  Appendix~\ref{sec:growth} discusses the growth rate of small-scale dark matter density fluctuations and how we account for this growth rate within Paper~I's halo-formation model.  Appendix~\ref{sec:sup} details how tidal suppression factors derived from Paper~II's tidal evolution model are aggregated over a host halo and presents a fitting function for future convenience.  Appendix~\ref{sec:rmodel} reanalyzes the simulations in Paper~II to present a new refinement to the tidal evolution model, while Appendix~\ref{sec:tidestar} uses a new array of $N$-body simulations to determine how to combine the effects of galactic tidal forces and stellar encounters.  Finally, Appendix~\ref{sec:profile} discusses how we estimate Draco's outer density profile.

\section{Early matter domination}\label{sec:EMDE}

\begin{figure}[t]
	\centering
	\includegraphics[width=\columnwidth]{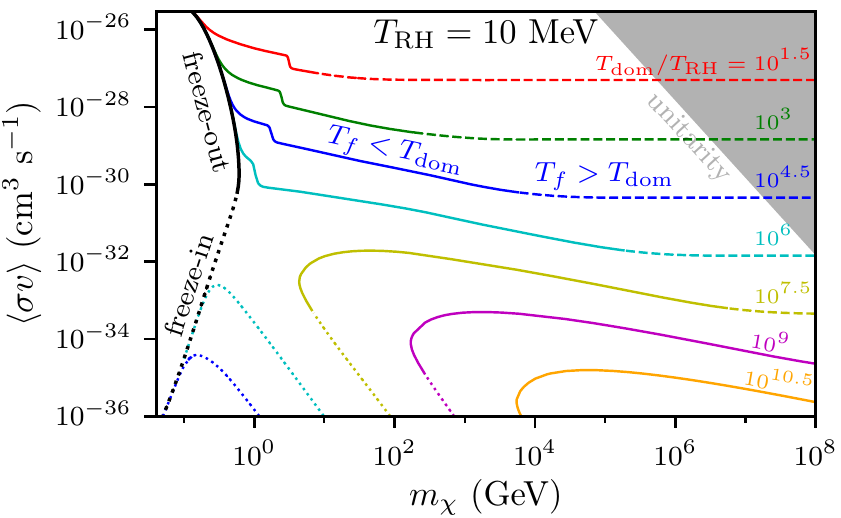}
	\caption{\label{fig:relic} Influence of an EMDE on the mass $m_\chi$ and annihilation cross section $\sigmav$ required for thermal-relic dark matter to achieve the observed relic abundance.  We show an array of EMDE scenarios with reheat temperature $\TRH=10$~MeV; one begins at temperature $\Tdom\gg\TRH$ (or has no preceding radiation-dominated epoch; thick black curve), while the others (colored curves) have $\xdom$ indicated by the numbers on the right.  If a colored curve is disconnected from the black curve, then the entire black curve is also viable for that $\xdom$.  ``Freeze-out'' (solid and dashed lines) indicates that dark matter dropped out of equilibrium with the thermal bath, while ``freeze-in'' (dotted lines) corresponds to dark matter never reaching equilibrium; in either case, production and/or annihilation ceases at a temperature $T_f$ closely related to $m_\chi$.  Entropy production during the EMDE dilutes the dark matter so rapidly that if there is no radiation from a prior epoch, the dark matter mass cannot exceed $\mathcal{O}(10^2)\TRH$ to have any hope of reaching the observed abundance.  On the other hand, the presence of leftover radiation from a prior epoch allows much larger $m_\chi$ to still achieve the observed abundance (colored solid curves).  Arbitrarily large $m_\chi$ can reach the observed abundance by freezing out before the EMDE (dashed curves).  Disjointed behavior occurs when the dark matter freezes out close to the QCD phase transition at temperature 170~MeV.  The shaded region marks where the dark matter's coupling constant exceeds unity \cite{griest1990}.}
\end{figure}

In this section, we review the implications of an EMDE for dark matter; further detail can be found in Refs.~\cite{kamionkowski1990thermal,giudice2001largest,gelmini2006neutralino,Kane_2016,Drees_2018,Drees:2018dsj,Bernal2019,Bernal:2018kcw,Cirelli:2018iax,allahverdi2019nonthermal,Evans:2019jcs,erickcek2011reheating,barenboim2014structure,fan2014nonthermal,erickcek2015dark,erickcek2016bringing,blanco2019annihilation}.  We denote by $\phi$ the heavy field that drives early matter domination.  The end of an EMDE is characterized by the reheat temperature $\TRH>3$ MeV at which $\phi$ domination gives way to radiation.  If the EMDE was preceded by another radiation-dominated epoch, then the transition to $\phi$ domination occurs at an even higher temperature $\Tdom$.

\subsection{Relic density of dark matter}

The relic density of a dark matter species $\chi$ with mass $m_\chi$ and annihilation cross section $\sigmav$ is set by $\TRH$ and $\Tdom$, and we determine this density by numerically integrating the Boltzmann equations in Ref.~\cite{erickcek2015dark}.  Figure~\ref{fig:relic} illustrates the ways dark matter in an EMDE cosmology can achieve the observed relic density $\rho_\chi/\rho_\mathrm{crit}=0.26$ today, where $\rho_\mathrm{crit}$ is the critical density.  In an EMDE scenario with no prior radiation (thick black curve), the annihilation cross section $\sigmav$ required to achieve the observed relic abundance depends strongly on $m_\chi$.  Dark matter freezes out from thermal equilibrium at a temperature $T_f$ that is approximately proportional to $m_\chi$, so higher $m_\chi$ means the dark matter freezes out earlier and consequently suffers more dilution by $\phi$ decay.  To compensate, $\sigmav$ must be smaller so that the dark matter freezes out at higher density.  However, $\sigmav$ can only become so small before the dark matter never achieves equilibrium in the first place.  Beyond this point, the dark matter is said to freeze in; further reducing $\sigmav$ now reduces the relic density, requiring smaller $m_\chi$ (later freeze-in) to achieve the observed abundance.  For $\TRH=10$~MeV, dark matter with $m_\chi> 100 \TRH$ suffers too much dilution to reach the observed abundance.

The presence of a prior radiation-dominated epoch changes the story considerably (colored curves in Fig.~\ref{fig:relic}).  While the $\phi$ decay sources radiation, this production remains subdominant to expansion-induced cooling until late in the EMDE at temperature $T\simeq \TRH^{4/5}\Tdom^{1/5}$.\footnote{Intuitively, newly sourced radiation may be viewed as remaining subdominant to prior radiation until late in the EMDE, although physically the two cannot be distinguished.}  Significant entropy production does not begin until that point.  Consequently, for $T_f>T_\mathrm{RH}$, there are three qualitatively different regimes for dark matter freeze-out.  If $T_f\lesssim \TRH^{4/5}\Tdom^{1/5}$, then the conditions required for dark matter to achieve the observed abundance are unaffected by the presence of prior radiation.  If instead $\TRH^{4/5}\Tdom^{1/5}\lesssim T_f\lesssim \Tdom$, then the dark matter experiences less dilution than if there were no prior radiation, so it can achieve the observed relic abundance for much larger $m_\chi$ than would be possible otherwise.  In this regime, the dark matter is diluted by the same factor regardless of its mass, but larger masses $m_\chi$ still require smaller $\sigmav$ to reach the observed abundance because of the influence of the dominant $\phi$ on the expansion rate.\footnote{$\phi$ domination boosts the expansion rate relative to the rate if only radiation were present, and this boost grows in time as $\phi$ becomes more dominant.  Faster expansion means $\sigmav$ must also be higher to achieve the observed relic abundance.  Heavier particles freeze out earlier, so they enjoy less of this boost to the expansion rate and require smaller $\sigmav$.}  Finally, if $T_f\gtrsim \Tdom$, then there is no $\phi$-induced boost to the expansion rate and all dark matter masses suffer the same dilution, so the required $\sigmav$ is independent of $m_\chi$.

\begin{figure}[t]
	\centering
	\includegraphics[width=\columnwidth]{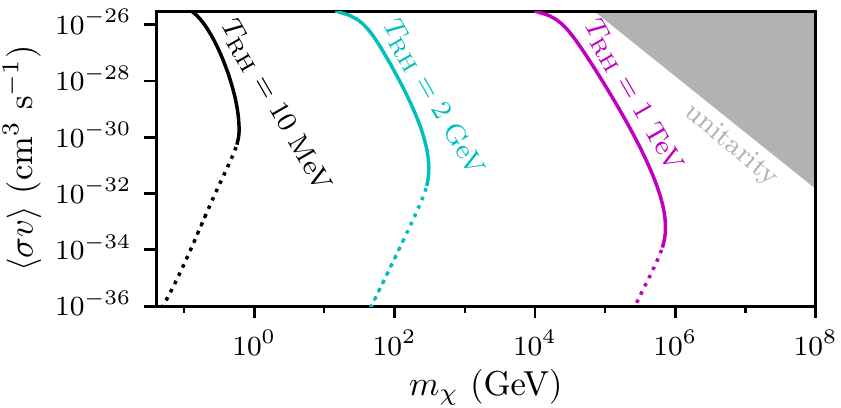}
	\caption{\label{fig:relic_TRH} Similar to Fig.~\ref{fig:relic}, but showing EMDE scenarios with different reheat temperatures $\TRH$ and $\Tdom\gg\TRH$ (or no prior radiation).  For a given $\TRH$, any dark matter candidate that lies to the right of the corresponding curve can be brought to the observed relic abundance by tuning $\Tdom$.}
\end{figure}

Evidently, by tuning $\Tdom$, any dark matter candidate that lies to the right of the $\Tdom\gg \TRH$ (black) curve in Fig.~\ref{fig:relic} can be brought to the observed relic abundance if $T_\mathrm{RH}=10$~MeV.  The story is similar with other reheat temperatures, and we show examples in Fig.~\ref{fig:relic_TRH}.  These figures clearly illustrate the breadth of the degeneracy between dark matter properties and the early thermal history.  The $\TRH>3$~MeV constraint limits $m_\chi\gtrsim 100$~MeV, but otherwise, almost any thermal relic with $\sigmav\lesssim 3\times 10^{-26}$~$\si{cm^3 s^{-1}}$ is viable.

\subsection{Growth of density perturbations}

When $\phi$ dominates, subhorizon dark matter density contrasts grow as $\delta\equiv\delta\rho_\chi/\bar\rho_\chi\propto a$, where $a$ is the scale factor, which is significantly faster than the $\delta\sim\log a$ \mbox{behavior} expected when radiation dominates.  Intuitively, the $\phi$ particles gravitationally cluster and carry the dark matter with them.  Reference~\cite{erickcek2011reheating} determined how the EMDE-boosted growth alters the power spectrum $\mathcal{P}(k)$ of dark matter density variations at later times.  $\mathcal{P}(k)$ is influenced by two main parameters: the reheat temperature $\TRH$ and the dark matter free-streaming scale, which sets a cutoff wave number $\kcut$.  The former is set by properties of the $\phi$ field, while the latter is determined by the microphysics of the dark matter, namely its interactions with relativistic particles and its residual velocity distribution \cite{green2004power,Loeb:2005pm,Profumo:2006bv,bertschinger2006effects,bringmann2007thermal,*bringmann2016erratum}.  The kinetic decoupling of dark matter during an EMDE is complicated by the entropy injected by $\phi$ decay \cite{Waldstein:2016blt}, but an EMDE generally leaves dark matter much colder than it would be in the EMDE's absence \cite{Gelmini:2008sh}.

\begin{figure}[t]
	\centering
	\includegraphics[width=\columnwidth]{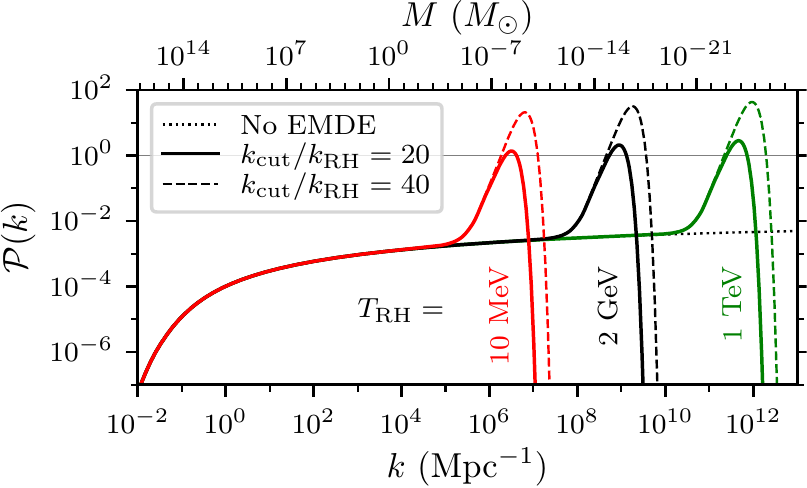}
	\caption{\label{fig:pk} The (dimensionless) power spectrum of dark matter density fluctuations at redshift $z=300$ in several EMDE cosmologies, as computed in linear theory using transfer functions from Ref.~\cite{erickcek2011reheating}.  Fluctuations are enhanced on the comoving scales $k$ that were inside the horizon during the EMDE, creating a ``bump'' in the power spectrum at small scales.  To supply intuition, we also plot the mass scale $M=(4\pi/3)\bar\rho k^{-3}$ associated with each wave number $k$, where $\bar\rho$ is the cosmological mean dark matter density; $M$ is of order the mass of the halo forming from a density variation of scale $k$.  In all scenarios fluctuations are already nonlinear (horizontal line) by $z=300$, implying microhalos have begun to form.}
\end{figure}

Figure~\ref{fig:pk} shows $\mathcal{P}(k)$ for several EMDE scenarios calculated using transfer functions from Ref.~\cite{erickcek2011reheating} as described in Ref.~\cite{erickcek2015dark}.  Fluctuations that were subhorizon during the EMDE are enhanced, so the reheat temperature $\TRH$ that marks the end of the EMDE sets the scales at which this enhancement occurs.  The smallest scales not suppressed by free streaming are enhanced the most; these are the modes near the wave number $\kcut$ associated with the dark matter's free-streaming cutoff.\footnote{As in Ref.~\cite{erickcek2015dark}, we define $\kcut$ such that the matter power spectrum is scaled by $\exp(-k^2/\kcut^2)$.}  The ratio $\xcut$ between this cutoff and the wave number entering the horizon at reheating is significant because it sets the maximum enhancement to $\mathcal{P}(k)$.

These power spectra were derived assuming no prior radiation.  To ensure their validity in scenarios with finite $\Tdom$, we require that modes entering the horizon prior to the EMDE, at temperature $T\gtrsim \Tdom$, lie below dark matter's free-streaming scale.  To be precise, we demand that the wave number $k_\mathrm{dom}$ entering the horizon at $\Tdom$ satisfy $k_\mathrm{dom} \gtrsim 3\kcut$, which ensures that the previous radiation epoch's imprint does not reduce the rms density variation in the dark matter by more than about 20\%.\footnote{We use the zero-baryon transfer function from Ref.~\cite{Eisenstein:1997ik} to approximate the power spectrum imprinted by an EMDE preceded by a radiation-dominated period, and we test how its rms density variation compares to that associated with a matter-only power spectrum if both have the same $\kcut$.  A 20\% decrease in the amplitude of a density contrast roughly corresponds to a factor of 2 drop in the corresponding collapsed halo's annihilation rate.}  This requirement is equivalent to
\begin{equation}\label{Tdom}
\xdom \gtrsim 5(\xcut)^{3/2}.
\end{equation}
Additionally, these power spectra assume that the dark matter froze out early enough before reheating that dark matter density perturbations were able to catch up to those in $\phi$.  Figure~4 of Ref.~\cite{erickcek2015dark} suggests that it takes roughly a factor of 5 in $a$, corresponding to a factor of 2 in $T$, for a dark matter density perturbation to settle into $\delta\propto a$ after freeze-out.  Thus, we demand
\begin{equation}\label{Tf}
T_f\gtrsim 2\TRH.
\end{equation}

\section{Microhalos and their properties}\label{sec:halos}

The density contrasts enhanced by an EMDE can collapse into dark matter microhalos long before the first halos would otherwise be expected to form, and their early formation makes these halos extremely small and dense.  In this work, we study scenarios with $20\leq \xcut\leq 40$, for which most microhalos form at redshift $200\lesssim z\lesssim 3000$.  Larger $ \xcut$ are theoretically plausible \cite{erickcek2016bringing}, especially for hidden-sector dark matter \cite{blanco2019annihilation}.  However, they enhance fluctuations enough to induce collapse prior to matter domination, the study of which is beyond this work's scope.\footnote{Sufficiently overdense regions can collapse when radiation dominates due to particle drift alone.  If these collapsed regions are locally matter dominated, they form bound halos long before matter-radiation equality; such early formation makes these halos much denser than any halo that forms during matter domination \cite{blanco2019annihilation}.  Regions that are even more overdense can collapse to form halos during an EMDE; these $\phi$-dominated halos gravitationally heat the dark matter so that after the $\phi$ decay destroys them, subsequent structure formation is suppressed \cite{blanco2019annihilation}.}  Additionally, we focus on EMDE scenarios with $\TRH=10$~MeV and $\TRH=2$~GeV: the former because it is close to the coldest reheat temperature possible without altering BBN, and the latter because it would bring certain supersymmetric dark matter candidates to the observed relic abundance (without assuming prior radiation) \cite{erickcek2016bringing}.  

The first microhalos form through the direct collapse of peaks in the density field, so they are expected to possess density profiles that scale as 
$\rho\propto r^{-3/2}$ at small radii \cite{ishiyama2010gamma,anderhalden2013density,*anderhalden2013erratum,ishiyama2014hierarchical,polisensky2015fingerprints,ogiya2017sets,delos2018ultracompact,delos2018density,angulo2017earth,delos2019predicting,ishiyama2019abundance}.  However, successive mergers drive their inner density cusps toward the shallower $\rho\propto r^{-1}$ scaling \cite{ogiya2016dynamical,angulo2017earth,gosenca20173d,delos2019predicting,ishiyama2019abundance}.  Thus, we will assume that microhalos eventually develop density profiles of the Navarro-Frenk-White (NFW) form,
\begin{equation}\label{NFW}
\rho(r) = \frac{\rho_s}{(r/r_s)(1+r/r_s)^{2}}
\end{equation}
which is a generic outcome of hierarchical halo clustering \cite{navarro1996structure,navarro1997universal}.

\subsection{Modeling the microhalo population}

We use the framework developed in Paper~I to characterize the microhalos that form after an EMDE.  This framework maps each peak in the (unfiltered) primordial density field to a collapsed halo, predicting the coefficient $A$ of that halo's inner $\rho=Ar^{-3/2}$ density asymptote, the radius $\rmax$ of maximum circular velocity, and the mass $\mmax$ that radius encloses.  We note that the model predicts a complete mass profile $M(r)$ for each halo, but this profile is only calibrated to agree with simulations at $r=\rmax$.  By sampling peaks from the density field as described in Appendix~C of Paper~I, we can thereby sample halos.  For simplicity, we use the turnaround model in Paper~I to predict $\rmax$ and $\mmax$, but as that work notes, alternate models yield very similar predictions.

We consider a variety of EMDE scenarios, and for each scenario we begin with the power spectrum of density fluctuations.  We first use the Boltzmann solver \textsc{CAMB Sources} \cite{lewis200721,challinor2011linear} to compute the power spectrum at redshift $z=500$ using Planck cosmological parameters \cite{aghanim2018planck}.  Subsequently, we apply the appropriate transfer function from Ref.~\cite{erickcek2015dark} to convert this power spectrum into one describing an EMDE scenario.  We then use this power spectrum to draw a halo population using the methods of Paper~I, but there is a complication.  Dark matter density contrasts $\delta$ at scales $k\gtrsim 10^2$ Mpc$^{-1}$ grow at the suppressed rate $\delta\propto a^{0.901}$ \cite{hu1996small}, where $a$ is the scale factor, because baryonic matter does not accrete into such small structures \cite{bertschinger2006effects}.  We describe in Appendix~\ref{sec:growth} how we adapt the Paper~I framework to this growth function.

Using the power spectrum, it is straightforward to apply the methods of Paper~I to predict the distribution of the asymptotic coefficient $A$.  However, the application to $\rmax$ and $\mmax$ demands some care since it requires sampling the profiles $\delta(q)$ of the precursor density peaks at a finite number of comoving radii $q$.  We sample the peaks at 300 radii distributed evenly in log space from $q=0.03\kcut^{-1}$ to $q=3\kRH^{-1}$; the minimum radius is well below the free-streaming cutoff, while the maximum radius is large enough to ensure that all EMDE-enhanced scales are sampled.  The advantage to using the turnaround model is that predictions are insensitive to the choice of maximum radius as long as the initial radius of the mass shell that collapses to $\rmax$ is sampled.

We additionally cut off the density profile of each peak when the average enclosed overdensity is either zero or begins to grow.  The former scenario implies that no farther mass shells are expected to accrete, while the latter suggests the presence of a denser neighboring structure; such a structure would cause the spherical collapse model that underlies the $\rmax$ and $\mmax$ predictions to break down.  The cutoff in the peak density profile imposes a cutoff in the predicted mass profile at the collapsed radius of the outermost mass shell, and if the circular-velocity maximum lies at that cutoff (implying it is not a local maximum), then we discard this halo; it is likely too rapidly accreted by a neighboring structure to be relevant.\footnote{We reanalyzed the EMDE scenario in Paper~I to test the impact of discarding peaks for which $\rmax$ lies beyond this cutoff in the mass profile.  Only 5\% of the peaks that matched to simulated halos satisfy the removal criterion, and of those halos, 87\% underwent a major merger during the simulation duration.}  Additionally, we discard peaks for which there is no ellipsoidal collapse solution using the approximation in Ref.~\cite{sheth2001ellipsoidal}.\footnote{Only highly aspherical peaks lack a collapse solution.  Increased deviation from spherical symmetry causes later collapse, so the absence of a solution may imply these peaks never collapse.  Spherical asymmetry is also anticorrelated with the amplitude of a density peak \cite{bardeen1986statistics}, so the collapse of highly aspherical peaks is delayed by both their small amplitude and their asphericity, implying that even if they do collapse they likely do so late enough that their contribution to observables is negligible.}

\begin{figure}[t]
	\centering
	\includegraphics[width=\columnwidth]{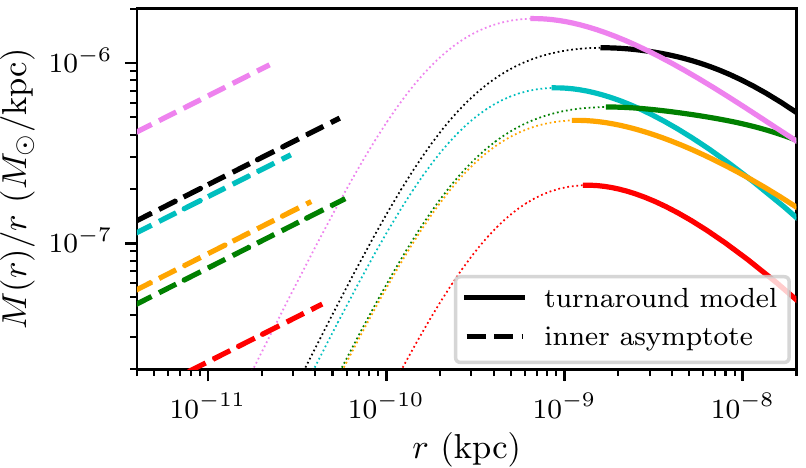}
	\caption{\label{fig:profiles} Mass profiles of six halos sampled using the methods of Paper~I in an EMDE scenario with $\TRH=2$ GeV and $\xcut=20$.  We separately show the inner asymptotes (dashed lines) and the greater mass profiles (solid lines).  The greater mass profiles are only tuned to match simulations at $r=\rmax$, which is where $M(r)/r$ is maximized.  Below $\rmax$ we plot them as dotted lines; since these profiles are derived assuming gradual mass accretion, they are not expected to be accurate at small $r$ because the mass there accreted rapidly.}
\end{figure}

We consider a variety of EMDE scenarios, and we use the above methods to sample $10^6$ peaks in each scenario and convert them into predicted halos.  As illustration, Fig.~\ref{fig:profiles} shows six example mass profiles $M(r)$ predicted for a cosmology with $\TRH=2$ GeV and $\xcut=20$.  We separately show the inner asymptote, set by the prediction of $A$, and the broader mass profile.  The latter is only tuned to be accurate at $\rmax$, which is where $M(r)/r$ peaks, and as Paper~I notes, it does not produce valid predictions at small $r$ because it is derived under the assumption that mass shells accrete adiabatically.

\subsection{Annihilation within microhalos}\label{sec:ann}

The dark matter annihilation rate within a microhalo is proportional to its $J$ factor,\footnote{We assume the dark matter annihilation cross section $\sigmav$ is velocity independent in the nonrelativistic limit.}
\begin{equation}\label{J}
J\equiv\int \rho^2 \diff V,
\end{equation}
and is largely set by the inner asymptote $A$ of its density profile.  If halos retain $\rho\propto r^{-3/2}$ inner density profiles, then the annihilation rate is proportional to $A^2$ if we neglect a logarithmic dependence on minimum radius\footnote{The annihilation rate from a $\rho\propto r^{-3/2}$ cusp diverges, which implies that the profile would shallow at some minimum radius due to these annihilations.} and maximum radius.  However, mergers between microhalos are expected to drive these cusps toward the $\rho\propto r^{-1}$ scaling of the NFW profile \cite{ogiya2016dynamical,angulo2017earth,gosenca20173d,delos2019predicting,ishiyama2019abundance}.  Mergers also deplete the microhalo count while making the survivors more dense.  To predict the annihilation signal from microhalos, it is necessary to account for these mergers' impact.

The precise impact of mergers between microhalos on their density profiles, and hence annihilation rates, is yet unclear.  However, we can make an estimate in the following way.  If a microhalo transitions from a profile with inner density asymptote $\rho=Ar^{-3/2}$ to the NFW profile with scale parameters $\rho_s$ and $r_s$, then $\rho_s^2 r_s^3\propto A^2$ from dimensional considerations.  Realistically, such a transition occurs simultaneously with mass increases caused by mergers, but we treat the two effects separately for simplicity.  The resulting NFW profile's $J$ factor is\footnote{We integrate the NFW profile to radius $\infty$, but the result is only marginally different if the profile is cut off at any radius $r\gtrsim r_s$.}
\begin{equation}\label{JA}
J=(4\pi/3) \omega A^2,
\end{equation}
where $\omega$ is the undetermined proportionality factor acquired in the transition, i.e., $\rho_s^2 r_s^3=\omega  A^2$.

We expect $\omega\gtrsim 1$ from the following examples.  If a density profile transitions from $\rho=Ar^{-3/2}$ to $\rho=\rho_s r_s/r$ with mass conserved within the scale radius $r_s$, then $\rho_s^2 r_s^3=(4/3)^2A^2$, i.e., $\omega=(4/3)^2\simeq 1.8$.  Alternatively, suppose a microhalo initially has density profile $\rho=Ar^{-3/2}(1+r/\tilde r)^{-3/2}$ \cite{delos2018ultracompact} or $\rho=Ar^{-3/2}[1+(r/\tilde r)^{3/2}]^{-1}$ \cite{moore1999cold} for some scale radius $\tilde r$, each of which appropriately obeys $\rho=Ar^{-3/2}$ when $r\ll \tilde r$.  If these profiles transition into NFW profiles while preserving the radius $\rmax$ of maximum circular velocity and the corresponding enclosed mass $\mmax$, then the transitions are characterized by $\omega=5.33$ and $\omega=8.05$, respectively.  We conservatively assume $\omega=(4/3)^2$ in line with the first example, but this factor carries cleanly through all calculations.

Paper~I found that the sum $\sum A^2$ over all halos is predicted by the model with reasonable accuracy even after mergers take place, which suggests that $J\propto A^2$ is additive in mergers.  This property can be understood in light of a conceptual argument.  For halos of fixed density profile (e.g., NFW), $J\propto \rho_s M$, where $M$ is halo mass.  If halo masses are additive in mergers and characteristic densities $\rho_s$ are not altered,\footnote{The central cusp can still become denser at a given radius when $\rho_s$ is fixed if $r_s$ grows.  Also, we assume that in mergers between halos of different $\rho_s$, the density of the merger remnant is the mass-weighted average of the densities of the progenitors.} then $J$ factors are also additive.  Indeed, Ref.~\cite{drakos2019major} found that in mergers between identical halos, $\rho_s$ tends to either be preserved or grow slightly, and $r_s$ grows roughly as would be expected from the doubling of mass.  This finding lends support to the notion that $J$ factors are approximately additive in mergers.

Further work is needed to tease out precisely how mergers alter the annihilation rates in halos, but we exploit the approximate conservation of the sum $\sum J$ to obtain an adequate estimate.  We compute the aggregate annihilation signal from microhalos by summing the $J$ factors given by Eq.~(\ref{JA}) over the previously predicted halo population.  If $N$ peaks are sampled to produce the halo population and $\bar n$ is the number density of peaks in the primordial density field (computed as in Paper~I or Ref.~\cite{bardeen1986statistics}), then the cosmologically averaged squared dark matter density within microhalos is
\begin{equation}\label{key}
\overline{\rho^2} = \frac{\bar n}{N}\sum_{i} J_i,
\end{equation}
where $J_i$ is the $J$ factor of the $i$th halo.  The mean gamma-ray luminosity expected from microhalos, per cosmological volume, is in turn 
\begin{equation}\label{Lmean}
\overline{\frac{\diff L}{\diff V}}=\overline{\rho^2}
\frac{\sigmav}{2m_\chi^2}
\int_{E_\mathrm{th}}^{m_\chi} E_\gamma\frac{\diff N_\gamma}{\diff E_\gamma} \diff E_\gamma,
\end{equation}
for threshold photon energy $E_\mathrm{th}$, where $m_\chi$ is the mass of the dark matter particle, $\sigmav$ is its annihilation cross section, and $\diff N_\gamma/\diff E_\gamma$ is the differential photon yield from annihilation events.

\begin{figure}[t]
	\centering
	\includegraphics[width=\columnwidth]{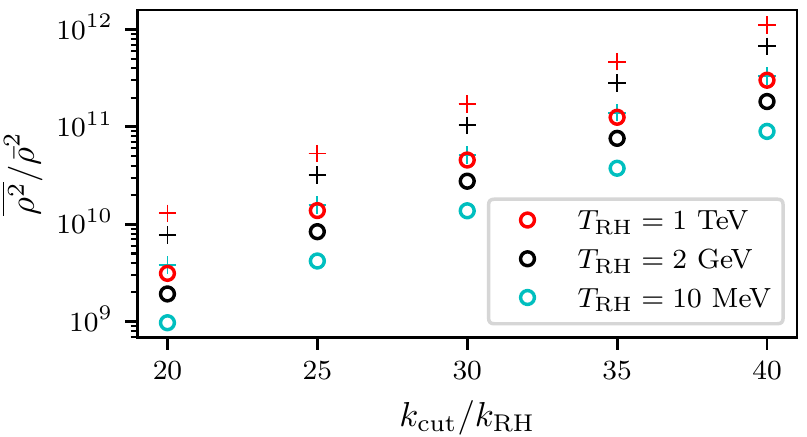}
	\caption{\label{fig:boost} The annihilation boost factor (circles) relative to uniform density as a function of EMDE scenario.  Annihilation rates are strongly sensitive to $\xcut$ and only weakly sensitive to $\TRH$.  For comparison, we also plot as crosses the annihilation boost factors computed using the procedure in Refs.~\cite{erickcek2015dark,erickcek2016bringing,blanco2019annihilation}, which uses Press-Schechter theory as described in the text.}
\end{figure}

The factor by which dark matter annihilation is boosted, relative to uniform density, is $\overline{\rho^2}/\bar\rho^2$, where $\bar\rho$ is the mean dark matter density.  In Fig.~\ref{fig:boost}, we plot this factor (as circles) as a function of EMDE scenario.  The annihilation boost is strongly sensitive to $\xcut$; this ratio sets the maximum enhancement to density variations, so it strongly influences halo formation times and hence the density within microhalos.  $\TRH$ also has a small influence on $\overline{\rho^2}/\bar\rho^2$ because it sets the duration of the last radiation-dominated epoch, during which the EMDE-enhanced density contrasts grow logarithmically.

For comparison, we also compute the annihilation boost factor using the procedure in Refs.~\cite{erickcek2015dark,erickcek2016bringing,blanco2019annihilation}.  Microhalos are counted using a Press-Schechter mass function, and each microhalo is assumed to have concentration $c\equiv r_\mathrm{vir}/r_s=2$ at some formation redshift $z_f$.  The annihilation boost factor is thus formally a function of $z_f$, but as the final step, $z_f$ is chosen such that the annihilation boost is maximized.  These estimates are plotted in Fig.~\ref{fig:boost} as crosses.  We find that this procedure overestimates annihilation rates by a factor of about 4 relative to the procedure described in this section.  This discrepancy likely owes to the time it takes for a halo to stabilize its NFW (or alternative $r^{-3/2}$) profile after formation.  A halo maintains $c\sim 2$ for a significant duration before its density profile (in physical coordinates) ceases to evolve, at which point $c$ begins to grow as the background density drops.  The assumption that $c$ begins to grow at halo formation consequently leads to overestimation of the halo's central density.

\subsection{Microhalo density profiles}\label{sec:profiles}

The procedure of the previous section suitably treats the impact of mergers between microhalos on their aggregate annihilation signals.  However, a subset of the microhalo population is further altered at late times ($z\lesssim 20$) by accretion onto much larger halos, such as those of galaxies.  Paper~II developed a model that predicts the evolution of subhalo $J$ factors due to a host halo's tidal forces, while Paper~III modeled the evolution of microhalos due to encounters with individual stars.  These models require the scale parameters $\rho_s$ and $r_s$ of the microhalo population, and in this section we use Paper~I's framework to estimate their distribution.  As a bonus, this distribution will assist in building an intuition for the microhalo population that the framework predicts.

For each halo, we obtain $\rho_s$ and $r_s$ from the structural parameters $\rmax$ and $\mmax$ predicted from the Paper~I framework by assuming an NFW density profile.  These parameters could be altered by mergers between microhalos, but Paper~I found that while mergers cause $\rmax$ and $\mmax$ to increase relative to their model predictions, this growth is relatively minor.\footnote{Paper~I found that as many as six major mergers only raised a halo's $\rmax$ by about 15\% relative to its model prediction.  $\mmax$ grew by as much as 50\% under the same conditions, but this change compensates the change in $\rmax$ to leave $\rho_s\propto\mmax/\rmax^3$ almost unaltered.  Note that $\rmax$ and $\mmax$ could be significantly altered by mergers even if they do not move appreciably relative to their predicted values; this would imply that the predictions already accounted for the mergers.}  Thus, lacking a precise understanding of how density profiles are influenced by mergers, we assume that the predictions of $\rho_s$ and $r_s$ remain accurate.  In this way, we obtain a distribution of microhalos in $\rho_s$ and $r_s$, and in the coming sections we will apply the dynamical evolution models of Papers II and~III to this distribution in order to predict the present-day aggregate annihilation rate.

\begin{figure}[t]
	\centering
	\includegraphics[width=\columnwidth]{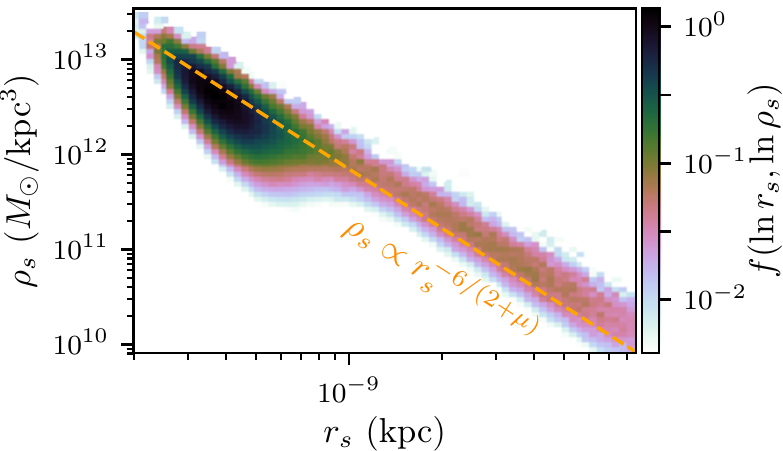}
	\caption{\label{fig:pop} Distribution of the annihilation signal in the NFW scale parameters $\rho_s$ and $r_s$ of source microhalos, as predicted using the framework in Paper~I.  The EMDE scenario represented has $\TRH=2$ GeV and $\xcut=20$.  The tail at large $r_s$ and small $\rho_s$ obeys $\rho_s\propto r_s^{-6/(2+\mu)}$, where $\mu=0.901$ is the linear growth exponent, and represents the rare halos that form from fluctuations much larger than the cutoff scale (see the text).  However, the dominant contribution to annihilation signals comes from the densest microhalos.}
\end{figure}

Figure~\ref{fig:pop} shows the microhalo distribution in $\rho_s$ and $r_s$, weighted by contribution to the annihilation signal, for an EMDE scenario with $\TRH=2$ GeV and $\xcut=20$.  The bulk of the microhalos form from density fluctuations near the free-streaming cutoff scale, and these halos comprise the dense clump in $r_s$-$\rho_s$ space depicted in Fig.~\ref{fig:pop}.  However, the halo distribution also includes a tail of halos with increasingly large $r_s$ and small $\rho_s$.  To understand this tail, we note that as a function of comoving scale $q$, initial density contrasts $\delta$ that enter the horizon during an EMDE, but are much larger than the free-streaming cutoff, scale as $\delta\propto q^{-2}$.  A microhalo's characteristic density is proportional to the background density at its formation time, so if a density contrast $\delta$ collapses to form a microhalo, then that microhalo has $\rho_s\propto \delta^{3/\mu}$, where $\mu=0.901$ is the linear growth exponent (see Appendix~\ref{sec:growth}).  Meanwhile, if $q$ is the comoving scale of that fluctuation, then the microhalo's characteristic size is $r_s\propto q\delta^{-1/\mu}$.  Consequently, for microhalos forming from density fluctuations that enter the horizon during an EMDE but are much larger than the free-streaming scale, $\rho_s\propto r_s^{-6/(2+\mu)}$.  The low-density halo tail in Fig.~\ref{fig:pop} follows this relationship.

\begin{figure}[t]
	\centering
	\includegraphics[width=\columnwidth]{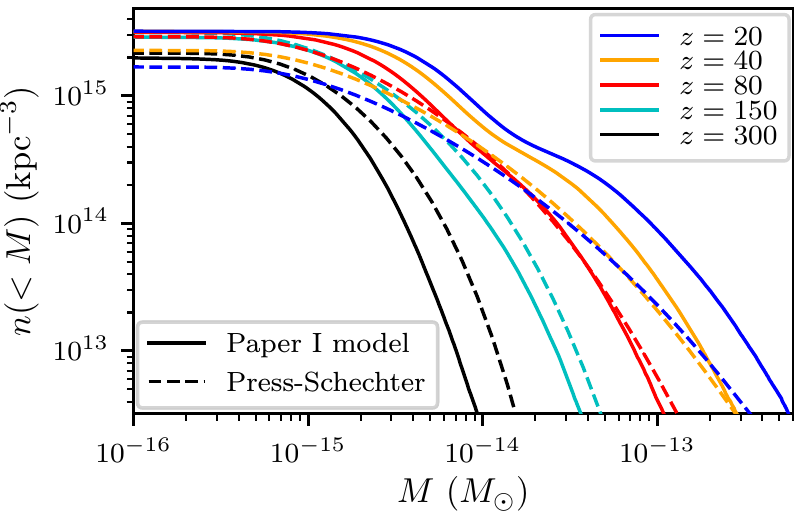}
	\caption{\label{fig:massfunc} Comparison between the halo population predicted using the methods of Paper I and that predicted by Press-Schechter theory.  We plot the number density $n(<M)$ of halos with mass smaller than $M$ at several redshifts as predicted by both methods for an EMDE cosmology with $\TRH=2$ GeV and $\xcut=20$.  The halo distributions predicted by the two methods are offset in mass by a factor as large as 2, but since the two frameworks predict different quantities, certain assumptions were necessary that may not be accurate (see the text).  Otherwise, the populations match reasonably well at early times.  At late times, the Paper I framework's neglect of halo mergers causes it to overpredict the halo count relative to the Press-Schechter calculation.}
\end{figure}

As another demonstration that the predicted population is sensible, we compare it to the population predicted by Press-Schechter theory \cite{press1974formation} using ellipsoidal collapse \cite{sheth2001ellipsoidal}.  Press-Schechter theory predicts a mass function $\diff n/\diff M$ describing the halo number density $n$ distributed in halo mass $M$.  In contrast, our halo population is distributed in the density profile parameters $\rho_s$ and $r_s$.  However, we can connect the two distributions by assuming $M=M_\mathrm{vir}$ is the mass enclosed within the virial radius $r_\mathrm{vir}$ that encloses an average density of 200 times the cosmological mean (which depends on the scale factor).  Meanwhile, given $\rho_s$ and $r_s$, the same virial mass $M_\mathrm{vir}$ can be estimated by assuming that the NFW profile is accurate out to $r_\mathrm{vir}$.  In Fig.~\ref{fig:massfunc}, we plot, for both methods and at several different redshifts, the number density $n(<M)$ of halos with mass smaller than $M$ in the EMDE scenario with $\TRH=2$ GeV and $\xcut=20$.  The halo distributions predicted by the two methods are offset in mass $M$ by up to a factor of 2, but this discrepancy is sufficiently minor and consistent that it is likely connected to the assumptions made above in order to compare the two prediction schemes.  Otherwise, the two populations match reasonably well at early redshifts $z\gtrsim 50$.  At late times, halo mergers start to become significant, causing the Paper I framework to overpredict the number of small halos relative to Press-Schechter.

\begin{figure}[t]
\centering
\includegraphics[width=\columnwidth]{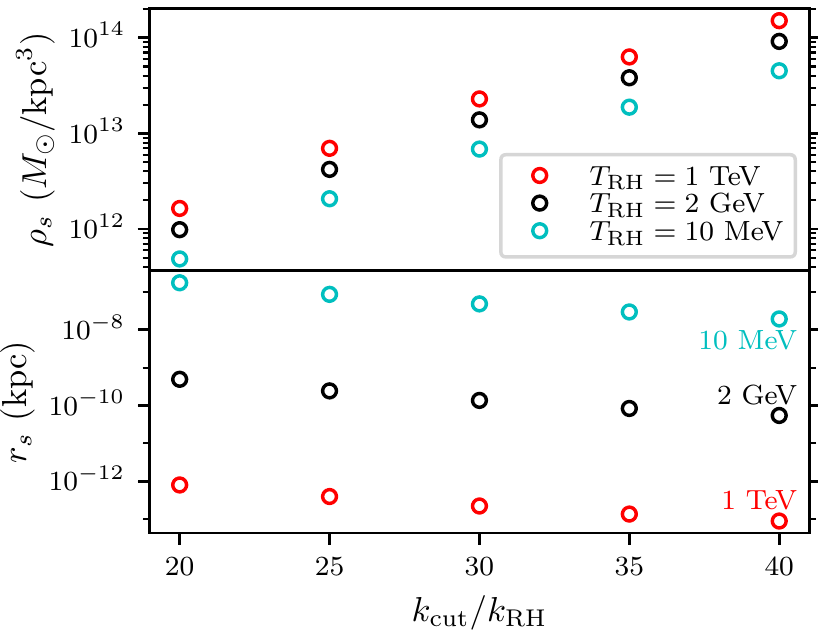}
\caption{\label{fig:microhalos} Median NFW scale density $\rho_s$ and radius $r_s$ of microhalos in a variety of EMDE scenarios.  The ratio $\xcut$ has a large impact on $\rho_s$ and a small impact on $r_s$, while the reverse is true of $\TRH$.}
\end{figure}

Finally, we discuss how the microhalo population is influenced by different EMDE cosmologies.  In Fig.~\ref{fig:microhalos}, we show the median scale density and radius values, $\rho_s$ and $r_s$, associated with a variety of EMDE scenarios.  As we noted in Sec.~\ref{sec:ann}, the ratio $\xcut$ has a large impact on $\rho_s$ because it sets the amplitudes of the most extreme density fluctuations, while $\TRH$ exerts only a minor influence by controlling the duration of the radiation-dominated epoch.  In contrast, $r_s$ is strongly sensitive to $\TRH$ because later reheating means larger-scale density fluctuations are enhanced, while $\xcut$ has a minor impact that owes to the fact that halos that form earlier, while the Universe was smaller, stabilize their density profiles at a smaller physical size.

\section{Isotropic gamma-ray background}\label{sec:igrb}

In an EMDE cosmology, dark matter annihilation within microhalos roughly traces the dark matter distribution, producing a signal similar to that of decaying dark matter.  Consequently, it would contribute substantially to the IGRB.  In this section, we use the Fermi Collaboration's measurement of the IGRB \cite{ackermann2015spectrum} to constrain annihilation within microhalos.

\subsection{Limits on annihilation}\label{sec:igrbann}

For a given annihilation channel, we can translate published bounds on the dark matter lifetime from the IGRB directly into bounds on the annihilation cross section using the procedure in Ref.~\cite{blanco2019annihilation}.  This translation is possible because annihilation within (unresolved) microhalos and dark matter decay both produce a gamma-ray signal that tracks the dark matter mass distribution.\footnote{The correspondence between annihilation within microhalos and decay breaks down at sufficiently high redshifts that the microhalos have not yet formed.  However, microhalos arising from the EMDE cosmologies we consider form at redshifts $z\gtrsim 200$, which lie well beyond the redshifts $z\lesssim 20$ relevant to any contribution to the IGRB from dark matter decay \cite{Cirelli:2009dv}.}  We equate the annihilation rate per mass, $\Gamma/M_\chi$, for particles with mass $m_\chi$ and cross section $\sigmav$ to the decay rate per mass of particles with mass $2m_\chi$ and effective lifetime $\tau_\mathrm{eff}$, obtaining
\begin{equation}\label{decay}
\frac{\Gamma}{M_\chi}=\frac{\sigmav}{2m_\chi^2}\frac{\overline{\rho^2}}{\bar\rho} = \frac{1}{2m_\chi\tau_\mathrm{eff}},
\end{equation}
where $\bar\rho$ and $\overline{\rho^2}$ are the mean and mean squared dark matter density, respectively, as in Sec.~\ref{sec:ann}.  Thus, a lower bound on $\tau_\mathrm{eff}$ for particles of mass $2m_\chi$ leads to an upper bound on $\sigmav$ for particles of mass $m_\chi$.  This procedure neglects disruption of microhalos within host halos, the impact of which we discuss soon.

We use Eq.~(\ref{decay}) to derive bounds on annihilating dark matter from two classes of limits on the dark matter lifetime.  Both employ the Fermi Collaboration's measurement of the IGRB \cite{ackermann2015spectrum}.  The first, from Ref.~\cite{liu2017constraints}, conservatively requires that the predicted gamma-ray flux from dark matter decay not exceed the flux reported by Fermi within any spectral bin.  The second, from Ref.~\cite{blanco2019constraints}, employs models of astrophysical background gamma-ray sources, such as star-forming galaxies and active galactic nuclei, to dramatically reduce the gamma-ray flux that can be attributed to dark matter.  Through Eq.~(\ref{decay}), these decay bounds (for particles of mass $2m_\chi$) translate into conservative and aggressive bounds on the dark matter annihilation cross section $\sigmav$ (for particles of mass $m_\chi$), respectively.

\begin{figure}[t]
	\includegraphics[width=\columnwidth]{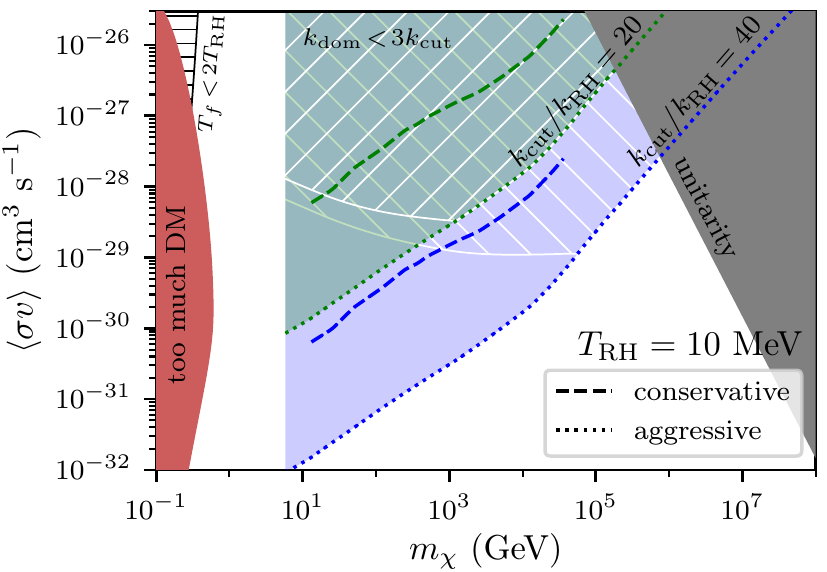}\\
	\includegraphics[width=\columnwidth]{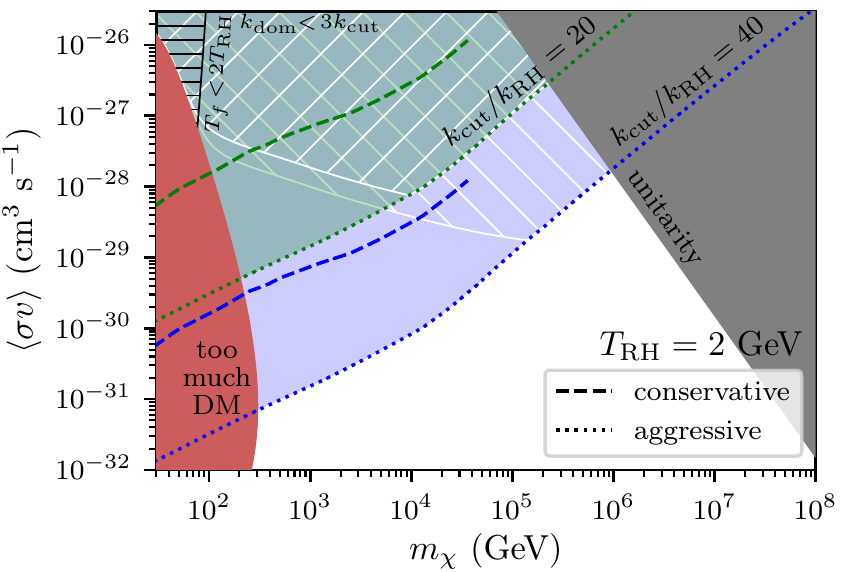}
	\caption{\label{fig:constraint_igrb} Upper bounds on the cross section for dark matter annihilating to $b\bar b$ for two reheat temperatures $\TRH=10$~MeV (top) and $\TRH=2$~GeV (bottom).  In each case, we consider both $\xcut=20$ (green) and $\xcut=40$ (blue) and plot both the conservative and aggressive bounds derived from Fermi-LAT's measurement of the IGRB; see the text.  The shaded region on the left is disallowed because it would overclose the Universe, while the shaded region on the right marks where the dark matter's coupling constant exceeds unity.  The black hatched region fails Eq.~(\ref{Tf}) while the white hatched regions (different for each $\xcut$) fail Eq.~(\ref{Tdom}).  The density fluctuation power spectra we employed do not apply within these regions, so constraints therein are tentative; further work is needed to account for the altered power spectra.}
\end{figure}

In Fig.~\ref{fig:constraint_igrb}, we plot the resulting constraints on dark matter annihilating into $b\bar b$ for several EMDE cosmologies.  Recall from Sec.~\ref{sec:EMDE} that dark matter with a vast range of parameters can be brought to the observed relic abundance by tuning $\TRH$ and $\Tdom$; the main requirement is that $m_\chi$ must exceed $\TRH$ by a sufficient margin to avoid overproducing dark matter for cross sections below the canonical $3\times 10^{-26}$~$\si{cm^3 s^{-1}}$.  If $\xcut=40$, the aggressive constraints can probe thermal-relic cross sections as small as $\sigmav\sim 10^{-32}$~$\si{cm^3s^{-1}}$.  Constraints are weaker for $\xcut=20$, but the aggressive constraints can still reach as far as $\sigmav\sim 10^{-30}$~$\si{cm^3s^{-1}}$.

Equations (\ref{Tdom}) and~(\ref{Tf}) set the conditions under which the density fluctuation power spectra we employed are expected to be accurate, and we mark on Fig.~\ref{fig:constraint_igrb} the regions that fail these conditions.  Only a small region with large $\sigmav$ and small $m_\chi$ fails $T_f>2\TRH$ (without also overproducing dark matter).  On the other hand, cross sections $\sigmav$ that are less than a few orders of magnitude below the canonical $3\times 10^{-26}$~$\si{cm^3 s^{-1}}$ tend to fail $\xdom > 5(\xcut)^{3/2}$ if $\Tdom$ is tuned to effect the observed relic abundance.  Constraints on thermal relics are tentative within these regions, and a more careful treatment is needed of the power spectra that arise therein.  Using the appropriate power spectra would delay microhalo formation, weakening the bounds on $\sigmav$.

\subsection{Tidal suppression}\label{sec:igrbtidal}

Compared to the gamma-ray signal from decaying dark matter, the signal from annihilation within microhalos is suppressed by tidal effects within host halos.  We neglected this effect when deriving the constraints shown in Fig.~\ref{fig:constraint_igrb}, and we now estimate its impact.  Dark matter's contribution to the IGRB comes from both the Galactic halo and extragalactic dark matter.  To estimate the tidal influence of extragalactic host halos, we use Press-Schechter theory with ellipsoidal collapse \cite{sheth2001ellipsoidal} to model the host halo mass function $\diff n/\diff  M$, where $n$ is the number density of hosts.  We exclude EMDE-boosted microhalos (as hosts) by only considering halos down to the mass scale associated with the reheating wave number $\kRH$.  We associate a concentration $c=R_\mathrm{vir}/R_s$ to each halo mass using the median concentration-mass relation $c(M)$ from Ref.~\cite{diemer2019accurate}, and we define $R_\mathrm{vir}$ as the radius enclosing 200 times the critical density to match that work.

With the host-halo population constructed, the next step is to calculate the suppression of microhalo annihilation signals within each host.  In Appendix~\ref{sec:sup}, we use the model in Paper II to compute the factor $S$ by which tidal evolution scales the aggregate annihilation rate within subhalos distributed throughout the host's phase space.  A convenient fitting function for $S(\rho_s/P_s,t\sqrt{GP_s},c)$ is presented in that appendix, where $\rho_s$ is the scale density of the microhalos,\footnote{$S$ can be averaged over a distribution of $\rho_s$, weighting by $J$ factors, to accommodate a distribution of microhalos.} $P_s$ is the scale density of the host, $c=R_\mathrm{vir}/R_s$ is the host's concentration, and $t$ is the duration of tidal evolution.  At a given redshift $z$, the host scale density is a function $P_s(c,M)$ of mass and concentration.  If we let $\bar S(\rho_s,t)$ be the global factor by which the annihilation rate within microhalos is scaled due to tidal evolution for the duration $t$, then
\begin{equation}\label{Sglob}
\bar S(\rho_s,t)=1-\frac{1}{\bar\rho}\int\!\diff M \left[1\!-\!S\!\left(\frac{\rho_s}{P_s},t\sqrt{GP_s},c\right)\right]M\frac{\diff n}{\diff  M}
\end{equation}
(with $P_s$ and $c$ functions of $M$).

\begin{figure}[t]
	\includegraphics[width=\columnwidth]{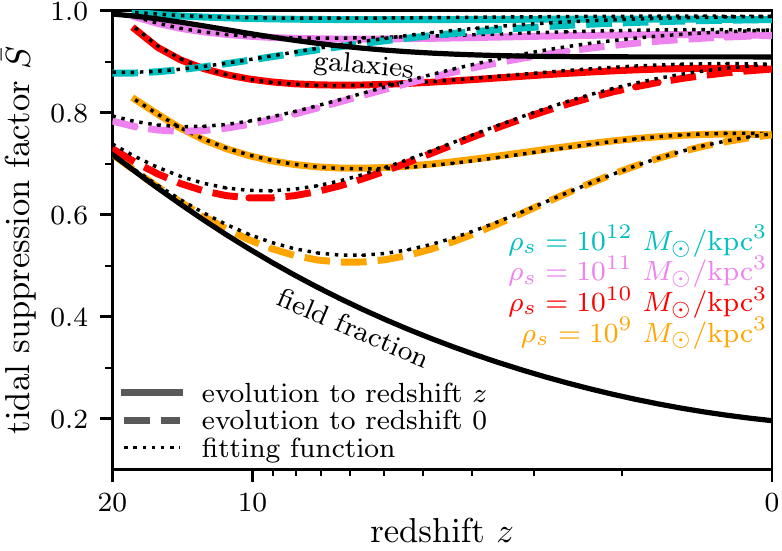}
	\caption{\label{fig:igrbsupp} Global scaling factor $\bar S$ for dark matter annihilation within microhalos of scale density $\rho_s$ due to tidal effects from host halos.  The host population is characterized with a Press-Schechter mass function at each redshift $z$.  Solid lines assume tidal evolution from redshift 20 to $z$, while dashed lines assume tidal evolution from $z=20$ to $z=0$ regardless of the redshift used to characterize the hosts.  We expect that the minimal value of $\bar S$ under the latter calculation is a reasonable estimate for the true tidal scaling factor today; see the text.  Dotted lines show the results using the fitting function in Appendix~\ref{sec:sup} instead of integrating the individual tidal scaling factors from Paper II over host halos' phase spaces as described in that appendix.  The lower black line indicates the fraction of microhalos that are not subhalos.  The upper black line marks the scaling of microhalo annihilation rates that results from an extreme estimate of disruption within galaxies.  For values $\rho_s\geq 10^{12}$ $M_\odot$/kpc$^3$ relevant to the EMDE scenarios we consider, all tidal suppression estimates are marginal.}
\end{figure}

We set the tidal evolution duration to be the time elapsed since $z=20$; this choice only marginally affects our results since any reasonable duration is essentially the age of the Universe.  Figure~\ref{fig:igrbsupp} shows $\bar S$ as a function of $z$ (solid lines) for several different $\rho_s$.  A limitation of this calculation, however, is that at any given time, microhalos are assumed to have resided within their current host since $z=20$.  Additionally, subhalos are neglected as possible hosts.  The largest host halos become less dense as time goes on, so these deficiencies explain why we improperly find that microhalo annihilation signals become less suppressed ($\bar S$ grows) over time even as the microhalos experience uninterrupted tidal evolution.

To approximately account for this effect, we carry out the same calculation except that at each redshift $z$, we assume the full duration of tidal evolution (from $z=20$ to $z=0$) instead of only assuming tidal evolution up until the redshift $z$.  This procedure means we can assume the microhalos continue to reside within the dense host population that existed at some redshift $z>0$ even if those hosts later accrete onto superhosts.  On the one hand, this procedure could underestimate the tidal suppression because the smaller number and size of host halos at high redshift implies that fewer microhalos are within hosts at all (see the field fraction curve in Fig.~\ref{fig:igrbsupp}).  On the other hand, it could overestimate the tidal suppression because we assume microhalos continue to reside within their dense original hosts even though many microhalos would be stripped onto a less dense superhost.  In Fig.~\ref{fig:igrbsupp}, we plot, as dashed lines, $\bar S$ computed using this procedure for several $\rho_s$.  Here, the redshift $z$ sets the host-halo population only.  We expect that the minimal value of $\bar S$, as a function of $z$, will be a reasonable estimate for the global tidal scaling factor.  For $\rho_s\geq 10^{12}$ $M_\odot/\si{kpc^3}$, $\bar S\gtrsim 0.9$, and as Fig.~\ref{fig:microhalos} indicates, $10^{12}$ $M_\odot/\si{kpc^3}$ is the median $\rho_s$ for $\xcut=20$ (with the effect of $\TRH$ being marginal).  If weighted by contribution to the annihilation signal, the average density would be even higher.  Thus, we conclude that the contribution to the IGRB from annihilation within extragalactic microhalos at $z\simeq 0$ is suppressed by less than 10\% due to tidal effects if $\xcut\gtrsim 20$.  Annihilation within high-redshift microhalos---which also contributes significantly to the IGRB---would be even less suppressed.

Microhalos can be also disrupted by baryonic structure within halos.  As a simple estimate, we assume that galaxies span their halos' scale radii; this is roughly true of the Milky Way and of the Draco dwarf (see Sec.~\ref{sec:draco}).  We further assume that any microhalo within its host's galactic extent, defined in this way, is destroyed; this is an extreme estimate, since as we see in Sec.~\ref{sec:draco}, annihilation from microhalos within the Draco dwarf is only slightly suppressed by encounters with stars.  Finally, we assume any halo larger than $10^5 M_\odot$ forms a galaxy.  The upper solid line in Fig.~\ref{fig:igrbsupp} shows the global scaling factor $\bar S$, as a function of redshift $z$, evaluated using these rules on the host-halo populations computed earlier.  We find that even in this extreme picture, dark matter annihilation within microhalos is only suppressed by about 10\% due to galactic disruption.  Neither tidal forces from host halos nor disruption due to galaxies significantly reduces extragalactic microhalos' contribution to the IGRB.

\begin{figure}[t]
	\includegraphics[width=\columnwidth]{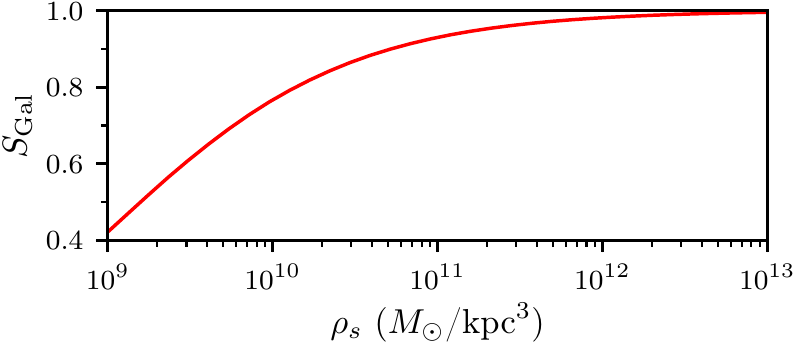}
	\caption{\label{fig:galsup} Tidal suppression of the contribution to the IGRB from microhalos within the Galactic halo.  We plot, as a function of microhalo scale density $\rho_s$, the factor $S_\mathrm{Gal}$ by which annihilation rates along a line of sight are scaled.  For values $\rho_s\geq 10^{12}$ $M_\odot$/kpc$^3$ relevant to the EMDE scenarios we consider, tidal effects reduce the annihilation rate by less than 2\%.}
\end{figure}

Finally, we estimate the tidal suppression of the contribution to the IGRB from microhalos within the Galactic halo.  To match the assumptions made in Ref.~\cite{blanco2019constraints}, we assume the Milky Way halo has an NFW density profile with scale radius 20 kpc and scale density set so that the local dark matter density at radius 8.25 kpc is 0.4 GeV/cm$^3$.  We integrate the tidal scaling factors as in Appendix~\ref{sec:sup}, but instead of integrating over the Milky Way's volume, we integrate along the line of sight perpendicular to the Galactic plane out to the 300-kpc virial radius.  The precise angle makes little difference.  The resulting tidal scaling factor $S_\mathrm{Gal}$ is plotted in Fig.~\ref{fig:galsup} as a function of microhalo scale density $\rho_s$.  For $\rho_s\geq 10^{12}$ $M_\odot$/kpc$^3$, the tidal suppression is negligible.

\section{Gamma rays from the Draco dwarf}\label{sec:draco}

Dwarf spheroidal galaxies represent some of the most promising targets for dark matter annihilation searches due to their high dark matter density and low astrophysical contamination \cite{strigari2007precise}.  Since the signal from annihilation within microhalos is similar to that from dark matter decay, we focus on the Draco dwarf, which among dwarf galaxies supplies the strongest constraints on the dark matter lifetime \cite{baring2016new}.  However, unlike the decaying dark matter signal, the signal from annihilation within microhalos is altered by the influence of tidal forces from the host halo and high-speed encounters with other microhalos.  Due to their small size, microhalos are also susceptible to encounters with individual stars.  We must account for these effects in order to characterize the microhalo-dominated annihilation signal from Draco.

To characterize Draco's dark matter halo, we assume its maximum circular velocity is $v_\mathrm{max}=18.2$ km/s \cite{martinez2015robust} and its density at radius 150 pc is $\rho(150\ \text{pc})=2.4\times 10^8$ $M_\odot/\text{kpc}^3$ \cite{read2018case}.  If the halo has an NFW density profile, then these constraints imply it has scale radius $R_s=0.435$ kpc and scale density $P_s=1.5\times 10^8$ $M_\odot/\text{kpc}^3$ (we use capital letters to distinguish these parameters from those of the microhalos).  While there is evidence that many galactic halos possess uniform-density cores instead of the NFW profile's cusp (e.g., Ref.~\cite{moore1994evidence}), Draco's halo appears to be cuspy \cite{read2018case}.  We assume the microhalos trace Draco's density profile with an isotropic velocity distribution.

As for its stellar content, we assume Draco has stellar luminosity $2.7\times 10^5 L_\odot$, projected half-light radius $0.22\ \text{kpc}$ \cite{martin2008comprehensive}, and a stellar mass-to-light ratio of 1.8 $M_\odot/L_\odot$ \cite{woo2008scaling}.  Additionally, we adopt a Plummer density profile \cite{plummer1911problem} for its stellar mass.  To model the masses of individual stars, we employ a Kroupa initial mass function with minimum mass $0.01M_\odot$ (we include brown dwarfs) and a high-mass index of 2.7 \cite{kroupa2002initial}.  Since Draco largely ceased star formation 10 Gyr ago \cite{aparicio2001star}, roughly the lifetime of a 1 $M_\odot$ star, we impose a maximum stellar mass of 1 $M_\odot$.  However, we still wish to include the white dwarf remnants of dead stars, so we assume any star initially heavier than 1 $M_\odot$ weighs 1 $M_\odot$.\footnote{\label{foot:mstar}This treatment is approximate; white dwarf masses vary and stellar masses change over their lifetimes.  However, the following scaling argument shows (and we verified) that the choice of star masses $M_*$ has little impact on microhalo evolution.  The energy injected by an encounter with impact parameter $b$ scales as $\Delta E\propto M_*^2 b^{-4}$.  Due to this scaling, the total energy injected by all encounters is dominated by the closest few encounters, which have $b\propto n_*^{-1/2}$.  But at fixed stellar mass density, $n_*\propto M_*^{-1}$.  Together, these relationships imply $\Delta E\propto M_*^0$.}  Finally, using the mean of the stellar mass distribution, we are able to fix the stellar number density profile $n_*(R)$.

\subsection{Suppression of annihilation rates}\label{sec:dracoJ}

Due to their small size, we expect microhalos to have essentially the same phase-space distribution as dark matter particles within Draco's halo.  Thus, at each radius $R$ within Draco, we sample 200 microhalo orbits using the isotropic distribution function given in Ref.~\cite{widrow2000distribution}.  To efficiently apply the models in Papers II and III to full distributions of microhalos, we use these sampled orbits to construct, at each radius, an interpolation table in $\rho_s$ for the orbit-averaged factor $J/J_\mathrm{init}$ by which annihilation within microhalos is suppressed.  We consider $\rho_s$ alone because the scale radius $r_s$ has no impact on tidal evolution, and we verified that for relevant microhalo parameters, $r_s$ also has no impact on the evolution by stellar encounters.  Subsequently, we use this interpolation table to find $J/J_\mathrm{init}$ for the full distribution of microhalos.  We then average $J/J_\mathrm{init}$ over all microhalos, weighting each halo by its initial $J$ factor given by Eq.~(\ref{JA}).

\begin{figure}[t]
	\includegraphics[width=\columnwidth]{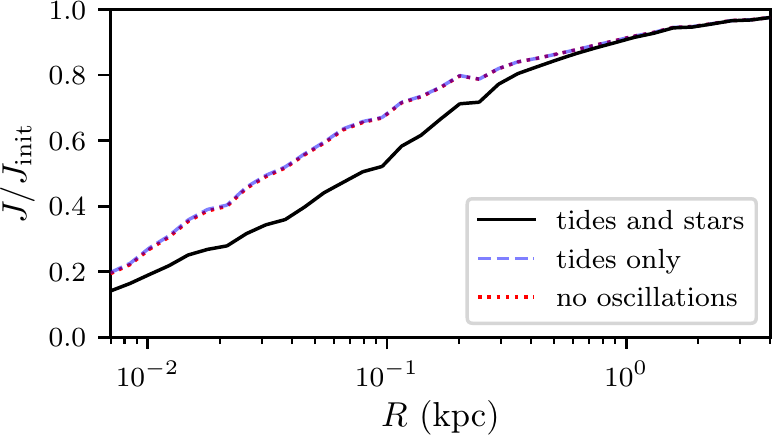}
	\caption{\label{fig:suppression} Suppression of dark matter annihilation rates in microhalos as a function of radius $R$ within the Draco dwarf.  At each radius we average over 200 randomly sampled orbits.  Additionally, for each orbit, we average over 200 randomly sampled stellar encounter histories.   The EMDE scenario represented has $\TRH=2$ GeV and $\xcut=20$.}
\end{figure}

It is straightforward to apply the model in Paper~II to each microhalo orbit to determine the suppression of the $J$ factor due to tidal forces from Draco's halo.  In Fig.~\ref{fig:suppression}, we show, as a function of radius within Draco, this model's prediction of $J/J_\mathrm{init}$ (dashed line) averaged as described above over the microhalo distribution.  Microhalo $J$ factors also oscillate over each orbital period, becoming largest near pericenter, and the model in Paper~II does not account for these oscillations since they do not affect the magnitude of an annihilation signal.  However, they can change the signal's morphology since they systematically bias it toward smaller radii.  In Appendix~\ref{sec:rmodel}, we use the simulations from Paper~II to explore these oscillations and present a simple model for their impact.  The results of this model are included in the scaling factor due to tidal forces depicted in Fig.~\ref{fig:suppression}, but we also show as dotted lines the scaling if these oscillations are neglected.  Evidently, their impact is negligible for the host-subhalo parameters relevant to our scenario.

Papers II and III describe how to account for tidal forces and stellar encounters separately.  However, it is not obvious how to combine the two effects.  To determine the appropriate procedure, we carry out additional simulations using the same procedures as Papers II and III.  In these simulations, a microhalo experiencing tidal forces is also subjected to stellar encounters.  Appendix~\ref{sec:tidestar} presents the model we build to describe this scenario.  The simulations indicate that it makes little difference when the stellar encounter occurs, so our model is conceptually based on the idea of applying tidal evolution first and stellar encounters afterward.

To determine the impact of stellar encounters using the framework of Paper~III and Appendix~\ref{sec:tidestar}, we must randomly sample stellar encounter histories for each microhalo orbit.  Let $f(R;V_*)$ be the stellar velocity distribution at radius $R$, which we assume to be Maxwell-Boltzmann with velocity dispersion equal to that of the dark halo at the same radius.  As before, we take $n_*(R)$ to be Draco's stellar number density profile.  We also take $V_\mathrm{h}(R)$ and $V_{r,\mathrm{h}}(R)$ to be the microhalo's total velocity and the radial component of its velocity, respectively, which depend on the orbit.  With these definitions, the differential number of stellar encounters per radius $R$, impact parameter $b$, stellar velocity $V_*$, and cosine $\mu\equiv\cos\theta$ of the angle between $V_\mathrm{h}$ and $V_*$ is
\begin{equation}\label{dNdRdb}
\frac{\diff^4 N_\mathrm{enc}}{\diff b\diff R\diff V_*\diff\mu}
=\pi b\frac{V_\mathrm{rel}[V_\mathrm{h}(R),V_*,\mu]}{V_{r,\mathrm{h}}(R)}n_*(R)f(R;V_*),
\end{equation}
where
\begin{equation}\label{Vrel}
V_\mathrm{rel}(V_\mathrm{h},V_*,\mu)\equiv \left[(V_\mathrm{h}\!-\!\mu V_*)^2\!+\!(1\!-\!\mu^2)V_*^2\right]^{1/2}
\end{equation}
is the relative velocity between the halo and the star.  We use Eq.~(\ref{dNdRdb}) to sample the stellar encounters for each orbit, evaluating the $\mu$ integral analytically and using a Markov chain Monte Carlo method \cite{Foreman_Mackey_2013} to sample $R$ and $V_*$.  For each encounter, we sample the stellar mass $M_*$ from the modified Kroupa distribution described earlier.  We assume that each orbit proceeds for 13.6 Gyr, roughly the time since $z=20$, and for each orbit we sample 200 encounter histories.  We determine the impact of stellar encounters for each history using the models in Paper~III and Appendix~\ref{sec:tidestar}, and Fig.~\ref{fig:suppression} shows the resulting $J$-factor suppression averaged over orbits and encounter histories.  We include the effect of the $J$-factor oscillations described in Appendix~\ref{sec:rmodel}, although, as noted above, their impact is negligible.

Finally, we explore the impact of microhalo-microhalo encounters.  We will see that their influence on microhalo $J$ factors is marginal.  For concreteness, we consider a microhalo on a circular orbit at Draco's 0.22-kpc half-light radius, and for its density profile we adopt separately the median parameters $\rho_s$ and $r_s$ associated with the reheat temperatures $\TRH=10$~MeV and 2~GeV and ratios $\xcut=20$ and 40.  We model the microhalos it encounters as point objects so that we can compute the energy each encounter injects similarly to stellar encounters; note that energy injections would only decrease if we were to model the microhalos as extended objects.  These microhalos are distributed along the density profile $P(R)$ of Draco's halo; i.e., the number density profile of microhalos is $n P(R)/\bar\rho$, where $n$ is the cosmological mean number density of microhalos.  We leave $n$ as a free parameter for now, although it is related to the known number density $n_\mathrm{peak}$ of peaks in the primordial density field.  Finally, to fix the masses $M$ of microhalos, we assume that they contain a fraction $f$ of the dark matter, so $M=f\bar\rho/n$.  Tidal effects likely reduce $f$ far below 1 inside Draco by stripping microhalos' massive but weakly bound outskirts, and we generously assume $f=0.25$.

\begin{figure}[t]
	\centering
	\includegraphics[width=\columnwidth]{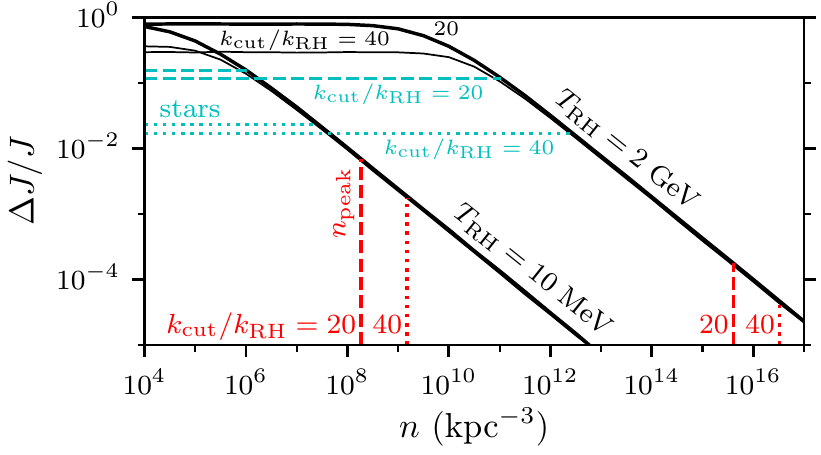}
	\caption{\label{fig:mhmh} Impact of encounters with other microhalos on a microhalo with the median density profile parameters $\rho_s$ and $r_s$ associated with the displayed $\TRH$ and $\xcut$.  This microhalo orbits Draco circularly with radius 0.22 kpc.  Black lines show the fractional change $\Delta J/J$ in the microhalo's $J$ factor due to all encounters over Draco's age; this ratio is averaged over $10^4$ random encounter histories and plotted as a function of the cosmological mean number density $n$ of microhalos.  We assume 25\% of dark matter is in microhalos.  As a reference, the number density $n_\mathrm{peak}$ of peaks in the primordial density field is marked for the relevant EMDE cosmologies with vertical lines; mergers deplete $n$ relative to $n_\mathrm{peak}$, but likely by much less than an order of magnitude.  Evidently, the impact of encounters with microhalos is negligible as long as $n$ is not much smaller than $n_\mathrm{peak}$.  For comparison, we also mark with horizontal lines the $\Delta J/J$ induced by stellar encounters.  For horizontal and vertical lines, the length indicates which $\TRH$ curve it matches.}
\end{figure}

Using a similar procedure to our treatment of stellar encounters above, we derive the fractional change $\Delta J/J$ in the subject microhalo's $J$ factor caused by microhalo encounters over Draco's age.  We seek only the relative impact of microhalo encounters, so we do not combine them with tidal evolution as in Appendix~\ref{sec:tidestar}, although as prescribed in that appendix we sum the encounters' energy injections instead of applying the encounters consecutively.  Figure~\ref{fig:mhmh} plots $\Delta J/J$ as a function of the mean number density $n$ of microhalos; smaller $n$ means more massive microhalos, which outweighs their reduced numbers for most $n$.\footnote{\label{foot:mhstar} The story is different for stars; increased number density there approximately compensates reduced mass (e.g., footnote~\ref{foot:mstar}).  This contrast is a consequence of the very different regimes that stellar and microhalo encounters occupy.  When the closest encounter is typically farther than the subject microhalo's scale radius $r_s$, as is the case with stellar encounters, the total energy injection is dominated by the closest few encounters due to the $\Delta E\propto b^{-4}$ scaling with impact parameter $b$.  As $b$ drops below $r_s$, this scaling shallows toward $\Delta E\propto b^0$ \cite{moore1993upper,van2017disruption}, so when there are many encounters with $b<r_s$, the total energy injection is broadly distributed across all such encounters.  The latter case applies to microhalo encounters with sufficiently large $n$.}  For each EMDE scenario, we also indicate $n_\mathrm{peak}$.  Evidently, $\Delta J/J\ll 1$ as long as $n$ is not smaller than $n_\mathrm{peak}$ by more than an order of magnitude.  While mergers cause $n < n_\mathrm{peak}$, they depleted $n$ by less than a factor of 2 by $z=50$ in the EMDE scenario simulated by Paper~I.  For this reason, and because the assumptions made in this calculation were broadly biased toward overestimating their impact, we conclude that microhalo-microhalo encounters can be neglected.  For comparison, we also mark in Fig.~\ref{fig:mhmh} the impact of stellar encounters on the same microhalo.

\subsection{Gamma-ray emission}\label{sec:dracogamma}

We denote by $S(R)$ the orbit-averaged tidal suppression of the $J$ factor, as a function of radius $R$ about Draco, computed in the last section.  With this quantity, we can predict the annihilation signal from microhalos within Draco.  If $P(R)$ is the density profile of Draco's halo and $\bar\rho$ is the cosmological mean dark matter density, then
\begin{equation}\label{Lhost}
\frac{\diff L}{\diff V}=\overline{\frac{\diff L}{\diff V}}\frac{P(R)}{\bar \rho} S(R)
\end{equation}
is the gamma-ray luminosity, per volume, from dark matter annihilation at radius $R$.  Here, $\overline{\diff L/\diff V}$ is the cosmological mean value given by Eq.~(\ref{Lmean}).  By integrating this emission over the line of sight, we obtain the differential flux per solid angle
\begin{equation}\label{flux}
\frac{\diff F}{\diff \Omega}=\frac{1}{4\pi}\int_{-x_\mathrm{max}}^{x_\mathrm{max}}\diff x \left.\frac{\diff L}{\diff V}\right|_{R=\sqrt{x^2+R_\mathrm{proj}^2}}
\end{equation}
at projected radius $R_\mathrm{proj}$ from Draco's center, where $x_\mathrm{max}\equiv \sqrt{R_\mathrm{max}^2-R_\mathrm{proj}^2}$ and $R_\mathrm{max}$ is the boundary radius of Draco's halo, which we fix shortly.

The gamma-ray flux from microhalos exhibits significant sensitivity to Draco's density profile at large radii, which cannot be constrained by stellar kinematics beyond the 1.9-kpc radius of its most distant observed star \cite{geringer2015dwarf}.  The density profile at large radii would be suppressed by tidal forces from the Milky Way and its halo, and we account for this effect by assuming the density profile
\begin{equation}\label{dracoprofile}
P(R)=P_s y^{-1}(1+y)^{-2}[1+(y/y_t)^{\delta}]^{-1},\ \ y\equiv R/R_s;
\end{equation}
the $[1+(y/y_t)^\delta]^{-1}$ suppression factor, with free parameters $y_t$ and $\delta$, is motivated by prior studies of tidal evolution in $N$-body simulations \cite{hayashi2003structural,penarrubia2010impact,green2019tidal}.  In Appendix~\ref{sec:profile}, we use the results of the tidal evolution simulations of Ref.~\cite{ogiya2019dash}, along with Draco's orbit and history, to fix $y_t=4.3$ and $\delta=3.9$.  We then set $R_\mathrm{max}=4$ kpc because this profile reaches the background density of the Galactic halo at this radius.

\begin{figure}[t]
	\centering
	\includegraphics[width=\columnwidth]{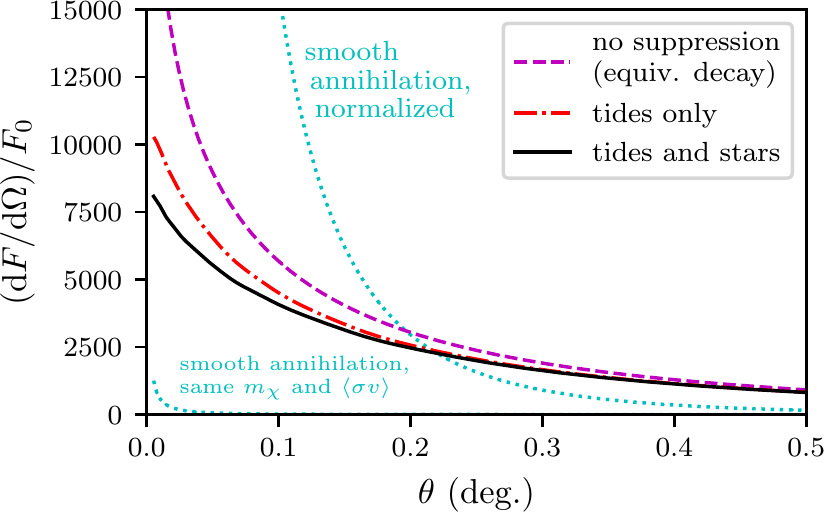}
	\caption{\label{fig:emission_st} Demonstration of the influence of tidal forces and stellar encounters on the microhalo-dominated annihilation signal from Draco.  We plot the projected emission profile (flux per solid angle) in the EMDE cosmology with $\TRH=2$ GeV and $\xcut=20$ if both, one, or neither of these effects are accounted for.  Note that without suppression, the signal is morphologically equivalent to that of dark matter decay.  All curves are normalized to the total flux $F_0$ (out to $3^\circ$) that would be expected in the absence of tidal evolution and stellar encounters.  We also show the signal from dark matter annihilation within Draco's smooth halo for comparison (dotted lines); this signal is plotted both normalized to its total flux and, in the bottom left, with the same dark matter particle as the microhalo curves.  The latter curve illustrates the extent to which microhalos dominate the annihilation signal.}
\end{figure}

For illustration, Fig.~\ref{fig:emission_st} shows the flux profile $\diff F/\diff \Omega$ for one EMDE scenario as a function of angle $\theta=R_\mathrm{proj}/D$.  We assume Draco lies at distance $D=76$ kpc \cite{bonanos2004rr,*bonanos2007erratum,*bonanos2008erratum}, so the emission extends out to $R_\mathrm{max}/D=3^\circ$ (not shown).  For comparison, we also show the flux profiles if stellar encounters are neglected and if tidal evolution is also neglected.  Evidently, tidal evolution and stellar encounters both influence the flux profile appreciably.  Note that the unsuppressed curve is equivalent to the signal from decaying dark matter.  As another comparison, we plot the flux profile from dark matter annihilation within Draco's smooth halo; it peaks much more sharply at small angles and drops off more quickly at large angles.  We also show the smooth annihilation profile normalized to the same dark matter properties as the microhalo flux profiles; this profile represents the contribution from dark matter outside of microhalos.\footnote{The annihilation rate outside of microhalos is scaled by the fraction $1-f$ of dark matter not in microhalos, but we assume tidal effects cause $1-f\simeq 1$ at the radii where smooth annihilation contributes nonnegligibly.}

\begin{figure}[t]
	\centering
	\includegraphics[width=\columnwidth]{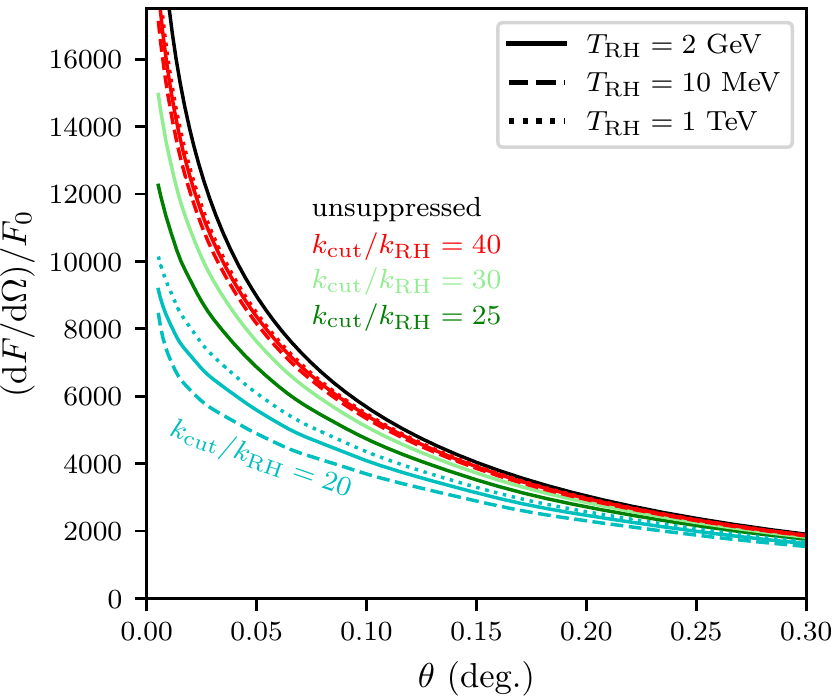}
	\caption{\label{fig:emission} Projected emission profiles (flux per solid angle) for the Draco dwarf in a variety of EMDE scenarios.  In each case, the emission profile is normalized to the total flux $F_0$ that would be expected in the absence of tidal evolution and stellar encounters.  There are clear differences in signal morphology between the different EMDE scenarios; these differences largely arise because denser microhalos are more resistant to disruption.  Thus, cosmologies that form microhalos earlier yield less-suppressed signals.}
\end{figure}

In Fig.~\ref{fig:emission}, we plot the flux profiles $\diff F/\diff \Omega$ in a variety of EMDE cosmologies, including the contribution from annihilation outside of microhalos.  Tidal evolution and stellar encounters evidently induce marked differences in signal morphology between the scenarios.  This diversity arises because denser microhalos are more resistant to these effects.  As Fig.~\ref{fig:microhalos} shows, scenarios with larger $\xcut$ result in denser halos, while $\TRH$ mostly controls halo size and only minimally affects density.  Thus, scenarios with larger $\xcut$ yield significantly less-suppressed annihilation signals.

\subsection{Limits on annihilation}\label{sec:dracoann}

We calculate the best-fitting flux and spectrum of the Draco dwarf using established techniques for the detection of dim, spatially extended sources, which were first developed for dwarf galaxies by Refs.~\cite{ackermann2014dark,ackermann2015searching,hooper2015gamma,albert2017searching}. We analyze 11 years of Fermi-LAT data taken between August 4, 2008 and September 3, 2019, extracting Pass 8 source-class photons recorded with an energy between 100~MeV and 1~TeV observed in a 30$^\circ \times$ 30$^\circ$ box centered on the position of Draco. We place standard cuts on the data quality and LAT configuration during events and divide the recorded events into 32 logarithmic energy bins and 0.1$^\circ$ angular bins over the region of interest.

In order to calculate the improvement to the log-likelihood generated by adding an extra degree of freedom at the position of the Draco dwarf, we first calculate the log-likelihood fit of a background model (which does not include the dwarf) to the Fermi-LAT data. We utilize the recently released 4FGL catalog~\cite{Fermi-LAT:2019yla}, the {\tt gll\_iem\_v07} diffuse model, and the {\tt iso\_P8R3\_SOURCE\_V2\_v1} isotropic background model. We independently fit the normalization of each diffuse source, as well as every point source with a detection significance exceeding 10, independently in every energy bin, fixing the spectra of these sources to their 4FGL default values.

\begin{figure}[t]
	\includegraphics[width=\columnwidth]{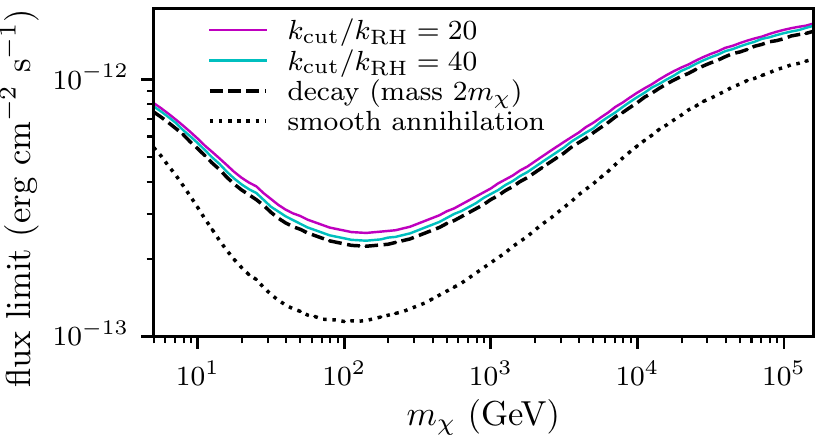}
	\caption{\label{fig:flux} Impact of the signal morphology on the bounds derived from Fermi-LAT observations on the energy flux from Draco in gamma rays above 100~MeV.  We plot $2\sigma$ upper limits assuming annihilation or decay into $b\bar b$; different curves assume different flux profiles as shown in Figs. \ref{fig:emission_st} and~\ref{fig:emission}.  For the EMDE-induced signals, we assume $T_\mathrm{RH}=2$~GeV.  Because it is so much less centrally concentrated, the gamma-ray flux from dark matter decay is constrained less strongly than the flux from annihilation within Draco's smooth profile by roughly a factor of 2.  Signals from annihilation within microhalos in EMDE cosmologies are even less centrally concentrated than decay signals due to the influence of tidal effects and stellar encounters, but this change in morphology is too small to significantly further weaken constraints.}
\end{figure}

We then fix the parameters of all sources in the background fit and add a new degree of freedom corresponding to the Draco dSph, appropriately employing a morphological model computed using Eq.~(\ref{flux}) (and depicted in Fig.~\ref{fig:emission}). Because the likelihood profile scans over the dark matter-induced flux in each model, this stage of the analysis is sensitive only to the morphology of each emission model and not to its overall normalization. We calculate the change in the likelihood of the fit as the flux from this component increases from an initial value of 0 in each independent energy bin. Using a spectral model based on the annihilation of dark matter particles of various masses to $b\bar{b}$ final states, we calculate the 2$\sigma$ combined upper limit by determining the Draco flux which worsens the log-likelihood of the fit by 2, compared to a model with no contribution from Draco.  Figure~\ref{fig:flux} shows the resulting limits on the total gamma-ray flux from Draco for several different signal morphologies.  More centrally concentrated signals are more strongly constrained, as illustrated by the strength of the flux limit for annihilation within Draco's smooth profile.  Consequently, naively applying a boost factor to account for the microhalos' increased annihilation rate relative to a smooth halo would produce constraints that are too strong by a factor of about 2.  However, relative to dark matter decay, the signal from microhalo-dominated annihilation in EMDE cosmologies is not sufficiently morphologically altered to significantly change the flux limits.

\begin{figure}[t]
	\includegraphics[width=\columnwidth]{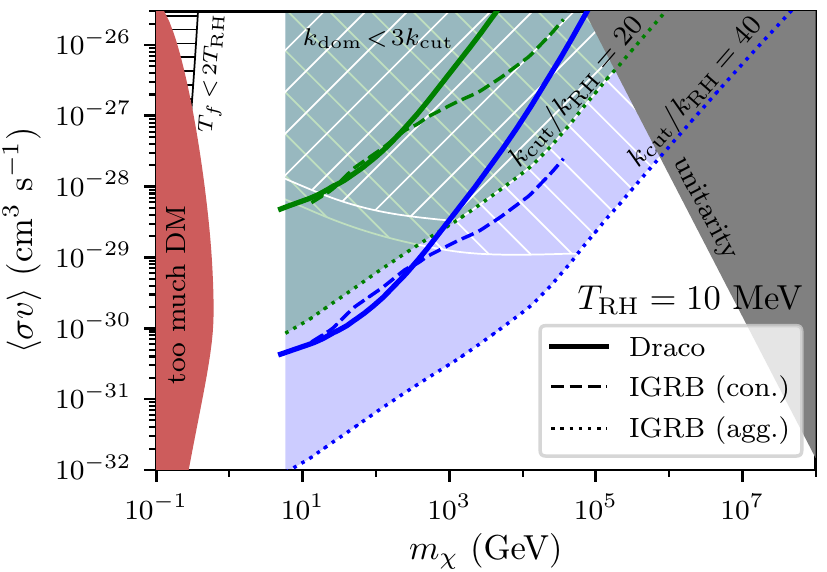}\\
	\includegraphics[width=\columnwidth]{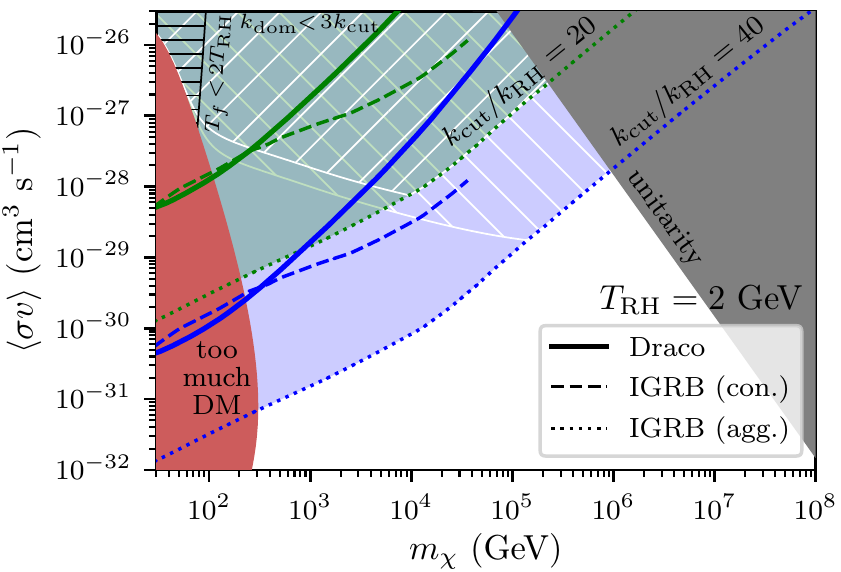}
	\caption{\label{fig:constraint_draco} Upper bounds on the cross section for dark matter annihilating into $b\bar b$ for two reheat temperatures $\TRH=10$~MeV (top) and $\TRH=2$~GeV (bottom), as derived from Fermi-LAT observations of the Draco dSph.  In each case, we consider both $\xcut=20$ (green) and $\xcut=40$ (blue).  In addition to the bounds from Draco, we also repeat from Fig.~\ref{fig:constraint_igrb} the conservative and aggressive bounds derived from the IGRB.  We find that the Draco-derived limits are comparable to the conservative limits from the IGRB and significantly weaker than the aggressive, background-subtracted limits.  The shaded region on the left is disallowed because it would overclose the Universe, while the shaded region on the right marks where the dark matter's coupling constant exceeds unity.  Constraints are tentative within the hatched regions because they fail Eq. (\ref{Tdom}) or~(\ref{Tf}), which implies that the density fluctuation power spectra we employed do not apply therein.}
\end{figure}

We translate the flux upper limit into an upper bound on the dark matter cross section by utilizing the expected net flux from Draco, which is obtained for each EMDE realization by integrating Eq.~(\ref{Lhost}) over Draco's volume.  In Fig.~\ref{fig:constraint_draco}, we plot the resulting limits on the dark matter annihilation cross section for EMDE scenarios with $\TRH=10$~MeV and $\TRH=2$~GeV.  For comparison, we also plot the constraints derived in Sec.~\ref{sec:igrb} using the IGRB.  We find that the limits from Draco are comparable to the conservative limits from the IGRB, where all of the gamma-ray flux is allowed to be attributed to dark matter, and are much weaker than the aggressive limits in which astrophysical sources are modeled and subtracted from the IGRB.  Evidently, for dark matter annihilation within microhalos, Draco produces bounds that are at best comparable to those from the IGRB.  We remark, however, that an advantage to searching for microhalo annihilation within galactic systems is the potential to distinguish it from dark matter decay through the presence of suppressive tidal effects.  We leave an exploration of this possibility to future work.  Another more general advantage to dark matter detection in regions with known overdensities is that any positive gamma-ray signal would be spatially correlated with that overdensity, making its attribution to dark matter more convincing.

We also note that recent analyses, such as Refs.~\cite{Calore:2018sdx, Hoof:2018hyn, Linden:2019soa}, have considered complex statistical issues which may arise due to systematic errors in the background modeling. While more precise treatments are possible, the systematic issues become most acute in three regimes: (1) when an analysis includes a joint-likelihood treatment of many dSphs, most of which have an expected dark matter content that falls far below the brightest few dSphs; (2) when a dwarf has a significant positive (or negative) flux associated with it, which may also be due to background mismodeling; and (3) when small changes in the constraint on the dark matter annihilation rate (at the order of 10\%) are highly relevant, such as in a comparison of standard annihilation constraints to dark matter models of the Galactic center gamma-ray excess~\cite{Daylan:2014rsa}. Our analysis of Draco does not fall into any of these regimes, so we do not produce a detailed calculation of the expected cross section constraints from multiple blank sky locations, an analysis which is computationally costly and would only mildly change our calculated limits.

\section{Conclusion}\label{sec:conc}

In this work, we developed a procedure to constrain thermal-relic dark matter that freezes out during or before an EMDE using existing indirect-detection probes.  These scenarios reduce the annihilation cross section required for dark matter to achieve the observed relic abundance, but they also induce the formation of abundant dark matter microhalos, which can bring these smaller cross sections into view.  As demonstration, we considered EMDE cosmologies with reheat temperatures $\TRH=10$ MeV and 2~GeV.  By comparing the annihilation signals from predicted microhalo populations to Fermi-LAT gamma-ray limits, we derived new constraints, shown in Fig.~\ref{fig:constraint_draco}, on thermal-relic dark matter in these cosmologies.

The principal challenge is to accurately model the microhalo population, and for that purpose we employed the recently developed models presented in Papers~I--III along with several new refinements.  These models describe the formation of microhalos and their evolution within larger systems.  Our refinements include accounting for the reduced growth rate of the smallest-scale dark matter density variations, modeling the combined impact of subhalo tidal evolution with stellar encounters, and accounting for transient tidal effects that trace a subhalo's orbital period.  We also devised a convenient fitting function to describe the aggregate tidal suppression of subhalo annihilation rates within a host.  These models allow precise tracking of the microhalo population through cosmic time with the caveat that the impact of mergers between microhalos remains unclear.  We employed an approximate model based on the results of Paper~I and Ref.~\cite{drakos2019major} to treat these mergers, and our results are subject to this approximation's accuracy.  Further study is needed to precisely understand how mergers influence a microhalo population.

We separately derived limits on dark matter annihilation in EMDE cosmologies using the IGRB and the Draco dwarf, and we found that the IGRB produces stronger bounds.  This result is unsurprising since the signal from microhalo annihilation roughly follows the dark matter mass distribution, so the sheer volume of the background outweighs the high density within dwarf galaxies.  The same property makes IGRB-derived bounds on the dark matter lifetime (e.g., Refs.~\cite{liu2017constraints,blanco2019constraints}) stronger than those derived from dSphs (e.g., Ref.~\cite{baring2016new}).  We note, however, that galactic systems can still be valuable in probing annihilation within microhalos because they can distinguish this process from dark matter decay.  The two produce morphologically similar signals, but microhalos within larger systems are subject to disruption by tidal effects and encounters with other objects, suppressing the annihilation signal near these systems' centers.  We leave further investigation of this possibility to future work.

The IGRB suffers significant astrophysical contamination, so the strength of our bounds depends strongly on the degree to which astrophysical gamma-ray sources are modeled and subtracted.  Using the aggressive subtraction program in Ref.~\cite{blanco2019constraints}, we are able to probe cross sections as small as $10^{-32}$~$\si{cm^3 s^{-1}}$ for dark matter annihilating into $b\bar b$, although the strength of constraints depends strongly on the dark matter mass, the reheat temperature, and the ratio $\xcut$ between the free-streaming scale and the wave number entering the horizon at reheating.  We explored only the range $20\leq \xcut\leq 40$, ratios plausible for certain supersymmetric dark matter candidates \cite{erickcek2016bringing}.  Larger values of $\xcut$ are also plausible and would be much more strongly constrained, but they require accurately modeling halo formation during the radiation-dominated epoch (e.g., \cite{blanco2019annihilation}), a problem we leave to future work.  Also, thermal relics with cross sections less than a few orders of magnitude below the canonical $\sigmav = 3\times 10^{-26}$ $\si{cm^3 s^{-1}}$ tend to be only tentatively constrained; the EMDE cosmologies needed to effect the observed dark matter abundance for these cross sections result in density fluctuation spectra different from those we assumed.  A more careful treatment is necessary to constrain thermal-relic dark matter candidates with cross sections in this regime.

This work represents an important step toward the development of robust constraints on thermal-relic dark matter that account for our ignorance of the Universe's early thermal history.  The possibility of early matter domination prior to BBN vastly broadens the range of dark matter properties that can produce the observed abundance, but we exploit the dark matter annihilation boost induced by the microhalo populations that arise in these cosmologies to considerably narrow the range of viable dark matter candidates.  We close by noting that while our constraints assume that dark matter is a thermal relic, the microhalo populations studied in this work could potentially be probed gravitationally through pulsar timing arrays \cite{Dror_2019,Jennings:2019qqz} and searches for lensing distortions in highly magnified stars \cite{dai2019gravitational}.  In this way, these microhalos can prospectively be used to constrain the early thermal history without assuming that dark matter has a thermal origin.

\begin{acknowledgments}
The simulations in Appendix~\ref{sec:tidestar} were carried out on the Dogwood computing cluster at the University of North Carolina at Chapel Hill.  This work was funded by NASA through the Fermi
Guest Investigator Cycle 10 Award No. 80NSSC17K0751.  M.\,S.\,D. was additionally supported by a Kenan Trust Graduate Student Research Grant and a Dissertation Completion Fellowship both from the University of North Carolina at Chapel Hill.
\end{acknowledgments}

\appendix

\section{Baryonic suppression of small-scale dark matter density fluctuations}\label{sec:growth}

Before recombination, baryons do not accrete into overdense regions because they are coupled to the photons.  Afterward, they still resist gravitational infall on small scales because a residual ionization fraction maintains the baryons at a temperature close to that of the cosmic microwave background \cite{bertschinger2006effects}.  In particular, dark matter structures of mass smaller than about $10^5 M_\odot$ are not expected to accrete baryons.  Consequently, dark matter density fluctuations at the microhalo scales we are concerned with grow more slowly than would otherwise be expected, and in this appendix we discuss how this effect influences the population of microhalos determined by the framework in Paper~I.

During matter domination, matter density contrasts $\delta\equiv\delta\rho/\bar\rho$ grow as $\delta(a)\propto a$ in the linear regime ($\delta\ll 1$) if both baryons and dark matter contribute (or if baryons are absent).  However, if baryons do not accrete into dark matter overdensities, then the dark matter density contrasts instead grow as $\delta(a)\propto a^\mu$ with \cite{hu1996small}
\begin{equation}\label{growthindex}
\mu = \frac{5}{4}\left(1-\frac{24}{25}\frac{\Omega_b}{\Omega_m}\right)^{1/2}-\frac{1}{4}.
\end{equation}
Here, $\Omega_b$ and $\Omega_m$ are the ratios of baryon and matter density today, respectively, to the critical density; if $\Omega_m=0.31$ and $\Omega_b=0.049$ \cite{aghanim2018planck}, then $\mu=0.901$.  This difference in growth rate is significant.  As an approximate example, a density contrast $\delta=0.17$ at $z=3000$ would reach the critical linear threshold $\delta_c=1.686$ and collapse at $z_c\simeq 300$ if $\delta\propto a$ or at $z_c\simeq 230$ if $\delta\propto a^\mu$.  Since the characteristic density of the resulting halo is proportional to $(1+z_c)^3$, incorrectly adopting $\delta\propto a$ would lead to overestimation of this halo's density, and hence annihilation rate, by a factor of 2.

By numerically integrating the spherical collapse model with baryons treated as a smooth background, we verified that the critical linear density contrast for collapse remains $\delta_c=1.686$.  That is, an initial spherical overdensity $\delta_i\ll 1$ at scale factor $a_i$ collapses at scale factor $(1.686/\delta_i)^{1/\mu}a_i$.  Additionally, the turnaround radius---the apocenter of the trajectory of the spherical shell enclosing the overdensity---is $r_\mathrm{ta}=(3/5)r_i/\delta_i^{1/\mu}$; the coefficient $3/5$ is unaltered from its standard value.

To account for the slower growth rate of $\delta$, we alter the definition in Paper~I of the scaled density field to become
\begin{equation}\label{delta}
\delta(\vec x)\equiv\delta(\vec x,a)/a^\mu.
\end{equation}
Applying this definition and the spherical collapse solution above, we find that Paper~I's prediction of the coefficient $A$ of a halo's $\rho=Ar^{-3/2}$ inner asymptote becomes
\begin{equation}\label{asy}
A=\alpha\delta_c^{\frac{3}{2}\left(1-\frac{1}{\mu}\right)} \bar\rho \delta^{\frac{3}{4}\left(\frac{2}{\mu}+1\right)}|\nabla^2\delta|^{-3/4}f_\mathrm{ec}^{-\frac{3}{2\mu}}(e,p),
\end{equation}
where $\alpha=12.1$ is the same proportionality constant as in Paper~I.  Here, $\bar\rho\simeq 33.1$ $M_\odot/\text{kpc}^3$ is the comoving mean dark matter density \cite{aghanim2018planck} and $f_\mathrm{ec}(e,p)\equiv \delta_\mathrm{ec}/\delta_\mathrm{c}$ is an ellipsoidal collapse correction \cite{sheth2001ellipsoidal}.  Meanwhile, in the turnaround model, the final radius of a mass shell at initial comoving radius $q$ is 
\begin{equation}\label{rta}
r=\beta q/\Delta(q)^{1/\mu},
\end{equation}
where $\Delta(q)=\Delta(q,a)/a^\mu$ is the fractional mass excess enclosed, in linear theory, and $\beta=0.131$ is the same proportionality constant as in Paper~I.  The enclosed mass
\begin{equation}\label{mta}
M(q)=\beta_M(4\pi/3)q^3\bar\rho
\end{equation}
($\beta_M=0.273$) is unaltered from its expression in Paper~I, leading to
\begin{equation}\label{Mslope}
\frac{\diff \ln M}{\diff \ln r}=\frac{3}{1+3\epsilon(q)/\mu}
\end{equation}
with $\epsilon(q)\equiv(-1/3)\diff \ln\Delta/\diff \ln q=1-\delta(q)/\Delta(q)$.  The radius $\rmax$ of maximum circular velocity and mass $\mmax$ enclosed are obtained by solving $\diff \ln M/\diff \ln r=1$ (or $\epsilon(q)=2\mu/3$); this computation yields an initial comoving radius $q_\mathrm{max}$ from which Eqs. (\ref{rta}) and~(\ref{mta}) yield the desired quantities.

\section{Aggregate tidal suppression of subhalo annihilation rates}\label{sec:sup}

In this appendix, we use the model in Paper~II to derive, and find a fitting function for, the overall factor by which annihilation rates from subhalos are scaled within a host due to tidal evolution.  For an individual subhalo with scale density $\rho_s$ orbiting with circular orbit radius $R_c$ and circularity $\eta$,\footnote{\label{foot:orbit}As defined in Paper~II and elsewhere, for an orbit with energy $E$ the circular orbit radius is the radius of the circular orbit with that energy.  Meanwhile, the circularity is the ratio $\eta=L/L_c$ between the orbit's angular momentum and that of the circular orbit with energy $E$.} the Paper~II model predicts the factor
\begin{equation}\label{s}
s(\rho_s/P_s,R_c/R_s,\eta,t/T)\equiv J/J_\mathrm{init}
\end{equation}
by which the subhalo's $J$ factor is scaled by tidal evolution.  Here, $R_s$ and $P_s$ are the scale radius and density of the host, $t$ is the duration of the tidal evolution, and $T$ is the subhalo's radial (apocenter-to-apocenter) orbit period.  To characterize the aggregate tidal suppression of all subhalos within the host, we seek the quantity
\begin{equation}\label{S}
S(\rho_s/P_s,t\sqrt{GP_s},c)\equiv\frac{\sum J}{\sum J_\mathrm{init}},
\end{equation}
where the sums proceed over subhalos distributed throughout the host's phase space.  Here, $c=R_\mathrm{host}/R_s$ is the radius of the host in units of $R_s$, which can be interpreted as its concentration $R_\mathrm{vir}/R_s$.  As Eq.~(\ref{S}) expresses, and we verified, $S$ only depends on system parameters in the combinations $\rho_s/P_s$, $t\sqrt{GP_s}$, and $R_\mathrm{host}/R_s$.  This property follows from the dependencies of $s$ in Eq.~(\ref{s}), noting that subhalo orbital periods are proportional to the host's dynamical timescale $(GP_s)^{-1/2}$.

We let $f(E,L,R)$ be the host halo's distribution function, where $E$ is energy (per mass), $L$ is angular momentum (per mass), and $R$ is radius.  The host's density profile $P(R)$ can be decomposed in the orbital parameters as
\begin{equation}\label{profile_int}
P(R)=\int_{R_{c,\mathrm{min}}(R)}^\infty\!\!\!\!\!\!\!\!\diff R_c\int_0^{\eta_\mathrm{max}(R,R_c)}\!\!\!\!\!\!\!\!\mathrm{d\eta}\ g(R,R_c,\eta),
\end{equation}
where
\begin{equation}\label{orbital_decomposition}
\begin{split}
g(R,R_c&,\eta)\equiv 
4\sqrt 2\pi f\!\left(K(R_c)+\Phi(R_c),\eta R_c\sqrt{2K(R_c)},R\right)
\\
&\hphantom{=}\times
\frac{\left[K(R_c)/R_c+2\pi G P(R_c)R_c\right] K(R_c) \eta R_c^2/R^2}
{\sqrt{K(R_c)(1-\eta^2 R_c^2/R^2)+\Phi(R_c)-\Phi(R)}}
\end{split}
\end{equation}
follows from the definitions of $R_c$ and $\eta$.  Here, $K(R_c)\equiv GM(R_c)/(2R_c)$ is the circular orbit kinetic energy (per mass), $\Phi(R)$ is the host's potential profile, and $M(R)$ is its enclosed mass profile.  The integration limits in Eq.~(\ref{profile_int}) are defined such that
\begin{gather}
K(R_{c,\mathrm{min}})+\Phi(R_{c,\mathrm{min}})=\Phi(R),
\\
\eta_\mathrm{max} \equiv (R/R_c)\sqrt{1+[\Phi(R_c)-\Phi(R)]/K(R_c)}.
\end{gather}
The decomposition in Eq.~(\ref{profile_int}) is valuable because we can insert the individual tidal scaling factor $s$ into the integrand.  The aggregate tidal scaling factor is thus
\begin{multline}\label{Sint}
S= \frac{1}{M_\mathrm{host}}
\int_0^{R_\mathrm{host}}\!\!\!\!\!\!4\pi R^2\diff R
\int_{R_{c,\mathrm{min}}(R)}^\infty\!\!\!\!\!\!\!\!\diff R_c
\int_0^{\eta_\mathrm{max}(R,R_c)}\!\!\!\!\!\!\!\!\mathrm{d\eta}
\\
\times g(R,R_c,\eta)s(\rho_s/P_s,R_c/R_s,\eta,t/T) ,
\end{multline}
where $T=T(P_s,R_c/R_s,\eta)$ and $M_\mathrm{host}$ is the host's mass (within $R_\mathrm{host}$).

For convenience, we supply the following fitting function for $S$.  We assume the host halo has an isotropic distribution function $f(E)$ and employ the fitting form for $f$ given in Ref.~\cite{widrow2000distribution} to evaluate $S$.  Let $p\equiv\rho_s/P_s$ and $\tau\equiv t\sqrt{GP_s}$.  For $1<c<10^2$, $1<p<10^8$, and $1<\tau<10^4$, the expression
\begin{align}\label{Sfit}
S(p,\tau,c)&=\exp\left[-(c/\alpha)^{-\beta}\right],
\nonumber\\
\alpha&=2.84p^{-0.698}\tau^{0.557}\exp(-6.67/\tau),
\nonumber\\
\beta&=0.577p^{0.0476}\tau^{-0.0526}\exp(-0.563/\tau)
\end{align}
is accurate to within 0.08, with better accuracy when $p>10$.  When predicting in Fig.~\ref{fig:igrbsupp} the suppression of annihilation signals from subhalos within the whole population of hosts, Eq.~(\ref{Sfit}) produces results that are accurate to within 2\%.  The predictions that use this fitting function are shown as thin dotted lines.

\section{$J$-factor oscillations during subhalo orbits}\label{sec:rmodel}

\begin{figure}[t]
	\centering
	\includegraphics[width=\columnwidth]{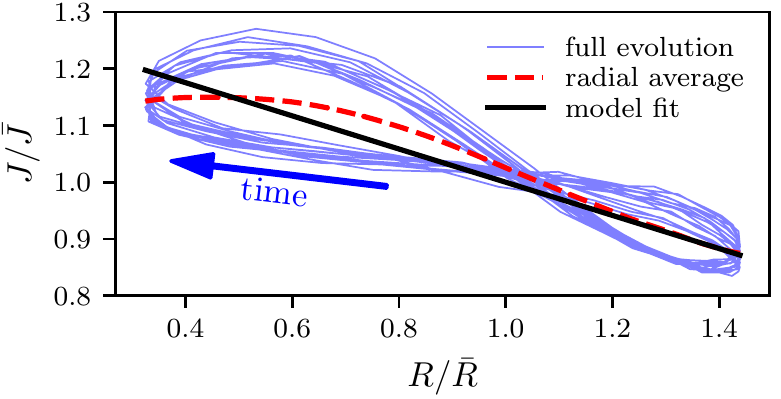}
	\caption{\label{fig:Jr} The trajectory of the $J$-factor oscillations, $J/\bar J$ (where $\bar J$ is the orbital period-averaged trajectory), in radius $R/\bar R$, where $\bar R$ is the time-averaged orbital radius.  The $J$ factor is larger at smaller radii due to compressive tides.  The red dashed line shows the average value of $J/\bar J$ at each radius, and the solid line shows our fit using Eq.~(\ref{Jr}).}
\end{figure}

It is useful to model the periodic oscillations in the $J$ factor observed in Paper~II.  While these oscillations do not alter the overall annihilation rate in subhalos, they still introduce a systematic biasing effect because subhalos at smaller radii have larger $J$, and this effect can alter the morphology of an annihilation signal.  In this appendix, we reanalyze the tidal evolution simulations in Paper~II to develop a simple model for the impact of these oscillations.  In particular, we model $J/\bar J$, where $\bar J$ is the orbital period-averaged $J$ factor, as a function of the ratio $R/\bar R$ of the instantaneous to the orbital period-averaged radius.  Figure~\ref{fig:Jr} plots $J/\bar J$ against $R/\bar R$ for one simulation.  We will also employ the relative energy parameter $x=|E_b|/\Delta E_\mathrm{imp}$ and the relative orbital radius parameter $y=\bar R/R_s$ defined in Paper~II that describe the host-subhalo system.

For each simulation, we obtain the average value of $J/\bar J$ at each radius starting at the pericenter of the fifth orbit about the host and ending at the final pericenter.  We also consider several additional simulations with $x<3$ in order to understand the small-$x$ behavior.  Next, we fit
\begin{equation}\label{Jr}
J/\bar J = 1-d(R/\bar R-1),
\end{equation}
the simplest possible relationship, to this average value.  Here, $d$ is the fitting parameter, and it is easy to see that this equation manifestly preserves the time-averaged $J$ factor.  Both the radial average and the fit are also depicted in Fig.~\ref{fig:Jr}.  One may worry that Eq.~(\ref{Jr}) unphysically allows $J<0$.  However, we will see shortly that $0\leq d\lesssim 0.5$, implying that $J>0$ as long as $R_a/\bar R < 3$, where $R_a$ is the radius of the subhalo's orbital apocenter.  If the host has an NFW profile, then $R_a/\bar R \leq 1.5$.

\begin{figure}[t]
	\centering
	\includegraphics[width=\columnwidth]{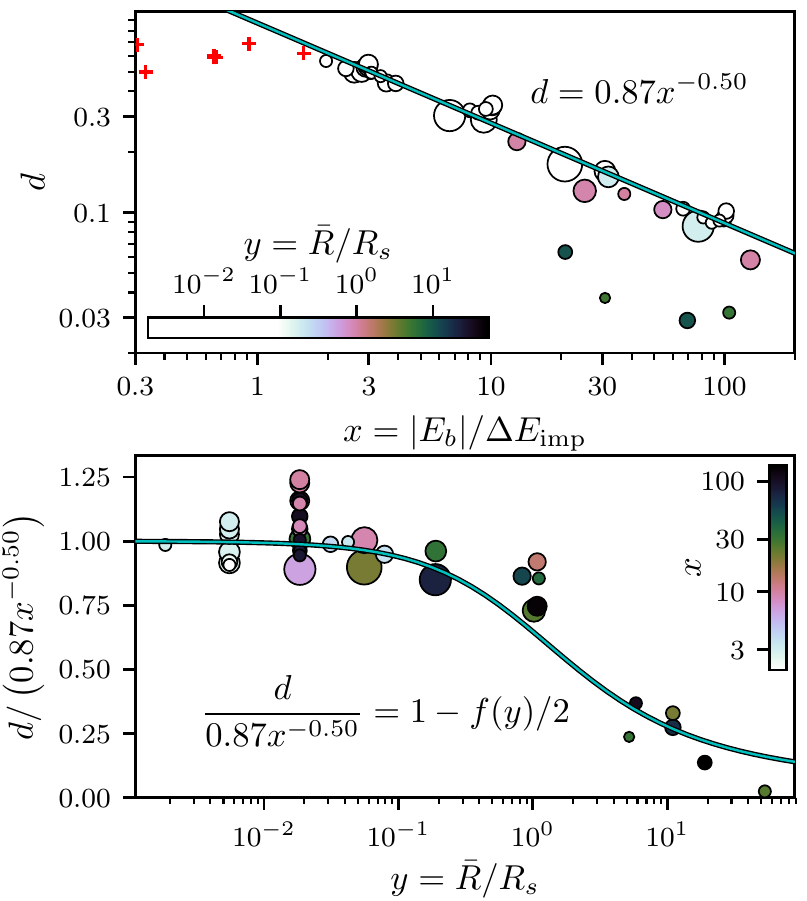}
	\caption{\label{fig:d} The dependence of the parameter $d$ that describes the $J$-factor oscillations on the host-subhalo system parameters $x$ and $y$ defined in Paper~II.  Top: in the $\bar R\ll R_s$ regime, $d$ follows a power law in $x$, the best fit of which is shown as a solid line.  At the small-$x$ end, we see that $x$ caps at roughly 0.5; the ``$+$'' markers are additional simulations at small $x$ not used in the fit.  Bottom: scaling of the normalization of $d$ with $y=\bar R/R_s$.  The solid line depicts the anticipated relationship (see the text) and is not a fit.  Each marker is a simulation, and the marker radius is proportional to the number of orbital periods, which ranges from 6 to 20.}
\end{figure}

Next, we relate $d$ to the system parameters $x$ and $y$.  As shown in the top panel of Fig.~\ref{fig:d}, $d$ follows a power law in $x$ in the $\bar R\ll R_s$ regime, but additionally, $d$ does not exceed roughly $0.5$.  We find that
\begin{equation}\label{d_}
d\simeq \min\{d_0 x^{-d_1},0.5\}, \ \text{if}\ \bar R\ll R_s
\end{equation}
with $d_0=0.87$ and $d_1=0.50$.  Beyond the $\bar R\ll R_s$ regime, $d$ is also sensitive to $y=\bar R/R_s$.  To understand this sensitivity, note that as discussed in Paper~II, the subhalo experiences compressive tidal forces proportional to $F/R$ along the axes perpendicular to the host-subhalo axis and stretching forces proportional to $f(R/R_s)F/R$ along the host-subhalo axis, where
\begin{equation}\label{asymf}
f(y) \equiv \frac{2\ln(1+y)-y(2+3y)/(1+y)^2}{\ln(1+y)-y/(1+y)}
\end{equation}
and $F$ is the host halo's gravitational force at radius $R$.  These forces cause the subhalo's size to scale as $1-A$ along two axes and $1+Af(R/R_s)$ along the other, where $A$ is a factor that is common to all axes.  Hence, the volume scales as $V\propto (1-A)^2 (1+f A)$, where $f=f(R/R_s)$, so taking $J\propto 1/V$ and expanding to linear order in $A$, we obtain $J\propto 1+2A(1-f/2)$.  Combined with Eq.~(\ref{d_}), this argument predicts the expression
\begin{equation}\label{d}
d = \min\{d_0 x^{-d_1},0.5\} \left[1-f(y)/2\right],
\end{equation}
where $y=\bar R/R_s$ as before.  The bottom panel of Fig.~\ref{fig:d} shows that this $y$-scaling works reasonably well.

\section{Combined impact of tidal forces and stellar encounters}\label{sec:tidestar}

\begin{figure}[t]
	\centering
	\includegraphics[width=\columnwidth]{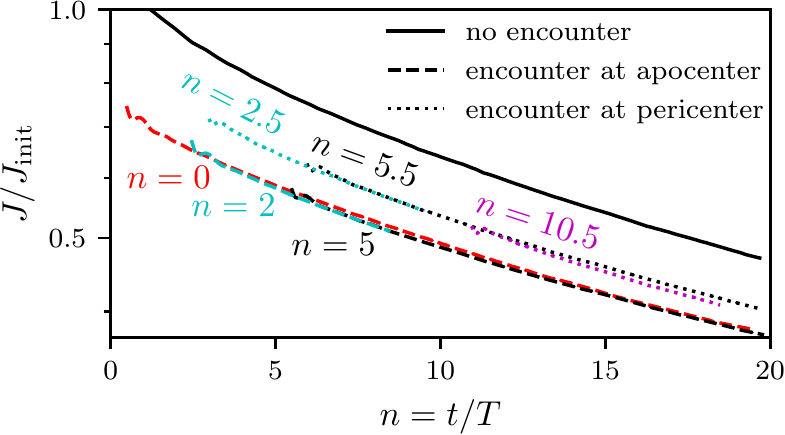}
	\caption{\label{fig:tidestar_t} Impact of the time of a stellar encounter for a subhalo undergoing tidal evolution.  This figure plots the $J$-factor trajectories of simulations of the same tidal evolution scenario where the subhalo is subjected to the same stellar encounter at different times; the labels indicate the time $n$ of the encounter.  The stellar encounter's impact is sensitive to the subhalo's position within its orbit during the encounter, but otherwise, the time of the encounter has minimal impact.}
\end{figure}

\begin{figure}[t]
	\centering
	\includegraphics[width=\columnwidth]{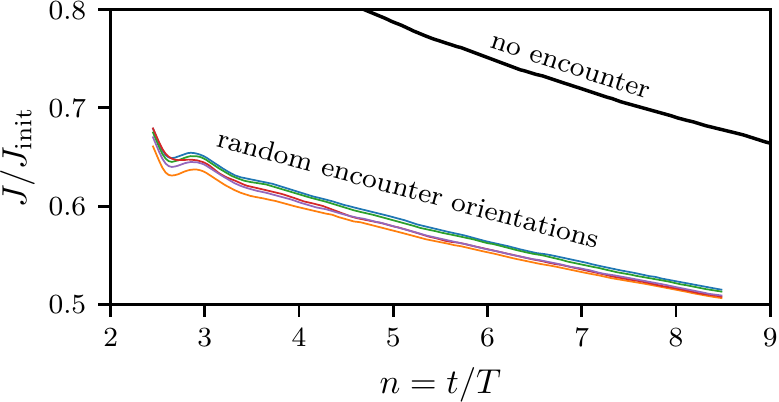}
	\caption{\label{fig:tidestar_o} Impact of the orientation of a stellar encounter for a subhalo undergoing tidal evolution.  This figure plots the $J$-factor trajectories of simulations of the same tidal evolution scenario where the subhalo is subjected to the same stellar encounter with five random orientations.  Evidently, the impact of encounter orientation is marginal.}
\end{figure}

\begin{figure}[t]
	\centering
	\includegraphics[width=\columnwidth]{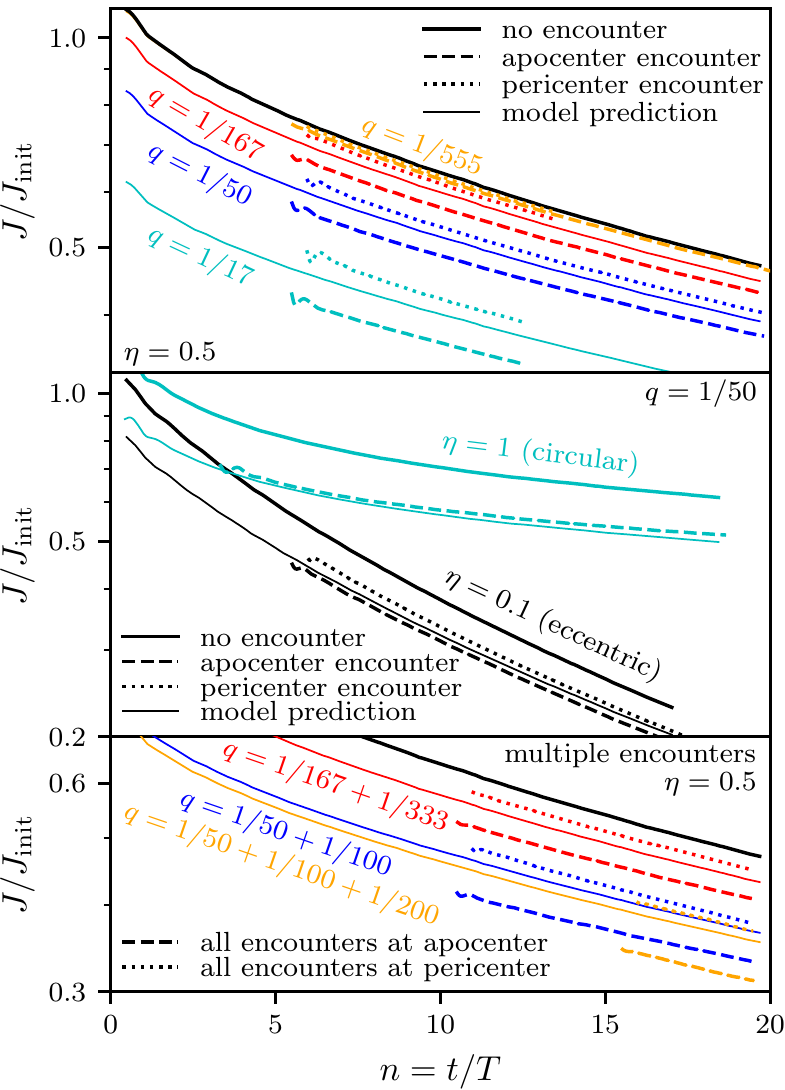}
	\caption{\label{fig:tidestar} Validation and tuning of the model developed in Appendix~\ref{sec:tidestar} for combining the impact of a host halo's tidal forces with those of stellar encounters.  We plot $J$-factor trajectories extracted from a variety of simulations (thick dashed and dotted lines) that include both tidal evolution and a stellar encounter; in the upper panel we vary the stellar encounter (labeling the relative energy parameter $q$), while in the middle panel we vary the subhalo orbit (labeling the orbit circularity $\eta$).  We also show simulations involving multiple stellar encounters in the lower panel.  Appropriately tuned, the model predictions (thin lines) match the simulations reasonably well.  Note that for the $q=1/555$ encounter in the upper panel, the model predicts no change in the $J$ factor.}
\end{figure}

\begin{figure}[t]
\centering
\includegraphics[width=\columnwidth]{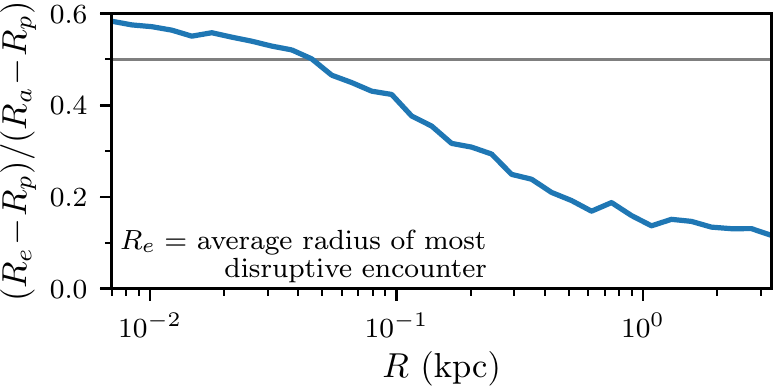}
\caption{\label{fig:encr} Average radius $R_e$ of the stellar encounter injecting the most energy for microhalos orbiting Draco.  We plot $R_e$ as a function of radius $R$ about Draco; it is averaged at each radius over microhalo orbits and stellar encounter histories.  $R_e$ is plotted relative to the subhalo's pericenter and apocenter, so because the curve mostly lies below 0.5 (horizontal line), disruption tends to occur closer to pericenter.}
\end{figure}

Paper~II explored the evolution of dark matter annihilation rates in subhalos due to the host halo's tidal forces, while Paper~III explored the evolution of microhalos subjected to encounters with individual stars.  In this appendix, we explore how to combine the two effects.  For this purpose, we simulated a variety of scenarios in which tidal evolution is combined with one or more stellar encounters.  These simulations were carried out using the methods of Papers II and III; in particular, velocity changes due to the passing star are applied directly and the star is not explicitly simulated.  We track the evolution of the subhalo's $J$ factor in each simulation as in Paper~II.  In all of these scenarios, the orbital radius is within the host's scale radius.  This region is most relevant for stellar encounters; for instance, Draco's stellar half-light radius of 0.22 kpc is within its 0.44-kpc scale radius (see Sec.~\ref{sec:draco}).

We first consider a tidal evolution scenario with ratio $\rho_s/P_s=1285$ between the subhalo and host scale density and ratio $R_c/R_s=0.055$ between the subhalo's circular orbit radius and the host's scale radius.  We take the subhalo's orbital circularity to be $\eta=0.5$.\footnote{See footnote~\ref{foot:orbit} for definitions of $R_c$ and $\eta$.}  In separate simulations, we applied the same stellar encounter at the beginning of the simulation, which is an apocenter passage, and also at the second and fifth subsequent apocenter passages.  For the subhalo's initial density profile, this encounter has relative energy injection $q\equiv\Delta E/|E_b|=1/50$, with $q$ as defined in Paper~III, and its impact parameter is much larger than the subhalo's scale radius.  Figure~\ref{fig:tidestar_t} shows the $J$-factor evolution (averaged over an orbital period as in Paper~II) that results from these scenarios.  Notably, the trajectory is essentially independent of the time of the encounter as long as it occurs at an apocenter.  We also show the trajectory if the encounter occurs during a pericenter passage.  As discussed in Appendix~\ref{sec:rmodel}, the subhalo is most compact near pericenter, and this compactness makes it more resistant to the stellar encounter.  However, the same time independence holds.

Tidal forces break the spherical symmetry that a subhalo would otherwise be expected to possess, so we also verify explicitly that the orientation of the stellar encounter is inconsequential.  We simulated the above tidal evolution scenario subjected to the same stellar encounter at a fixed time but with five random orientations.  We plot the resulting $J$-factor trajectories in Fig.~\ref{fig:tidestar_o}.  The scatter between these trajectories is only about 5\% of the change in $J$ caused by the encounters, which confirms that the encounter orientation does not have a significant impact.

Since the time of a stellar encounter has little impact on the resulting $J$-factor evolution, we can build a model for the impact of stellar encounters based on the idea of inserting the encounters after the tidal evolution.  Let $s$ be the orbital period-averaged tidal scaling factor for the subhalo's $J$ factor due to the host's tidal forces; that is, $s=J/J_\mathrm{init}$ is the quantity predicted by the model in Paper~II.  To build our model, we make the ansatz that for the purpose of stellar encounters, the subhalo's scale parameters are related to their initial NFW values by
\begin{equation}\label{stmodel}
r_s/r_s^\mathrm{init} = f s^{\zeta}
\ \ \text{and}\ \ 
\rho_s/\rho_s^\mathrm{init} = g s^{-\xi}.
\end{equation}
We will tune the parameters $f$, $g$, $\zeta$, and $\xi$ to simulations, but we enforce $3\zeta-2\xi=1$ to ensure $J\propto \rho_s^2 r_s^3$.  For the purpose of our model, we assume that $\rho_s$ and $r_s$ are the parameters of the density profile
\begin{equation}\label{stellarprofile}
\rho(r)=\rho_s \frac{r_s}{r} \exp\left[-\frac{1}{\alpha}\left(\frac{r}{r_s}\right)^\alpha\right], \ \ \alpha=0.78,
\end{equation}
which Paper~III found to be a universal outcome of stellar encounters.  We can now apply the model of Paper~III to find the scale parameters $\rho_s^\prime$ and $r_s^\prime$ after stellar encounters have taken place, and by integrating the density profile we find that the subhalo's final $J$ factor is
\begin{equation}\label{Js}
J = 4.34\rho_s^{\prime 2} r_s^{\prime 3}.
\end{equation}
Finally, if stellar encounters would be predicted to increase $J$, we instead assume they have no effect.

To validate and tune our model, we carried out simulations with different stellar encounters and different subhalo orbits.  The upper panel of Fig.~\ref{fig:tidestar} shows the impact of different stellar encounters, while the middle panel shows different orbits.  We also plot the predictions of the above model using the parameters $f=0.875$, $g=1.245$, $\zeta=0.71$, and $\xi=0.57$, and we find that it matches the simulation results reasonably well.  In tuning these parameters, we aim for a trajectory closer to the simulation results for encounters at pericenter than to the results for encounters at apocenter.  This preference is justified by Fig.~\ref{fig:encr}, which plots the average radius of the most disruptive stellar encounter, relative to apocenter and pericenter, for microhalos orbiting Draco in one of the EMDE scenarios studied in Sec.~\ref{sec:draco}.  This quantity is plotted as a function of radius about Draco and is averaged over microhalo orbits at that radius and over stellar encounter histories.  We see that the most disruptive encounter generally occurs closer to pericenter than to apocenter.

We also simulated scenarios involving two to three stellar encounters, the results of which are shown in the lower panel of Fig.~\ref{fig:tidestar}.  These encounters occur at intervals of five orbital periods, and each encounter injects half the energy of the previous one.\footnote{The $b^{-4}$ scaling of energy injection with impact parameter implies that the three most disruptive encounters typically inject roughly 62\%, 14\%, and 6\%, respectively, of the total energy injected by all encounters.}  In order to match the predictions from the model to these results, we find it necessary to add encounter energy injections directly rather than apply the encounters consecutively.  In the language of Paper III, we set $\lambda=\infty$; the conceptual interpretation is that a tidally evolving halo does not relax after an encounter.

\section{Draco's outer density profile}\label{sec:profile}

We assume Draco's density profile asymptotes to ${\rho\propto r^{-1}}$ at small radii as observed in halos that form in cosmological dark matter simulations \cite{navarro1996structure,navarro1997universal}.  However, as a subhalo of the Milky Way, Draco's profile at large radii is altered by tidal forces.  We assume Draco's tidally evolved density profile follows the form
\begin{equation}\tag{\ref{dracoprofile}}
P(R)=P_s y^{-1}(1+y)^{-2}[1+(y/y_t)^{\delta}]^{-1},\ \ y\equiv R/R_s,
\end{equation}
which begins to diverge from the NFW profile near the radius $y_t R_s$.  To determine the parameters $y_t$ and $\delta$, we analyze the Dynamical Aspects of SubHaloes (DASH) library of tidal evolution simulations published by Ref.~\cite{ogiya2019dash}.

\begin{figure}[t]
	\centering
	\includegraphics[width=\columnwidth]{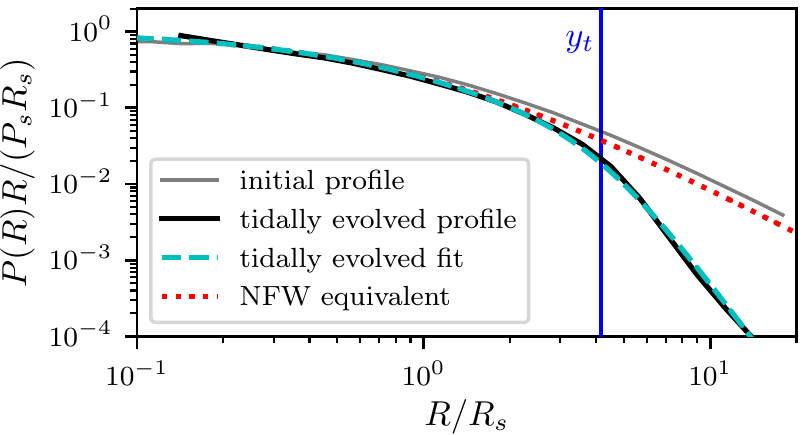}
	\caption{\label{fig:draco_profile} Density profile $P(R)$ of a subhalo simulated by Ref.~\cite{ogiya2019dash} after 7 Gyr of tidal evolution; the host, subhalo, and orbital properties are similar to those of the Milky Way-Draco system.  The initial and final profiles in the simulation are shown as solid lines, while the dashed lines show the fit to the tidally evolved profile using Eq.~(\ref{dracoprofile}).  The dotted line shows the NFW profile with the same scale parameters $R_s$ and $P_s$ as obtained in the fit to the evolved profile; it is not a fit to the initial profile.}
\end{figure}

\begin{figure}[t]
	\centering
	\includegraphics[width=\columnwidth]{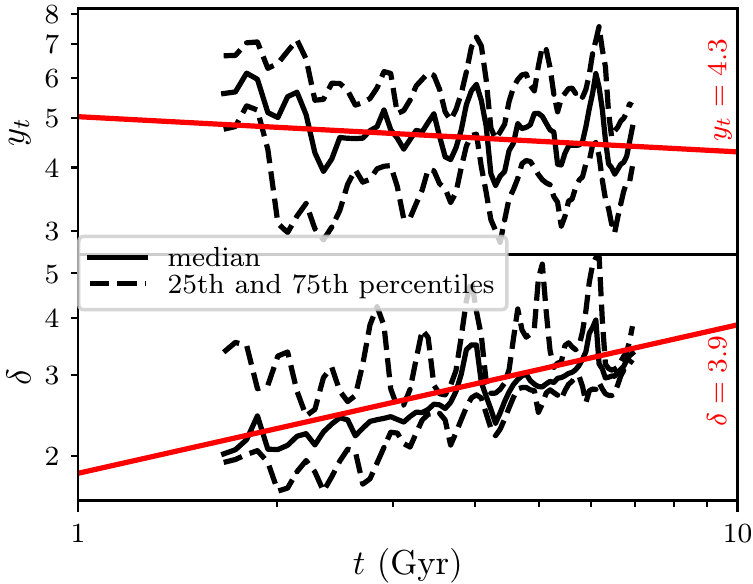}
	\caption{\label{fig:yt_delta} Evolution of the parameters $y_t$ (top) and $\delta$ (bottom) of the subhalo density profile given by Eq.~(\ref{dracoprofile}) for tidal evolution simulations with host-subhalo parameters similar to those of the Draco-Milky Way system.  The median (solid curve) and 25th and 75th percentiles (dashed curves) are plotted at each time.  We extrapolate each parameter's evolution to 10 Gyr by fitting a power law (red line) to the full dataset; the extrapolated parameters at 10 Gyr are shown.}
\end{figure}

In this library, a host-subhalo system is parametrized by the host and subhalo concentration parameters $c_\mathrm{host}$ and $c_\mathrm{sub}$, where a halo's concentration $c\equiv r_\mathrm{vir}/r_s$ is the ratio between virial and scale radii, and the subhalo's relative circular orbit radius $x_c\equiv r_c/r_\mathrm{vir,host}$ and circularity $\eta$.\footnote{See footnote~\ref{foot:orbit} for definitions of $r_c$ and $\eta$.}  These properties are determined at the time of subhalo accretion, which we assume to be roughly 10 Gyr ago at redshift $z\simeq 2$.\footnote{While $c_\mathrm{sub}$, $c_\mathrm{host}$, and $x_c$ exhibit significant sensitivity to redshift $z$ through the host and subhalo virial radii, they vary together in a way that does not alter the tidal evolution.  This property is a consequence of the insensitivity of tidal evolution to the virial radii (e.g., Paper II).  Thus, it is not necessary to precisely tune the accretion time or redshift; it only controls the duration of tidal evolution, which varies marginally for large changes in redshift.}  Draco's scale radius $R_s=0.435$ kpc and scale density $P_s=1.5\times 10^{8}$ $M_\odot$/kpc$^3$ imply that $c_\mathrm{sub}\simeq 15$ at $z=2$.  Meanwhile, various Milky Way mass models place its concentration $c_\mathrm{host}$ in the range 3--7 at redshift 2 \cite{battaglia2005radial,read2008thin,mcmillan2011mass,sohn2013space,nesti2013dark}.  For the same mass models, kinematic data \cite{helmi2018gaia,fritz2018gaia} put Draco on an orbit with $x_c$ ranging from 0.7 to 1.1 at accretion and $\eta\simeq 0.8$.

To match the Milky Way-Draco system, we consider 27 DASH simulations spanning the parameter range $3.2\leq c_\mathrm{host}\leq 6.3$, $12.6\leq c_\mathrm{sub}\leq 20.0$, $0.87\leq x_c\leq 1.15$,\footnote{$x_c\lesssim 0.8$ at accretion is atypical \cite{ogiya2019dash} and is consequently not included in the DASH simulations.} and $\eta=0.8$.  With time units rescaled to $z=2$, these simulations run for 7 Gyr, and in Fig.~\ref{fig:draco_profile} we plot the density profile of the subhalo after 7 Gyr of tidal evolution for an example set of parameters.  We fit Eq.~(\ref{dracoprofile}) to the evolved density profile in each simulation, generally obtaining $y_t\simeq 5$ and $\delta\simeq 3$.  This fit is plotted in Fig.~\ref{fig:draco_profile}, and to illustrate the effect of $y_t$ and $\delta$, we also show the NFW profile with the same $R_s$ and $P_s$.  Since the DASH simulations only extend to 7 Gyr, we must extrapolate the evolution of $y_t$ and $\delta$, and we do so by fitting power laws to the evolution of these parameters over all 27 simulations.  Figure~\ref{fig:yt_delta} shows this evolution and the power-law fits; we obtain $\delta\simeq 3.9$ and $y_t\simeq 4.3$ at 10 Gyr.

We use the tidally altered density profile determined in this section to calculate the gamma-ray flux profile from microhalos in Sec.~\ref{sec:dracogamma}.  Note that with the chosen $y_t$ and $\delta$, the modified profile alters the determination of $R_s$ and $P_s$ described in Sec.~\ref{sec:draco} by less than 0.1\%.  Also, for simplicity, we do not account for tidal alteration of the profile when computing the suppression of microhalo $J$ factors in Sec.~\ref{sec:dracoJ}.  Properly accounting for the tidally evolved profile in this calculation would be complicated because the profile changes over the course of tidal evolution.  Fortunately, the suppression of the annihilation rates within microhalos is already minimal at the radii at which Draco's density profile is tidally altered.  Since using the modified density profile would further reduce this suppression, using the original profile is a simplification that does not significantly impact our results.

\clearpage

\bibliography{references}

 \newcommand{\noop}[1]{}
\begin{thebibliography}{150}%
\makeatletter
\providecommand \@ifxundefined [1]{%
 \@ifx{#1\undefined}
}%
\providecommand \@ifnum [1]{%
 \ifnum #1\expandafter \@firstoftwo
 \else \expandafter \@secondoftwo
 \fi
}%
\providecommand \@ifx [1]{%
 \ifx #1\expandafter \@firstoftwo
 \else \expandafter \@secondoftwo
 \fi
}%
\providecommand \natexlab [1]{#1}%
\providecommand \enquote  [1]{``#1''}%
\providecommand \bibnamefont  [1]{#1}%
\providecommand \bibfnamefont [1]{#1}%
\providecommand \citenamefont [1]{#1}%
\providecommand \href@noop [0]{\@secondoftwo}%
\providecommand \href [0]{\begingroup \@sanitize@url \@href}%
\providecommand \@href[1]{\@@startlink{#1}\@@href}%
\providecommand \@@href[1]{\endgroup#1\@@endlink}%
\providecommand \@sanitize@url [0]{\catcode `\\12\catcode `\$12\catcode
  `\&12\catcode `\#12\catcode `\^12\catcode `\_12\catcode `\%12\relax}%
\providecommand \@@startlink[1]{}%
\providecommand \@@endlink[0]{}%
\providecommand \url  [0]{\begingroup\@sanitize@url \@url }%
\providecommand \@url [1]{\endgroup\@href {#1}{\urlprefix }}%
\providecommand \urlprefix  [0]{URL }%
\providecommand \Eprint [0]{\href }%
\providecommand \doibase [0]{http://dx.doi.org/}%
\providecommand \selectlanguage [0]{\@gobble}%
\providecommand \bibinfo  [0]{\@secondoftwo}%
\providecommand \bibfield  [0]{\@secondoftwo}%
\providecommand \translation [1]{[#1]}%
\providecommand \BibitemOpen [0]{}%
\providecommand \bibitemStop [0]{}%
\providecommand \bibitemNoStop [0]{.\EOS\space}%
\providecommand \EOS [0]{\spacefactor3000\relax}%
\providecommand \BibitemShut  [1]{\csname bibitem#1\endcsname}%
\let\auto@bib@innerbib\@empty
\bibitem [{\citenamefont {Kawasaki}\ \emph {et~al.}(1999)\citenamefont
  {Kawasaki}, \citenamefont {Kohri},\ and\ \citenamefont
  {Sugiyama}}]{kawasaki1999cosmological}%
  \BibitemOpen
  \bibfield  {author} {\bibinfo {author} {\bibfnamefont {M.}~\bibnamefont
  {Kawasaki}}, \bibinfo {author} {\bibfnamefont {K.}~\bibnamefont {Kohri}}, \
  and\ \bibinfo {author} {\bibfnamefont {N.}~\bibnamefont {Sugiyama}},\ }\href
  {\doibase 10.1103/physrevlett.82.4168} {\bibfield  {journal} {\bibinfo
  {journal} {Phys. Rev. Lett.}\ }\textbf {\bibinfo {volume} {82}},\ \bibinfo
  {pages} {4168} (\bibinfo {year} {1999})},\ \Eprint
  {http://arxiv.org/abs/astro-ph/9811437} {arXiv:astro-ph/9811437} \BibitemShut
  {NoStop}%
\bibitem [{\citenamefont {Kawasaki}\ \emph {et~al.}(2000)\citenamefont
  {Kawasaki}, \citenamefont {Kohri},\ and\ \citenamefont
  {Sugiyama}}]{kawasaki2000mev}%
  \BibitemOpen
  \bibfield  {author} {\bibinfo {author} {\bibfnamefont {M.}~\bibnamefont
  {Kawasaki}}, \bibinfo {author} {\bibfnamefont {K.}~\bibnamefont {Kohri}}, \
  and\ \bibinfo {author} {\bibfnamefont {N.}~\bibnamefont {Sugiyama}},\ }\href
  {\doibase 10.1103/physrevd.62.023506} {\bibfield  {journal} {\bibinfo
  {journal} {Phys. Rev. D}\ }\textbf {\bibinfo {volume} {62}},\ \bibinfo
  {pages} {023506} (\bibinfo {year} {2000})},\ \Eprint
  {http://arxiv.org/abs/astro-ph/0002127} {arXiv:astro-ph/0002127} \BibitemShut
  {NoStop}%
\bibitem [{\citenamefont {Hannestad}(2004)}]{hannestad2004lowest}%
  \BibitemOpen
  \bibfield  {author} {\bibinfo {author} {\bibfnamefont {S.}~\bibnamefont
  {Hannestad}},\ }\href {\doibase 10.1103/physrevd.70.043506} {\bibfield
  {journal} {\bibinfo  {journal} {Phys. Rev. D}\ }\textbf {\bibinfo {volume}
  {70}},\ \bibinfo {pages} {043506} (\bibinfo {year} {2004})},\ \Eprint
  {http://arxiv.org/abs/astro-ph/0403291} {arXiv:astro-ph/0403291} \BibitemShut
  {NoStop}%
\bibitem [{\citenamefont {Ichikawa}\ \emph {et~al.}(2005)\citenamefont
  {Ichikawa}, \citenamefont {Kawasaki},\ and\ \citenamefont
  {Takahashi}}]{ichikawa2005oscillation}%
  \BibitemOpen
  \bibfield  {author} {\bibinfo {author} {\bibfnamefont {K.}~\bibnamefont
  {Ichikawa}}, \bibinfo {author} {\bibfnamefont {M.}~\bibnamefont {Kawasaki}},
  \ and\ \bibinfo {author} {\bibfnamefont {F.}~\bibnamefont {Takahashi}},\
  }\href {\doibase 10.1103/physrevd.72.043522} {\bibfield  {journal} {\bibinfo
  {journal} {Phys. Rev. D}\ }\textbf {\bibinfo {volume} {72}},\ \bibinfo
  {pages} {043522} (\bibinfo {year} {2005})},\ \Eprint
  {http://arxiv.org/abs/astro-ph/0505395} {arXiv:astro-ph/0505395} \BibitemShut
  {NoStop}%
\bibitem [{\citenamefont {Ichikawa}\ \emph {et~al.}(2007)\citenamefont
  {Ichikawa}, \citenamefont {Kawasaki},\ and\ \citenamefont
  {Takahashi}}]{ichikawa2007constraint}%
  \BibitemOpen
  \bibfield  {author} {\bibinfo {author} {\bibfnamefont {K.}~\bibnamefont
  {Ichikawa}}, \bibinfo {author} {\bibfnamefont {M.}~\bibnamefont {Kawasaki}},
  \ and\ \bibinfo {author} {\bibfnamefont {F.}~\bibnamefont {Takahashi}},\
  }\href {\doibase 10.1088/1475-7516/2007/05/007} {\bibfield  {journal}
  {\bibinfo  {journal} {J. Cosmol. Astropart. Phys.}\ }\textbf {\bibinfo
  {volume} {05}},\ \bibinfo {pages} {007} (\bibinfo {year} {2007})},\ \Eprint
  {http://arxiv.org/abs/astro-ph/0611784} {arXiv:astro-ph/0611784} \BibitemShut
  {NoStop}%
\bibitem [{\citenamefont {De~Bernardis}\ \emph {et~al.}(2008)\citenamefont
  {De~Bernardis}, \citenamefont {Pagano},\ and\ \citenamefont
  {Melchiorri}}]{de2008new}%
  \BibitemOpen
  \bibfield  {author} {\bibinfo {author} {\bibfnamefont {F.}~\bibnamefont
  {De~Bernardis}}, \bibinfo {author} {\bibfnamefont {L.}~\bibnamefont
  {Pagano}}, \ and\ \bibinfo {author} {\bibfnamefont {A.}~\bibnamefont
  {Melchiorri}},\ }\href {\doibase 10.1016/j.astropartphys.2008.09.005}
  {\bibfield  {journal} {\bibinfo  {journal} {Astropart. Phys.}\ }\textbf
  {\bibinfo {volume} {30}},\ \bibinfo {pages} {192} (\bibinfo {year}
  {2008})}\BibitemShut {NoStop}%
\bibitem [{\citenamefont {Guth}(1981)}]{guth1981inflationary}%
  \BibitemOpen
  \bibfield  {author} {\bibinfo {author} {\bibfnamefont {A.~H.}\ \bibnamefont
  {Guth}},\ }\href {\doibase 10.1103/physrevd.23.347} {\bibfield  {journal}
  {\bibinfo  {journal} {Phys. Rev. D}\ }\textbf {\bibinfo {volume} {23}},\
  \bibinfo {pages} {347} (\bibinfo {year} {1981})}\BibitemShut {NoStop}%
\bibitem [{\citenamefont {Albrecht}\ and\ \citenamefont
  {Steinhardt}(1982)}]{albrecht1982cosmology}%
  \BibitemOpen
  \bibfield  {author} {\bibinfo {author} {\bibfnamefont {A.}~\bibnamefont
  {Albrecht}}\ and\ \bibinfo {author} {\bibfnamefont {P.~J.}\ \bibnamefont
  {Steinhardt}},\ }\href {\doibase 10.1103/physrevlett.48.1220} {\bibfield
  {journal} {\bibinfo  {journal} {Phys. Rev. Lett.}\ }\textbf {\bibinfo
  {volume} {48}},\ \bibinfo {pages} {1220} (\bibinfo {year}
  {1982})}\BibitemShut {NoStop}%
\bibitem [{\citenamefont {Linde}(1982)}]{linde1982new}%
  \BibitemOpen
  \bibfield  {author} {\bibinfo {author} {\bibfnamefont {A.~D.}\ \bibnamefont
  {Linde}},\ }\href {\doibase 10.1007/978-1-4613-2701-1_13} {\bibfield
  {journal} {\bibinfo  {journal} {Phys. Lett. B}\ }\textbf {\bibinfo {volume}
  {108}},\ \bibinfo {pages} {389} (\bibinfo {year} {1982})}\BibitemShut
  {NoStop}%
\bibitem [{\citenamefont {Liddle}(1994)}]{liddle1994inflationary}%
  \BibitemOpen
  \bibfield  {author} {\bibinfo {author} {\bibfnamefont {A.~R.}\ \bibnamefont
  {Liddle}},\ }\href {\doibase 10.1103/physrevd.49.739} {\bibfield  {journal}
  {\bibinfo  {journal} {Phys. Rev. D}\ }\textbf {\bibinfo {volume} {49}},\
  \bibinfo {pages} {739} (\bibinfo {year} {1994})},\ \Eprint
  {http://arxiv.org/abs/astro-ph/9307020} {arXiv:astro-ph/9307020} \BibitemShut
  {NoStop}%
\bibitem [{\citenamefont {Guo}\ \emph {et~al.}(2011)\citenamefont {Guo},
  \citenamefont {Schwarz},\ and\ \citenamefont {Zhang}}]{guo2011observational}%
  \BibitemOpen
  \bibfield  {author} {\bibinfo {author} {\bibfnamefont {Z.-K.}\ \bibnamefont
  {Guo}}, \bibinfo {author} {\bibfnamefont {D.~J.}\ \bibnamefont {Schwarz}}, \
  and\ \bibinfo {author} {\bibfnamefont {Y.-Z.}\ \bibnamefont {Zhang}},\ }\href
  {\doibase 10.1103/physrevd.83.083522} {\bibfield  {journal} {\bibinfo
  {journal} {Phys. Rev. D}\ }\textbf {\bibinfo {volume} {83}},\ \bibinfo
  {pages} {083522} (\bibinfo {year} {2011})},\ \Eprint
  {http://arxiv.org/abs/1008.5258} {arXiv:1008.5258} \BibitemShut {NoStop}%
\bibitem [{\citenamefont {Amin}\ \emph {et~al.}(2015)\citenamefont {Amin},
  \citenamefont {Hertzberg}, \citenamefont {Kaiser},\ and\ \citenamefont
  {Karouby}}]{Amin_2014}%
  \BibitemOpen
  \bibfield  {author} {\bibinfo {author} {\bibfnamefont {M.~A.}\ \bibnamefont
  {Amin}}, \bibinfo {author} {\bibfnamefont {M.~P.}\ \bibnamefont {Hertzberg}},
  \bibinfo {author} {\bibfnamefont {D.~I.}\ \bibnamefont {Kaiser}}, \ and\
  \bibinfo {author} {\bibfnamefont {J.}~\bibnamefont {Karouby}},\ }\href
  {\doibase 10.1142/s0218271815300037} {\bibfield  {journal} {\bibinfo
  {journal} {Int. J. Mod. Phys. D}\ }\textbf {\bibinfo {volume} {24}},\
  \bibinfo {pages} {1530003} (\bibinfo {year} {2015})},\ \Eprint
  {http://arxiv.org/abs/1410.3808} {arXiv:1410.3808} \BibitemShut {NoStop}%
\bibitem [{\citenamefont {Pospelov}\ \emph {et~al.}(2008)\citenamefont
  {Pospelov}, \citenamefont {Ritz},\ and\ \citenamefont
  {Voloshin}}]{Pospelov_2008}%
  \BibitemOpen
  \bibfield  {author} {\bibinfo {author} {\bibfnamefont {M.}~\bibnamefont
  {Pospelov}}, \bibinfo {author} {\bibfnamefont {A.}~\bibnamefont {Ritz}}, \
  and\ \bibinfo {author} {\bibfnamefont {M.}~\bibnamefont {Voloshin}},\ }\href
  {\doibase 10.1016/j.physletb.2008.02.052} {\bibfield  {journal} {\bibinfo
  {journal} {Phys. Lett. B}\ }\textbf {\bibinfo {volume} {662}},\ \bibinfo
  {pages} {53} (\bibinfo {year} {2008})},\ \Eprint
  {http://arxiv.org/abs/0711.4866} {arXiv:0711.4866} \BibitemShut {NoStop}%
\bibitem [{\citenamefont {Arkani-Hamed}\ \emph {et~al.}(2009)\citenamefont
  {Arkani-Hamed}, \citenamefont {Finkbeiner}, \citenamefont {Slatyer},\ and\
  \citenamefont {Weiner}}]{Arkani_Hamed_2009}%
  \BibitemOpen
  \bibfield  {author} {\bibinfo {author} {\bibfnamefont {N.}~\bibnamefont
  {Arkani-Hamed}}, \bibinfo {author} {\bibfnamefont {D.~P.}\ \bibnamefont
  {Finkbeiner}}, \bibinfo {author} {\bibfnamefont {T.~R.}\ \bibnamefont
  {Slatyer}}, \ and\ \bibinfo {author} {\bibfnamefont {N.}~\bibnamefont
  {Weiner}},\ }\href {\doibase 10.1103/PhysRevD.79.015014} {\bibfield
  {journal} {\bibinfo  {journal} {Phys. Rev. D}\ }\textbf {\bibinfo {volume}
  {79}},\ \bibinfo {pages} {015014} (\bibinfo {year} {2009})},\ \Eprint
  {http://arxiv.org/abs/0810.0713} {arXiv:0810.0713} \BibitemShut {NoStop}%
\bibitem [{\citenamefont {Hooper}\ \emph {et~al.}(2012)\citenamefont {Hooper},
  \citenamefont {Weiner},\ and\ \citenamefont {Xue}}]{Hooper_2012}%
  \BibitemOpen
  \bibfield  {author} {\bibinfo {author} {\bibfnamefont {D.}~\bibnamefont
  {Hooper}}, \bibinfo {author} {\bibfnamefont {N.}~\bibnamefont {Weiner}}, \
  and\ \bibinfo {author} {\bibfnamefont {W.}~\bibnamefont {Xue}},\ }\href
  {\doibase 10.1103/PhysRevD.86.056009} {\bibfield  {journal} {\bibinfo
  {journal} {Phys. Rev. D}\ }\textbf {\bibinfo {volume} {86}},\ \bibinfo
  {pages} {056009} (\bibinfo {year} {2012})},\ \Eprint
  {http://arxiv.org/abs/1206.2929} {arXiv:1206.2929} \BibitemShut {NoStop}%
\bibitem [{\citenamefont {Abdullah}\ \emph {et~al.}(2014)\citenamefont
  {Abdullah}, \citenamefont {DiFranzo}, \citenamefont {Rajaraman},
  \citenamefont {Tait}, \citenamefont {Tanedo},\ and\ \citenamefont
  {Wijangco}}]{Abdullah_2014}%
  \BibitemOpen
  \bibfield  {author} {\bibinfo {author} {\bibfnamefont {M.}~\bibnamefont
  {Abdullah}}, \bibinfo {author} {\bibfnamefont {A.}~\bibnamefont {DiFranzo}},
  \bibinfo {author} {\bibfnamefont {A.}~\bibnamefont {Rajaraman}}, \bibinfo
  {author} {\bibfnamefont {T.~M.~P.}\ \bibnamefont {Tait}}, \bibinfo {author}
  {\bibfnamefont {P.}~\bibnamefont {Tanedo}}, \ and\ \bibinfo {author}
  {\bibfnamefont {A.~M.}\ \bibnamefont {Wijangco}},\ }\href {\doibase
  10.1103/PhysRevD.90.035004} {\bibfield  {journal} {\bibinfo  {journal} {Phys.
  Rev. D}\ }\textbf {\bibinfo {volume} {90}},\ \bibinfo {pages} {035004}
  (\bibinfo {year} {2014})},\ \Eprint {http://arxiv.org/abs/1404.6528}
  {arXiv:1404.6528} \BibitemShut {NoStop}%
\bibitem [{\citenamefont {Berlin}\ \emph {et~al.}(2014)\citenamefont {Berlin},
  \citenamefont {Gratia}, \citenamefont {Hooper},\ and\ \citenamefont
  {McDermott}}]{Berlin_2014}%
  \BibitemOpen
  \bibfield  {author} {\bibinfo {author} {\bibfnamefont {A.}~\bibnamefont
  {Berlin}}, \bibinfo {author} {\bibfnamefont {P.}~\bibnamefont {Gratia}},
  \bibinfo {author} {\bibfnamefont {D.}~\bibnamefont {Hooper}}, \ and\ \bibinfo
  {author} {\bibfnamefont {S.~D.}\ \bibnamefont {McDermott}},\ }\href {\doibase
  10.1103/PhysRevD.90.015032} {\bibfield  {journal} {\bibinfo  {journal} {Phys.
  Rev. D}\ }\textbf {\bibinfo {volume} {90}},\ \bibinfo {pages} {015032}
  (\bibinfo {year} {2014})}\BibitemShut {NoStop}%
\bibitem [{\citenamefont {Martin}\ \emph {et~al.}(2014)\citenamefont {Martin},
  \citenamefont {Shelton},\ and\ \citenamefont {Unwin}}]{Martin_2014}%
  \BibitemOpen
  \bibfield  {author} {\bibinfo {author} {\bibfnamefont {A.}~\bibnamefont
  {Martin}}, \bibinfo {author} {\bibfnamefont {J.}~\bibnamefont {Shelton}}, \
  and\ \bibinfo {author} {\bibfnamefont {J.}~\bibnamefont {Unwin}},\ }\href
  {\doibase 10.1103/PhysRevD.90.103513} {\bibfield  {journal} {\bibinfo
  {journal} {Phys. Rev. D}\ }\textbf {\bibinfo {volume} {90}},\ \bibinfo
  {pages} {103513} (\bibinfo {year} {2014})},\ \Eprint
  {http://arxiv.org/abs/1405.0272} {arXiv:1405.0272} \BibitemShut {NoStop}%
\bibitem [{\citenamefont {Zhang}(2015)}]{zhang2015long}%
  \BibitemOpen
  \bibfield  {author} {\bibinfo {author} {\bibfnamefont {Y.}~\bibnamefont
  {Zhang}},\ }\href {\doibase 10.1088/1475-7516/2015/05/008} {\bibfield
  {journal} {\bibinfo  {journal} {J. Cosmol. Astropart. Phys.}\ }\textbf
  {\bibinfo {volume} {05}},\ \bibinfo {pages} {008} (\bibinfo {year} {2015})},\
  \Eprint {http://arxiv.org/abs/1502.06983} {arXiv:1502.06983} \BibitemShut
  {NoStop}%
\bibitem [{\citenamefont {Berlin}\ \emph
  {et~al.}(2016{\natexlab{a}})\citenamefont {Berlin}, \citenamefont {Hooper},\
  and\ \citenamefont {Krnjaic}}]{Berlin_2016a}%
  \BibitemOpen
  \bibfield  {author} {\bibinfo {author} {\bibfnamefont {A.}~\bibnamefont
  {Berlin}}, \bibinfo {author} {\bibfnamefont {D.}~\bibnamefont {Hooper}}, \
  and\ \bibinfo {author} {\bibfnamefont {G.}~\bibnamefont {Krnjaic}},\ }\href
  {\doibase 10.1016/j.physletb.2016.06.037} {\bibfield  {journal} {\bibinfo
  {journal} {Phys. Lett. B}\ }\textbf {\bibinfo {volume} {760}},\ \bibinfo
  {pages} {106} (\bibinfo {year} {2016}{\natexlab{a}})},\ \Eprint
  {http://arxiv.org/abs/1602.08490} {arXiv:1602.08490} \BibitemShut {NoStop}%
\bibitem [{\citenamefont {Berlin}\ \emph
  {et~al.}(2016{\natexlab{b}})\citenamefont {Berlin}, \citenamefont {Hooper},\
  and\ \citenamefont {Krnjaic}}]{Berlin_2016b}%
  \BibitemOpen
  \bibfield  {author} {\bibinfo {author} {\bibfnamefont {A.}~\bibnamefont
  {Berlin}}, \bibinfo {author} {\bibfnamefont {D.}~\bibnamefont {Hooper}}, \
  and\ \bibinfo {author} {\bibfnamefont {G.}~\bibnamefont {Krnjaic}},\ }\href
  {\doibase 10.1103/PhysRevD.94.095019} {\bibfield  {journal} {\bibinfo
  {journal} {Phys. Rev. D}\ }\textbf {\bibinfo {volume} {94}},\ \bibinfo
  {pages} {095019} (\bibinfo {year} {2016}{\natexlab{b}})},\ \Eprint
  {http://arxiv.org/abs/1609.02555} {arXiv:1609.02555} \BibitemShut {NoStop}%
\bibitem [{\citenamefont {Dror}\ \emph {et~al.}(2016)\citenamefont {Dror},
  \citenamefont {Kuflik},\ and\ \citenamefont {Ng}}]{Dror_2016}%
  \BibitemOpen
  \bibfield  {author} {\bibinfo {author} {\bibfnamefont {J.~A.}\ \bibnamefont
  {Dror}}, \bibinfo {author} {\bibfnamefont {E.}~\bibnamefont {Kuflik}}, \ and\
  \bibinfo {author} {\bibfnamefont {W.~H.}\ \bibnamefont {Ng}},\ }\href
  {\doibase 10.1103/PhysRevLett.117.211801} {\bibfield  {journal} {\bibinfo
  {journal} {Phys. Rev. Lett.}\ }\textbf {\bibinfo {volume} {117}},\ \bibinfo
  {pages} {211801} (\bibinfo {year} {2016})},\ \Eprint
  {http://arxiv.org/abs/1607.03110} {arXiv:1607.03110} \BibitemShut {NoStop}%
\bibitem [{\citenamefont {Tenkanen}\ and\ \citenamefont
  {Vaskonen}(2016)}]{Tenkanen:2016jic}%
  \BibitemOpen
  \bibfield  {author} {\bibinfo {author} {\bibfnamefont {T.}~\bibnamefont
  {Tenkanen}}\ and\ \bibinfo {author} {\bibfnamefont {V.}~\bibnamefont
  {Vaskonen}},\ }\href {\doibase 10.1103/PhysRevD.94.083516} {\bibfield
  {journal} {\bibinfo  {journal} {Phys. Rev. D}\ }\textbf {\bibinfo {volume}
  {94}},\ \bibinfo {pages} {083516} (\bibinfo {year} {2016})},\ \Eprint
  {http://arxiv.org/abs/1606.00192} {arXiv:1606.00192} \BibitemShut {NoStop}%
\bibitem [{\citenamefont {Dror}\ \emph {et~al.}(2018)\citenamefont {Dror},
  \citenamefont {Kuflik}, \citenamefont {Melcher},\ and\ \citenamefont
  {Watson}}]{Dror_2018}%
  \BibitemOpen
  \bibfield  {author} {\bibinfo {author} {\bibfnamefont {J.~A.}\ \bibnamefont
  {Dror}}, \bibinfo {author} {\bibfnamefont {E.}~\bibnamefont {Kuflik}},
  \bibinfo {author} {\bibfnamefont {B.}~\bibnamefont {Melcher}}, \ and\
  \bibinfo {author} {\bibfnamefont {S.}~\bibnamefont {Watson}},\ }\href
  {\doibase 10.1103/PhysRevD.97.063524} {\bibfield  {journal} {\bibinfo
  {journal} {Phys. Rev. D}\ }\textbf {\bibinfo {volume} {97}},\ \bibinfo
  {pages} {063524} (\bibinfo {year} {2018})},\ \Eprint
  {http://arxiv.org/abs/1711.04773} {arXiv:1711.04773} \BibitemShut {NoStop}%
\bibitem [{\citenamefont {Tenkanen}(2019)}]{Tenkanen:2019cik}%
  \BibitemOpen
  \bibfield  {author} {\bibinfo {author} {\bibfnamefont {T.}~\bibnamefont
  {Tenkanen}},\ }\href {\doibase 10.1103/PhysRevD.100.083515} {\bibfield
  {journal} {\bibinfo  {journal} {Phys. Rev. D}\ }\textbf {\bibinfo {volume}
  {100}},\ \bibinfo {pages} {083515} (\bibinfo {year} {2019})},\ \Eprint
  {http://arxiv.org/abs/1905.11737} {arXiv:1905.11737} \BibitemShut {NoStop}%
\bibitem [{\citenamefont {Coughlan}\ \emph {et~al.}(1983)\citenamefont
  {Coughlan}, \citenamefont {Fischler}, \citenamefont {Kolb}, \citenamefont
  {Raby},\ and\ \citenamefont {Ross}}]{coughlan1983cosmological}%
  \BibitemOpen
  \bibfield  {author} {\bibinfo {author} {\bibfnamefont {G.}~\bibnamefont
  {Coughlan}}, \bibinfo {author} {\bibfnamefont {W.}~\bibnamefont {Fischler}},
  \bibinfo {author} {\bibfnamefont {E.~W.}\ \bibnamefont {Kolb}}, \bibinfo
  {author} {\bibfnamefont {S.}~\bibnamefont {Raby}}, \ and\ \bibinfo {author}
  {\bibfnamefont {G.~G.}\ \bibnamefont {Ross}},\ }\href {\doibase
  10.1016/0370-2693(83)91091-2} {\bibfield  {journal} {\bibinfo  {journal}
  {Phys. Lett. B}\ }\textbf {\bibinfo {volume} {131}},\ \bibinfo {pages} {59}
  (\bibinfo {year} {1983})}\BibitemShut {NoStop}%
\bibitem [{\citenamefont {De~Carlos}\ \emph {et~al.}(1993)\citenamefont
  {De~Carlos}, \citenamefont {Casas}, \citenamefont {Quevedo},\ and\
  \citenamefont {Roulet}}]{de1993model}%
  \BibitemOpen
  \bibfield  {author} {\bibinfo {author} {\bibfnamefont {B.}~\bibnamefont
  {De~Carlos}}, \bibinfo {author} {\bibfnamefont {J.}~\bibnamefont {Casas}},
  \bibinfo {author} {\bibfnamefont {F.}~\bibnamefont {Quevedo}}, \ and\
  \bibinfo {author} {\bibfnamefont {E.}~\bibnamefont {Roulet}},\ }\href
  {\doibase 10.1016/0370-2693(93)91538-x} {\bibfield  {journal} {\bibinfo
  {journal} {Phys. Lett. B}\ }\textbf {\bibinfo {volume} {318}},\ \bibinfo
  {pages} {447} (\bibinfo {year} {1993})},\ \Eprint
  {http://arxiv.org/abs/hep-ph/9308325} {arXiv:hep-ph/9308325} \BibitemShut
  {NoStop}%
\bibitem [{\citenamefont {Banks}\ \emph {et~al.}(1994)\citenamefont {Banks},
  \citenamefont {Kaplan},\ and\ \citenamefont
  {Nelson}}]{banks1994cosmological}%
  \BibitemOpen
  \bibfield  {author} {\bibinfo {author} {\bibfnamefont {T.}~\bibnamefont
  {Banks}}, \bibinfo {author} {\bibfnamefont {D.~B.}\ \bibnamefont {Kaplan}}, \
  and\ \bibinfo {author} {\bibfnamefont {A.~E.}\ \bibnamefont {Nelson}},\
  }\href {\doibase 10.1103/physrevd.49.779} {\bibfield  {journal} {\bibinfo
  {journal} {Phys. Rev. D}\ }\textbf {\bibinfo {volume} {49}},\ \bibinfo
  {pages} {779} (\bibinfo {year} {1994})},\ \Eprint
  {http://arxiv.org/abs/hep-ph/9308292} {arXiv:hep-ph/9308292} \BibitemShut
  {NoStop}%
\bibitem [{\citenamefont {Banks}\ \emph
  {et~al.}(1995{\natexlab{a}})\citenamefont {Banks}, \citenamefont {Berkooz},\
  and\ \citenamefont {Steinhardt}}]{banks1995cosmological}%
  \BibitemOpen
  \bibfield  {author} {\bibinfo {author} {\bibfnamefont {T.}~\bibnamefont
  {Banks}}, \bibinfo {author} {\bibfnamefont {M.}~\bibnamefont {Berkooz}}, \
  and\ \bibinfo {author} {\bibfnamefont {P.~J.}\ \bibnamefont {Steinhardt}},\
  }\href {\doibase 10.1103/physrevd.52.705} {\bibfield  {journal} {\bibinfo
  {journal} {Phys. Rev. D}\ }\textbf {\bibinfo {volume} {52}},\ \bibinfo
  {pages} {705} (\bibinfo {year} {1995}{\natexlab{a}})},\ \Eprint
  {http://arxiv.org/abs/hep-th/9501053} {arXiv:hep-th/9501053} \BibitemShut
  {NoStop}%
\bibitem [{\citenamefont {Banks}\ \emph
  {et~al.}(1995{\natexlab{b}})\citenamefont {Banks}, \citenamefont {Berkooz},
  \citenamefont {Shenker}, \citenamefont {Moore},\ and\ \citenamefont
  {Steinhardt}}]{banks1995modular}%
  \BibitemOpen
  \bibfield  {author} {\bibinfo {author} {\bibfnamefont {T.}~\bibnamefont
  {Banks}}, \bibinfo {author} {\bibfnamefont {M.}~\bibnamefont {Berkooz}},
  \bibinfo {author} {\bibfnamefont {S.~H.}\ \bibnamefont {Shenker}}, \bibinfo
  {author} {\bibfnamefont {G.}~\bibnamefont {Moore}}, \ and\ \bibinfo {author}
  {\bibfnamefont {P.~J.}\ \bibnamefont {Steinhardt}},\ }\href {\doibase
  10.1103/physrevd.52.3548} {\bibfield  {journal} {\bibinfo  {journal} {Phys.
  Rev. D}\ }\textbf {\bibinfo {volume} {52}},\ \bibinfo {pages} {3548}
  (\bibinfo {year} {1995}{\natexlab{b}})},\ \Eprint
  {http://arxiv.org/abs/hep-th/9503114} {arXiv:hep-th/9503114} \BibitemShut
  {NoStop}%
\bibitem [{\citenamefont {Acharya}\ \emph {et~al.}(2014)\citenamefont
  {Acharya}, \citenamefont {Kane},\ and\ \citenamefont
  {Kuflik}}]{acharya2014bounds}%
  \BibitemOpen
  \bibfield  {author} {\bibinfo {author} {\bibfnamefont {B.~S.}\ \bibnamefont
  {Acharya}}, \bibinfo {author} {\bibfnamefont {G.}~\bibnamefont {Kane}}, \
  and\ \bibinfo {author} {\bibfnamefont {E.}~\bibnamefont {Kuflik}},\ }\href
  {\doibase 10.1142/s0217751x14500730} {\bibfield  {journal} {\bibinfo
  {journal} {Int. J. Mod. Phys. A}\ }\textbf {\bibinfo {volume} {29}},\
  \bibinfo {pages} {1450073} (\bibinfo {year} {2014})},\ \Eprint
  {http://arxiv.org/abs/1006.3272} {arXiv:1006.3272} \BibitemShut {NoStop}%
\bibitem [{\citenamefont {Kane}\ \emph {et~al.}(2015)\citenamefont {Kane},
  \citenamefont {Sinha},\ and\ \citenamefont {Watson}}]{Kane_2015}%
  \BibitemOpen
  \bibfield  {author} {\bibinfo {author} {\bibfnamefont {G.}~\bibnamefont
  {Kane}}, \bibinfo {author} {\bibfnamefont {K.}~\bibnamefont {Sinha}}, \ and\
  \bibinfo {author} {\bibfnamefont {S.}~\bibnamefont {Watson}},\ }\href
  {\doibase 10.1142/s0218271815300220} {\bibfield  {journal} {\bibinfo
  {journal} {Int. J. Mod. Phys. D}\ }\textbf {\bibinfo {volume} {24}},\
  \bibinfo {pages} {1530022} (\bibinfo {year} {2015})},\ \Eprint
  {http://arxiv.org/abs/1502.07746} {arXiv:1502.07746} \BibitemShut {NoStop}%
\bibitem [{\citenamefont {Giblin}\ \emph {et~al.}(2017)\citenamefont {Giblin},
  \citenamefont {Kane}, \citenamefont {Nesbit}, \citenamefont {Watson},\ and\
  \citenamefont {Zhao}}]{Giblin_2017}%
  \BibitemOpen
  \bibfield  {author} {\bibinfo {author} {\bibfnamefont {J.~T.}\ \bibnamefont
  {Giblin}}, \bibinfo {author} {\bibfnamefont {G.}~\bibnamefont {Kane}},
  \bibinfo {author} {\bibfnamefont {E.}~\bibnamefont {Nesbit}}, \bibinfo
  {author} {\bibfnamefont {S.}~\bibnamefont {Watson}}, \ and\ \bibinfo {author}
  {\bibfnamefont {Y.}~\bibnamefont {Zhao}},\ }\href {\doibase
  10.1103/PhysRevD.96.043525} {\bibfield  {journal} {\bibinfo  {journal} {Phys.
  Rev. D}\ }\textbf {\bibinfo {volume} {96}},\ \bibinfo {pages} {043525}
  (\bibinfo {year} {2017})},\ \Eprint {http://arxiv.org/abs/1706.08536}
  {arXiv:1706.08536} \BibitemShut {NoStop}%
\bibitem [{\citenamefont {Mollerach}(1990)}]{mollerach1990isocurvature}%
  \BibitemOpen
  \bibfield  {author} {\bibinfo {author} {\bibfnamefont {S.}~\bibnamefont
  {Mollerach}},\ }\href {\doibase 10.1103/physrevd.42.313} {\bibfield
  {journal} {\bibinfo  {journal} {Phys. Rev. D}\ }\textbf {\bibinfo {volume}
  {42}},\ \bibinfo {pages} {313} (\bibinfo {year} {1990})}\BibitemShut
  {NoStop}%
\bibitem [{\citenamefont {Linde}\ and\ \citenamefont
  {Mukhanov}(1997)}]{linde1997non}%
  \BibitemOpen
  \bibfield  {author} {\bibinfo {author} {\bibfnamefont {A.}~\bibnamefont
  {Linde}}\ and\ \bibinfo {author} {\bibfnamefont {V.}~\bibnamefont
  {Mukhanov}},\ }\href {\doibase 10.1103/physrevd.56.r535} {\bibfield
  {journal} {\bibinfo  {journal} {Phys. Rev. D}\ }\textbf {\bibinfo {volume}
  {56}},\ \bibinfo {pages} {R535(R)} (\bibinfo {year} {1997})},\ \Eprint
  {http://arxiv.org/abs/astro-ph/9610219} {arXiv:astro-ph/9610219} \BibitemShut
  {NoStop}%
\bibitem [{\citenamefont {Lyth}\ and\ \citenamefont
  {Wands}(2002)}]{lyth2002generating}%
  \BibitemOpen
  \bibfield  {author} {\bibinfo {author} {\bibfnamefont {D.~H.}\ \bibnamefont
  {Lyth}}\ and\ \bibinfo {author} {\bibfnamefont {D.}~\bibnamefont {Wands}},\
  }\href {\doibase 10.1016/s0370-2693(01)01366-1} {\bibfield  {journal}
  {\bibinfo  {journal} {Phys. Lett. B}\ }\textbf {\bibinfo {volume} {524}},\
  \bibinfo {pages} {5} (\bibinfo {year} {2002})},\ \Eprint
  {http://arxiv.org/abs/hep-ph/0110002} {arXiv:hep-ph/0110002} \BibitemShut
  {NoStop}%
\bibitem [{\citenamefont {Moroi}\ and\ \citenamefont
  {Takahashi}(2001)}]{moroi2001effects}%
  \BibitemOpen
  \bibfield  {author} {\bibinfo {author} {\bibfnamefont {T.}~\bibnamefont
  {Moroi}}\ and\ \bibinfo {author} {\bibfnamefont {T.}~\bibnamefont
  {Takahashi}},\ }\href {\doibase 10.1016/s0370-2693(01)01295-3} {\bibfield
  {journal} {\bibinfo  {journal} {Phys. Lett. B}\ }\textbf {\bibinfo {volume}
  {522}},\ \bibinfo {pages} {215} (\bibinfo {year} {2001})},\ \Eprint
  {http://arxiv.org/abs/hep-ph/0110096} {arXiv:hep-ph/0110096} \BibitemShut
  {NoStop}%
\bibitem [{\citenamefont {Moroi}\ and\ \citenamefont
  {Takahashi}(2002)}]{moroi2002erratum}%
  \BibitemOpen
  \bibfield  {author} {\bibinfo {author} {\bibfnamefont {T.}~\bibnamefont
  {Moroi}}\ and\ \bibinfo {author} {\bibfnamefont {T.}~\bibnamefont
  {Takahashi}},\ }\href {\doibase 10.1016/S0370-2693(02)02070-1} {\ \textbf
  {\bibinfo {volume} {539}},\ \bibinfo {pages} {303} (\bibinfo {year}
  {2002})}\BibitemShut {NoStop}%
\bibitem [{\citenamefont {Albrecht}\ \emph {et~al.}(1982)\citenamefont
  {Albrecht}, \citenamefont {Steinhardt}, \citenamefont {Turner},\ and\
  \citenamefont {Wilczek}}]{albrecht1982reheating}%
  \BibitemOpen
  \bibfield  {author} {\bibinfo {author} {\bibfnamefont {A.}~\bibnamefont
  {Albrecht}}, \bibinfo {author} {\bibfnamefont {P.~J.}\ \bibnamefont
  {Steinhardt}}, \bibinfo {author} {\bibfnamefont {M.~S.}\ \bibnamefont
  {Turner}}, \ and\ \bibinfo {author} {\bibfnamefont {F.}~\bibnamefont
  {Wilczek}},\ }\href {\doibase 10.1103/physrevlett.48.1437} {\bibfield
  {journal} {\bibinfo  {journal} {Phys. Rev. Lett.}\ }\textbf {\bibinfo
  {volume} {48}},\ \bibinfo {pages} {1437} (\bibinfo {year}
  {1982})}\BibitemShut {NoStop}%
\bibitem [{\citenamefont {Turner}(1983)}]{turner1983coherent}%
  \BibitemOpen
  \bibfield  {author} {\bibinfo {author} {\bibfnamefont {M.~S.}\ \bibnamefont
  {Turner}},\ }\href {\doibase 10.1103/physrevd.28.1243} {\bibfield  {journal}
  {\bibinfo  {journal} {Phys. Rev. D}\ }\textbf {\bibinfo {volume} {28}},\
  \bibinfo {pages} {1243} (\bibinfo {year} {1983})}\BibitemShut {NoStop}%
\bibitem [{\citenamefont {Traschen}\ and\ \citenamefont
  {Brandenberger}(1990)}]{traschen1990particle}%
  \BibitemOpen
  \bibfield  {author} {\bibinfo {author} {\bibfnamefont {J.~H.}\ \bibnamefont
  {Traschen}}\ and\ \bibinfo {author} {\bibfnamefont {R.~H.}\ \bibnamefont
  {Brandenberger}},\ }\href {\doibase 10.1103/physrevd.42.2491} {\bibfield
  {journal} {\bibinfo  {journal} {Phys. Rev. D}\ }\textbf {\bibinfo {volume}
  {42}},\ \bibinfo {pages} {2491} (\bibinfo {year} {1990})}\BibitemShut
  {NoStop}%
\bibitem [{\citenamefont {Kofman}\ \emph {et~al.}(1994)\citenamefont {Kofman},
  \citenamefont {Linde},\ and\ \citenamefont
  {Starobinsky}}]{kofman1994reheating}%
  \BibitemOpen
  \bibfield  {author} {\bibinfo {author} {\bibfnamefont {L.}~\bibnamefont
  {Kofman}}, \bibinfo {author} {\bibfnamefont {A.}~\bibnamefont {Linde}}, \
  and\ \bibinfo {author} {\bibfnamefont {A.~A.}\ \bibnamefont {Starobinsky}},\
  }\href {\doibase 10.1103/physrevlett.73.3195} {\bibfield  {journal} {\bibinfo
   {journal} {Phys. Rev. Lett.}\ }\textbf {\bibinfo {volume} {73}},\ \bibinfo
  {pages} {3195} (\bibinfo {year} {1994})},\ \Eprint
  {http://arxiv.org/abs/hep-th/9405187} {arXiv:hep-th/9405187} \BibitemShut
  {NoStop}%
\bibitem [{\citenamefont {Kofman}\ \emph {et~al.}(1997)\citenamefont {Kofman},
  \citenamefont {Linde},\ and\ \citenamefont
  {Starobinsky}}]{kofman1997towards}%
  \BibitemOpen
  \bibfield  {author} {\bibinfo {author} {\bibfnamefont {L.}~\bibnamefont
  {Kofman}}, \bibinfo {author} {\bibfnamefont {A.}~\bibnamefont {Linde}}, \
  and\ \bibinfo {author} {\bibfnamefont {A.~A.}\ \bibnamefont {Starobinsky}},\
  }\href {\doibase 10.1103/physrevd.56.3258} {\bibfield  {journal} {\bibinfo
  {journal} {Phys. Rev. D}\ }\textbf {\bibinfo {volume} {56}},\ \bibinfo
  {pages} {3258} (\bibinfo {year} {1997})},\ \Eprint
  {http://arxiv.org/abs/hep-ph/9704452} {arXiv:hep-ph/9704452} \BibitemShut
  {NoStop}%
\bibitem [{\citenamefont {Dufaux}\ \emph {et~al.}(2006)\citenamefont {Dufaux},
  \citenamefont {Felder}, \citenamefont {Kofman}, \citenamefont {Peloso},\ and\
  \citenamefont {Podolsky}}]{dufaux2006preheating}%
  \BibitemOpen
  \bibfield  {author} {\bibinfo {author} {\bibfnamefont {J.~F.}\ \bibnamefont
  {Dufaux}}, \bibinfo {author} {\bibfnamefont {G.~N.}\ \bibnamefont {Felder}},
  \bibinfo {author} {\bibfnamefont {L.}~\bibnamefont {Kofman}}, \bibinfo
  {author} {\bibfnamefont {M.}~\bibnamefont {Peloso}}, \ and\ \bibinfo {author}
  {\bibfnamefont {D.}~\bibnamefont {Podolsky}},\ }\href {\doibase
  10.1088/1475-7516/2006/07/006} {\bibfield  {journal} {\bibinfo  {journal} {J.
  Cosmol. Astropart. Phys.}\ }\textbf {\bibinfo {volume} {07}},\ \bibinfo
  {pages} {006} (\bibinfo {year} {2006})},\ \Eprint
  {http://arxiv.org/abs/hep-ph/0602144} {arXiv:hep-ph/0602144} \BibitemShut
  {NoStop}%
\bibitem [{\citenamefont {Allahverdi}\ \emph {et~al.}(2010)\citenamefont
  {Allahverdi}, \citenamefont {Brandenberger}, \citenamefont {Cyr-Racine},\
  and\ \citenamefont {Mazumdar}}]{allahverdi2010reheating}%
  \BibitemOpen
  \bibfield  {author} {\bibinfo {author} {\bibfnamefont {R.}~\bibnamefont
  {Allahverdi}}, \bibinfo {author} {\bibfnamefont {R.}~\bibnamefont
  {Brandenberger}}, \bibinfo {author} {\bibfnamefont {F.-Y.}\ \bibnamefont
  {Cyr-Racine}}, \ and\ \bibinfo {author} {\bibfnamefont {A.}~\bibnamefont
  {Mazumdar}},\ }\href {\doibase 10.1146/annurev.nucl.012809.104511} {\bibfield
   {journal} {\bibinfo  {journal} {Ann. Rev. Nucl. Part. Sci.}\ }\textbf
  {\bibinfo {volume} {60}},\ \bibinfo {pages} {27} (\bibinfo {year} {2010})},\
  \Eprint {http://arxiv.org/abs/1001.2600} {arXiv:1001.2600} \BibitemShut
  {NoStop}%
\bibitem [{\citenamefont {Jedamzik}\ \emph {et~al.}(2010)\citenamefont
  {Jedamzik}, \citenamefont {Lemoine},\ and\ \citenamefont
  {Martin}}]{jedamzik2010collapse}%
  \BibitemOpen
  \bibfield  {author} {\bibinfo {author} {\bibfnamefont {K.}~\bibnamefont
  {Jedamzik}}, \bibinfo {author} {\bibfnamefont {M.}~\bibnamefont {Lemoine}}, \
  and\ \bibinfo {author} {\bibfnamefont {J.}~\bibnamefont {Martin}},\ }\href
  {\doibase 10.1088/1475-7516/2010/09/034} {\bibfield  {journal} {\bibinfo
  {journal} {J. Cosmol. Astropart. Phys.}\ }\textbf {\bibinfo {volume} {09}},\
  \bibinfo {pages} {034} (\bibinfo {year} {2010})},\ \Eprint
  {http://arxiv.org/abs/1002.3039} {arXiv:1002.3039} \BibitemShut {NoStop}%
\bibitem [{\citenamefont {Easther}\ \emph {et~al.}(2011)\citenamefont
  {Easther}, \citenamefont {Flauger},\ and\ \citenamefont
  {Gilmore}}]{easther2011delayed}%
  \BibitemOpen
  \bibfield  {author} {\bibinfo {author} {\bibfnamefont {R.}~\bibnamefont
  {Easther}}, \bibinfo {author} {\bibfnamefont {R.}~\bibnamefont {Flauger}}, \
  and\ \bibinfo {author} {\bibfnamefont {J.~B.}\ \bibnamefont {Gilmore}},\
  }\href {\doibase 10.1088/1475-7516/2011/04/027} {\bibfield  {journal}
  {\bibinfo  {journal} {J. Cosmol. Astropart. Phys.}\ }\textbf {\bibinfo
  {volume} {04}},\ \bibinfo {pages} {027} (\bibinfo {year} {2011})},\ \Eprint
  {http://arxiv.org/abs/1003.3011} {arXiv:1003.3011} \BibitemShut {NoStop}%
\bibitem [{\citenamefont {Musoke}\ \emph {et~al.}()\citenamefont {Musoke},
  \citenamefont {Hotchkiss},\ and\ \citenamefont
  {Easther}}]{musoke2019lighting}%
  \BibitemOpen
  \bibfield  {author} {\bibinfo {author} {\bibfnamefont {N.}~\bibnamefont
  {Musoke}}, \bibinfo {author} {\bibfnamefont {S.}~\bibnamefont {Hotchkiss}}, \
  and\ \bibinfo {author} {\bibfnamefont {R.}~\bibnamefont {Easther}},\
  }\href@noop {} {\ }\Eprint {http://arxiv.org/abs/1909.11678}
  {arXiv:1909.11678} \BibitemShut {NoStop}%
\bibitem [{\citenamefont {Ackermann}\ \emph
  {et~al.}(2015{\natexlab{a}})\citenamefont {Ackermann} \emph
  {et~al.}}]{ackermann2015searching}%
  \BibitemOpen
  \bibfield  {author} {\bibinfo {author} {\bibfnamefont {M.}~\bibnamefont
  {Ackermann}} \emph {et~al.} (\bibinfo {collaboration} {Fermi-LAT
  Collaboration}),\ }\href {\doibase 10.1103/physrevlett.115.231301} {\bibfield
   {journal} {\bibinfo  {journal} {Phys. Rev. Lett.}\ }\textbf {\bibinfo
  {volume} {115}},\ \bibinfo {pages} {231301} (\bibinfo {year}
  {2015}{\natexlab{a}})},\ \Eprint {http://arxiv.org/abs/1503.02641}
  {arXiv:1503.02641} \BibitemShut {NoStop}%
\bibitem [{\citenamefont {Ahnen}\ \emph {et~al.}(2016)\citenamefont {Ahnen}
  \emph {et~al.}}]{ahnen2016qkx}%
  \BibitemOpen
  \bibfield  {author} {\bibinfo {author} {\bibfnamefont {M.~L.}\ \bibnamefont
  {Ahnen}} \emph {et~al.} (\bibinfo {collaboration} {MAGIC Collaboration}),\
  }\href {\doibase 10.1088/1475-7516/2016/02/039} {\bibfield  {journal}
  {\bibinfo  {journal} {J. Cosmol. Astropart. Phys.}\ }\textbf {\bibinfo
  {volume} {02}},\ \bibinfo {pages} {039} (\bibinfo {year} {2016})},\ \Eprint
  {http://arxiv.org/abs/1601.06590} {arXiv:1601.06590} \BibitemShut {NoStop}%
\bibitem [{\citenamefont {Albert}\ \emph {et~al.}(2017)\citenamefont {Albert}
  \emph {et~al.}}]{albert2017searching}%
  \BibitemOpen
  \bibfield  {author} {\bibinfo {author} {\bibfnamefont {A.}~\bibnamefont
  {Albert}} \emph {et~al.} (\bibinfo {collaboration} {Fermi-LAT and DES
  Collaborations}),\ }\href {\doibase 10.3847/1538-4357/834/2/110} {\bibfield
  {journal} {\bibinfo  {journal} {Astrophys. J.}\ }\textbf {\bibinfo {volume}
  {834}},\ \bibinfo {pages} {110} (\bibinfo {year} {2017})},\ \Eprint
  {http://arxiv.org/abs/1611.03184} {arXiv:1611.03184} \BibitemShut {NoStop}%
\bibitem [{\citenamefont {Abdallah}\ \emph {et~al.}(2016)\citenamefont
  {Abdallah} \emph {et~al.}}]{hess2016}%
  \BibitemOpen
  \bibfield  {author} {\bibinfo {author} {\bibfnamefont {H.}~\bibnamefont
  {Abdallah}} \emph {et~al.} (\bibinfo {collaboration} {H.E.S.S.
  Collaboration}),\ }\href {\doibase 10.1103/physrevlett.117.111301} {\bibfield
   {journal} {\bibinfo  {journal} {Phys. Rev. Lett.}\ }\textbf {\bibinfo
  {volume} {117}},\ \bibinfo {pages} {111301} (\bibinfo {year}
  {2016})}\BibitemShut {NoStop}%
\bibitem [{\citenamefont {Kamionkowski}\ and\ \citenamefont
  {Turner}(1990)}]{kamionkowski1990thermal}%
  \BibitemOpen
  \bibfield  {author} {\bibinfo {author} {\bibfnamefont {M.}~\bibnamefont
  {Kamionkowski}}\ and\ \bibinfo {author} {\bibfnamefont {M.~S.}\ \bibnamefont
  {Turner}},\ }\href {\doibase 10.1103/physrevd.42.3310} {\bibfield  {journal}
  {\bibinfo  {journal} {Phys. Rev. D}\ }\textbf {\bibinfo {volume} {42}},\
  \bibinfo {pages} {3310} (\bibinfo {year} {1990})}\BibitemShut {NoStop}%
\bibitem [{\citenamefont {Giudice}\ \emph {et~al.}(2001)\citenamefont
  {Giudice}, \citenamefont {Kolb},\ and\ \citenamefont
  {Riotto}}]{giudice2001largest}%
  \BibitemOpen
  \bibfield  {author} {\bibinfo {author} {\bibfnamefont {G.~F.}\ \bibnamefont
  {Giudice}}, \bibinfo {author} {\bibfnamefont {E.~W.}\ \bibnamefont {Kolb}}, \
  and\ \bibinfo {author} {\bibfnamefont {A.}~\bibnamefont {Riotto}},\ }\href
  {\doibase 10.1103/physrevd.64.023508} {\bibfield  {journal} {\bibinfo
  {journal} {Phys. Rev. D}\ }\textbf {\bibinfo {volume} {64}},\ \bibinfo
  {pages} {023508} (\bibinfo {year} {2001})},\ \Eprint
  {http://arxiv.org/abs/hep-ph/0005123} {arXiv:hep-ph/0005123} \BibitemShut
  {NoStop}%
\bibitem [{\citenamefont {Gelmini}\ and\ \citenamefont
  {Gondolo}(2006)}]{gelmini2006neutralino}%
  \BibitemOpen
  \bibfield  {author} {\bibinfo {author} {\bibfnamefont {G.}~\bibnamefont
  {Gelmini}}\ and\ \bibinfo {author} {\bibfnamefont {P.}~\bibnamefont
  {Gondolo}},\ }\href {\doibase 10.1103/physrevd.74.023510} {\bibfield
  {journal} {\bibinfo  {journal} {Phys. Rev. D}\ }\textbf {\bibinfo {volume}
  {74}},\ \bibinfo {pages} {023510} (\bibinfo {year} {2006})},\ \Eprint
  {http://arxiv.org/abs/hep-ph/0602230} {arXiv:hep-ph/0602230} \BibitemShut
  {NoStop}%
\bibitem [{\citenamefont {Kane}\ \emph {et~al.}(2016)\citenamefont {Kane},
  \citenamefont {Kumar}, \citenamefont {Nelson},\ and\ \citenamefont
  {Zheng}}]{Kane_2016}%
  \BibitemOpen
  \bibfield  {author} {\bibinfo {author} {\bibfnamefont {G.~L.}\ \bibnamefont
  {Kane}}, \bibinfo {author} {\bibfnamefont {P.}~\bibnamefont {Kumar}},
  \bibinfo {author} {\bibfnamefont {B.~D.}\ \bibnamefont {Nelson}}, \ and\
  \bibinfo {author} {\bibfnamefont {B.}~\bibnamefont {Zheng}},\ }\href
  {\doibase 10.1103/PhysRevD.93.063527} {\bibfield  {journal} {\bibinfo
  {journal} {Phys. Rev. D}\ }\textbf {\bibinfo {volume} {93}},\ \bibinfo
  {pages} {063527} (\bibinfo {year} {2016})},\ \Eprint
  {http://arxiv.org/abs/1502.05406} {arXiv:1502.05406} \BibitemShut {NoStop}%
\bibitem [{\citenamefont {Drees}\ and\ \citenamefont
  {Hajkarim}(2018{\natexlab{a}})}]{Drees_2018}%
  \BibitemOpen
  \bibfield  {author} {\bibinfo {author} {\bibfnamefont {M.}~\bibnamefont
  {Drees}}\ and\ \bibinfo {author} {\bibfnamefont {F.}~\bibnamefont
  {Hajkarim}},\ }\href {http://dx.doi.org/10.1088/1475-7516/2018/02/057}
  {\bibfield  {journal} {\bibinfo  {journal} {J. Cosmol. Astropart. Phys.}\
  }\textbf {\bibinfo {volume} {02}},\ \bibinfo {pages} {057} (\bibinfo {year}
  {2018}{\natexlab{a}})},\ \Eprint {http://arxiv.org/abs/1711.05007}
  {arXiv:1711.05007} \BibitemShut {NoStop}%
\bibitem [{\citenamefont {Drees}\ and\ \citenamefont
  {Hajkarim}(2018{\natexlab{b}})}]{Drees:2018dsj}%
  \BibitemOpen
  \bibfield  {author} {\bibinfo {author} {\bibfnamefont {M.}~\bibnamefont
  {Drees}}\ and\ \bibinfo {author} {\bibfnamefont {F.}~\bibnamefont
  {Hajkarim}},\ }\href {\doibase 10.1007/JHEP12(2018)042} {\bibfield  {journal}
  {\bibinfo  {journal} {J. High Energy Phys.}\ }\textbf {\bibinfo {volume}
  {12}},\ \bibinfo {pages} {042} (\bibinfo {year} {2018}{\natexlab{b}})},\
  \Eprint {http://arxiv.org/abs/1808.05706} {arXiv:1808.05706} \BibitemShut
  {NoStop}%
\bibitem [{\citenamefont {Bernal}\ \emph
  {et~al.}(2019{\natexlab{a}})\citenamefont {Bernal}, \citenamefont {Cosme},\
  and\ \citenamefont {Tenkanen}}]{Bernal2019}%
  \BibitemOpen
  \bibfield  {author} {\bibinfo {author} {\bibfnamefont {N.}~\bibnamefont
  {Bernal}}, \bibinfo {author} {\bibfnamefont {C.}~\bibnamefont {Cosme}}, \
  and\ \bibinfo {author} {\bibfnamefont {T.}~\bibnamefont {Tenkanen}},\ }\href
  {\doibase 10.1140/epjc/s10052-019-6608-8} {\bibfield  {journal} {\bibinfo
  {journal} {Eur. Phys. J. C}\ }\textbf {\bibinfo {volume} {79}},\ \bibinfo
  {pages} {99} (\bibinfo {year} {2019}{\natexlab{a}})},\ \Eprint
  {http://arxiv.org/abs/1803.08064} {arXiv:1803.08064} \BibitemShut {NoStop}%
\bibitem [{\citenamefont {Bernal}\ \emph
  {et~al.}(2019{\natexlab{b}})\citenamefont {Bernal}, \citenamefont {Cosme},
  \citenamefont {Tenkanen},\ and\ \citenamefont {Vaskonen}}]{Bernal:2018kcw}%
  \BibitemOpen
  \bibfield  {author} {\bibinfo {author} {\bibfnamefont {N.}~\bibnamefont
  {Bernal}}, \bibinfo {author} {\bibfnamefont {C.}~\bibnamefont {Cosme}},
  \bibinfo {author} {\bibfnamefont {T.}~\bibnamefont {Tenkanen}}, \ and\
  \bibinfo {author} {\bibfnamefont {V.}~\bibnamefont {Vaskonen}},\ }\href
  {\doibase 10.1140/epjc/s10052-019-6550-9} {\bibfield  {journal} {\bibinfo
  {journal} {Eur. Phys. J. C}\ }\textbf {\bibinfo {volume} {79}},\ \bibinfo
  {pages} {30} (\bibinfo {year} {2019}{\natexlab{b}})},\ \Eprint
  {http://arxiv.org/abs/1806.11122} {arXiv:1806.11122} \BibitemShut {NoStop}%
\bibitem [{\citenamefont {Cirelli}\ \emph {et~al.}(2019)\citenamefont
  {Cirelli}, \citenamefont {Gouttenoire}, \citenamefont {Petraki},\ and\
  \citenamefont {Sala}}]{Cirelli:2018iax}%
  \BibitemOpen
  \bibfield  {author} {\bibinfo {author} {\bibfnamefont {M.}~\bibnamefont
  {Cirelli}}, \bibinfo {author} {\bibfnamefont {Y.}~\bibnamefont
  {Gouttenoire}}, \bibinfo {author} {\bibfnamefont {K.}~\bibnamefont
  {Petraki}}, \ and\ \bibinfo {author} {\bibfnamefont {F.}~\bibnamefont
  {Sala}},\ }\href {\doibase 10.1088/1475-7516/2019/02/014} {\bibfield
  {journal} {\bibinfo  {journal} {J. Cosmol. Astropart. Phys.}\ }\textbf
  {\bibinfo {volume} {02}},\ \bibinfo {pages} {014} (\bibinfo {year} {2019})},\
  \Eprint {http://arxiv.org/abs/1811.03608} {arXiv:1811.03608} \BibitemShut
  {NoStop}%
\bibitem [{\citenamefont {Allahverdi}\ and\ \citenamefont
  {Osi{\'n}ski}()}]{allahverdi2019nonthermal}%
  \BibitemOpen
  \bibfield  {author} {\bibinfo {author} {\bibfnamefont {R.}~\bibnamefont
  {Allahverdi}}\ and\ \bibinfo {author} {\bibfnamefont {J.~K.}\ \bibnamefont
  {Osi{\'n}ski}},\ }\href@noop {} {\ }\Eprint {http://arxiv.org/abs/1909.01457}
  {arXiv:1909.01457} \BibitemShut {NoStop}%
\bibitem [{\citenamefont {Evans}\ \emph {et~al.}()\citenamefont {Evans},
  \citenamefont {Ghalsasi}, \citenamefont {Gori}, \citenamefont {Tammaro},\
  and\ \citenamefont {Zupan}}]{Evans:2019jcs}%
  \BibitemOpen
  \bibfield  {author} {\bibinfo {author} {\bibfnamefont {J.~A.}\ \bibnamefont
  {Evans}}, \bibinfo {author} {\bibfnamefont {A.}~\bibnamefont {Ghalsasi}},
  \bibinfo {author} {\bibfnamefont {S.}~\bibnamefont {Gori}}, \bibinfo {author}
  {\bibfnamefont {M.}~\bibnamefont {Tammaro}}, \ and\ \bibinfo {author}
  {\bibfnamefont {J.}~\bibnamefont {Zupan}},\ }\href@noop {} {\ }\Eprint
  {http://arxiv.org/abs/1910.06319} {arXiv:1910.06319} \BibitemShut {NoStop}%
\bibitem [{\citenamefont {Erickcek}\ and\ \citenamefont
  {Sigurdson}(2011)}]{erickcek2011reheating}%
  \BibitemOpen
  \bibfield  {author} {\bibinfo {author} {\bibfnamefont {A.~L.}\ \bibnamefont
  {Erickcek}}\ and\ \bibinfo {author} {\bibfnamefont {K.}~\bibnamefont
  {Sigurdson}},\ }\href {\doibase 10.1103/physrevd.84.083503} {\bibfield
  {journal} {\bibinfo  {journal} {Phys. Rev. D}\ }\textbf {\bibinfo {volume}
  {84}},\ \bibinfo {pages} {083503} (\bibinfo {year} {2011})},\ \Eprint
  {http://arxiv.org/abs/1106.0536} {arXiv:1106.0536} \BibitemShut {NoStop}%
\bibitem [{\citenamefont {Barenboim}\ and\ \citenamefont
  {Rasero}(2014)}]{barenboim2014structure}%
  \BibitemOpen
  \bibfield  {author} {\bibinfo {author} {\bibfnamefont {G.}~\bibnamefont
  {Barenboim}}\ and\ \bibinfo {author} {\bibfnamefont {J.}~\bibnamefont
  {Rasero}},\ }\href {\doibase 10.1007/jhep04(2014)138} {\bibfield  {journal}
  {\bibinfo  {journal} {J. High Energy Phys.}\ }\textbf {\bibinfo {volume}
  {04}},\ \bibinfo {pages} {138} (\bibinfo {year} {2014})},\ \Eprint
  {http://arxiv.org/abs/1311.4034} {arXiv:1311.4034} \BibitemShut {NoStop}%
\bibitem [{\citenamefont {Fan}\ \emph {et~al.}(2014)\citenamefont {Fan},
  \citenamefont {{\"O}zsoy},\ and\ \citenamefont {Watson}}]{fan2014nonthermal}%
  \BibitemOpen
  \bibfield  {author} {\bibinfo {author} {\bibfnamefont {J.~J.}\ \bibnamefont
  {Fan}}, \bibinfo {author} {\bibfnamefont {O.}~\bibnamefont {{\"O}zsoy}}, \
  and\ \bibinfo {author} {\bibfnamefont {S.}~\bibnamefont {Watson}},\ }\href
  {\doibase 10.1103/physrevd.90.043536} {\bibfield  {journal} {\bibinfo
  {journal} {Phys. Rev. D}\ }\textbf {\bibinfo {volume} {90}},\ \bibinfo
  {pages} {043536} (\bibinfo {year} {2014})},\ \Eprint
  {http://arxiv.org/abs/1405.7373} {arXiv:1405.7373} \BibitemShut {NoStop}%
\bibitem [{\citenamefont {Erickcek}(2015)}]{erickcek2015dark}%
  \BibitemOpen
  \bibfield  {author} {\bibinfo {author} {\bibfnamefont {A.~L.}\ \bibnamefont
  {Erickcek}},\ }\href {\doibase 10.1103/physrevd.92.103505} {\bibfield
  {journal} {\bibinfo  {journal} {Phys. Rev. D}\ }\textbf {\bibinfo {volume}
  {92}},\ \bibinfo {pages} {103505} (\bibinfo {year} {2015})},\ \Eprint
  {http://arxiv.org/abs/1504.03335} {arXiv:1504.03335} \BibitemShut {NoStop}%
\bibitem [{\citenamefont {Erickcek}\ \emph {et~al.}(2016)\citenamefont
  {Erickcek}, \citenamefont {Sinha},\ and\ \citenamefont
  {Watson}}]{erickcek2016bringing}%
  \BibitemOpen
  \bibfield  {author} {\bibinfo {author} {\bibfnamefont {A.~L.}\ \bibnamefont
  {Erickcek}}, \bibinfo {author} {\bibfnamefont {K.}~\bibnamefont {Sinha}}, \
  and\ \bibinfo {author} {\bibfnamefont {S.}~\bibnamefont {Watson}},\ }\href
  {\doibase 10.1103/physrevd.94.063502} {\bibfield  {journal} {\bibinfo
  {journal} {Phys. Rev. D}\ }\textbf {\bibinfo {volume} {94}},\ \bibinfo
  {pages} {063502} (\bibinfo {year} {2016})},\ \Eprint
  {http://arxiv.org/abs/1510.04291} {arXiv:1510.04291} \BibitemShut {NoStop}%
\bibitem [{\citenamefont {Blanco}\ \emph {et~al.}(2019)\citenamefont {Blanco},
  \citenamefont {Delos}, \citenamefont {Erickcek},\ and\ \citenamefont
  {Hooper}}]{blanco2019annihilation}%
  \BibitemOpen
  \bibfield  {author} {\bibinfo {author} {\bibfnamefont {C.}~\bibnamefont
  {Blanco}}, \bibinfo {author} {\bibfnamefont {M.~S.}\ \bibnamefont {Delos}},
  \bibinfo {author} {\bibfnamefont {A.~L.}\ \bibnamefont {Erickcek}}, \ and\
  \bibinfo {author} {\bibfnamefont {D.}~\bibnamefont {Hooper}},\ }\href
  {\doibase 10.1103/PhysRevD.100.103010} {\bibfield  {journal} {\bibinfo
  {journal} {Phys. Rev. D}\ }\textbf {\bibinfo {volume} {100}},\ \bibinfo
  {pages} {103010} (\bibinfo {year} {2019})},\ \Eprint
  {http://arxiv.org/abs/1906.00010} {arXiv:1906.00010} \BibitemShut {NoStop}%
\bibitem [{\citenamefont {Delos}\ \emph {et~al.}(2019)\citenamefont {Delos},
  \citenamefont {Bruff},\ and\ \citenamefont {Erickcek}}]{delos2019predicting}%
  \BibitemOpen
  \bibfield  {author} {\bibinfo {author} {\bibfnamefont {M.~S.}\ \bibnamefont
  {Delos}}, \bibinfo {author} {\bibfnamefont {M.}~\bibnamefont {Bruff}}, \ and\
  \bibinfo {author} {\bibfnamefont {A.~L.}\ \bibnamefont {Erickcek}},\ }\href
  {\doibase 10.1103/physrevd.100.023523} {\bibfield  {journal} {\bibinfo
  {journal} {Phys. Rev. D}\ }\textbf {\bibinfo {volume} {100}},\ \bibinfo
  {pages} {023523} (\bibinfo {year} {2019})},\ \Eprint
  {http://arxiv.org/abs/1905.05766} {arXiv:1905.05766} \BibitemShut {NoStop}%
\bibitem [{\citenamefont {Delos}(2019{\natexlab{a}})}]{delos2019tidal}%
  \BibitemOpen
  \bibfield  {author} {\bibinfo {author} {\bibfnamefont {M.~S.}\ \bibnamefont
  {Delos}},\ }\href {\doibase 10.1103/physrevd.100.063505} {\bibfield
  {journal} {\bibinfo  {journal} {Phys. Rev. D}\ }\textbf {\bibinfo {volume}
  {100}},\ \bibinfo {pages} {063505} (\bibinfo {year} {2019}{\natexlab{a}})},\
  \Eprint {http://arxiv.org/abs/1906.10690} {arXiv:1906.10690} \BibitemShut
  {NoStop}%
\bibitem [{\citenamefont {Delos}(2019{\natexlab{b}})}]{delos2019evolution}%
  \BibitemOpen
  \bibfield  {author} {\bibinfo {author} {\bibfnamefont {M.~S.}\ \bibnamefont
  {Delos}},\ }\href {\doibase 10.1103/PhysRevD.100.083529} {\bibfield
  {journal} {\bibinfo  {journal} {Phys. Rev. D}\ }\textbf {\bibinfo {volume}
  {100}},\ \bibinfo {pages} {083529} (\bibinfo {year} {2019}{\natexlab{b}})},\
  \Eprint {http://arxiv.org/abs/1907.13133} {arXiv:1907.13133} \BibitemShut
  {NoStop}%
\bibitem [{\citenamefont {Atwood}\ \emph {et~al.}(2009)\citenamefont {Atwood}
  \emph {et~al.}}]{atwood2009large}%
  \BibitemOpen
  \bibfield  {author} {\bibinfo {author} {\bibfnamefont {W.}~\bibnamefont
  {Atwood}} \emph {et~al.} (\bibinfo {collaboration} {Fermi-LAT
  Collaboration}),\ }\href {\doibase 10.1088/0004-637X/697/2/1071} {\bibfield
  {journal} {\bibinfo  {journal} {Astrophys. J.}\ }\textbf {\bibinfo {volume}
  {697}},\ \bibinfo {pages} {1071} (\bibinfo {year} {2009})},\ \Eprint
  {http://arxiv.org/abs/0902.1089} {arXiv:0902.1089} \BibitemShut {NoStop}%
\bibitem [{\citenamefont {Liu}\ \emph {et~al.}(2017)\citenamefont {Liu},
  \citenamefont {Bi}, \citenamefont {Lin},\ and\ \citenamefont
  {Yin}}]{liu2017constraints}%
  \BibitemOpen
  \bibfield  {author} {\bibinfo {author} {\bibfnamefont {W.}~\bibnamefont
  {Liu}}, \bibinfo {author} {\bibfnamefont {X.-J.}\ \bibnamefont {Bi}},
  \bibinfo {author} {\bibfnamefont {S.-J.}\ \bibnamefont {Lin}}, \ and\
  \bibinfo {author} {\bibfnamefont {P.-F.}\ \bibnamefont {Yin}},\ }\href
  {\doibase 10.1088/1674-1137/41/4/045104} {\bibfield  {journal} {\bibinfo
  {journal} {Chin. Phys. C}\ }\textbf {\bibinfo {volume} {41}},\ \bibinfo
  {pages} {045104} (\bibinfo {year} {2017})},\ \Eprint
  {http://arxiv.org/abs/1602.01012} {arXiv:1602.01012} \BibitemShut {NoStop}%
\bibitem [{\citenamefont {Blanco}\ and\ \citenamefont
  {Hooper}(2019)}]{blanco2019constraints}%
  \BibitemOpen
  \bibfield  {author} {\bibinfo {author} {\bibfnamefont {C.}~\bibnamefont
  {Blanco}}\ and\ \bibinfo {author} {\bibfnamefont {D.}~\bibnamefont
  {Hooper}},\ }\href {\doibase 10.1088/1475-7516/2019/03/019} {\bibfield
  {journal} {\bibinfo  {journal} {J. Cosmol. Astropart. Phys.}\ }\textbf
  {\bibinfo {volume} {03}},\ \bibinfo {pages} {019} (\bibinfo {year} {2019})},\
  \Eprint {http://arxiv.org/abs/1811.05988} {arXiv:1811.05988} \BibitemShut
  {NoStop}%
\bibitem [{\citenamefont {Griest}\ and\ \citenamefont
  {Kamionkowski}(1990)}]{griest1990}%
  \BibitemOpen
  \bibfield  {author} {\bibinfo {author} {\bibfnamefont {K.}~\bibnamefont
  {Griest}}\ and\ \bibinfo {author} {\bibfnamefont {M.}~\bibnamefont
  {Kamionkowski}},\ }\href {\doibase 10.1103/PhysRevLett.64.615} {\bibfield
  {journal} {\bibinfo  {journal} {Phys. Rev. Lett.}\ }\textbf {\bibinfo
  {volume} {64}},\ \bibinfo {pages} {615} (\bibinfo {year} {1990})}\BibitemShut
  {NoStop}%
\bibitem [{\citenamefont {Green}\ \emph {et~al.}(2004)\citenamefont {Green},
  \citenamefont {Hofmann},\ and\ \citenamefont {Schwarz}}]{green2004power}%
  \BibitemOpen
  \bibfield  {author} {\bibinfo {author} {\bibfnamefont {A.~M.}\ \bibnamefont
  {Green}}, \bibinfo {author} {\bibfnamefont {S.}~\bibnamefont {Hofmann}}, \
  and\ \bibinfo {author} {\bibfnamefont {D.~J.}\ \bibnamefont {Schwarz}},\
  }\href {\doibase 10.1111/j.1365-2966.2004.08232.x} {\bibfield  {journal}
  {\bibinfo  {journal} {Mon. Not. R. Astron. Soc.}\ }\textbf {\bibinfo {volume}
  {353}},\ \bibinfo {pages} {L23} (\bibinfo {year} {2004})},\ \Eprint
  {http://arxiv.org/abs/astro-ph/0309621} {arXiv:astro-ph/0309621} \BibitemShut
  {NoStop}%
\bibitem [{\citenamefont {Loeb}\ and\ \citenamefont
  {Zaldarriaga}(2005)}]{Loeb:2005pm}%
  \BibitemOpen
  \bibfield  {author} {\bibinfo {author} {\bibfnamefont {A.}~\bibnamefont
  {Loeb}}\ and\ \bibinfo {author} {\bibfnamefont {M.}~\bibnamefont
  {Zaldarriaga}},\ }\href {\doibase 10.1103/PhysRevD.71.103520} {\bibfield
  {journal} {\bibinfo  {journal} {Phys. Rev. D}\ }\textbf {\bibinfo {volume}
  {71}},\ \bibinfo {pages} {103520} (\bibinfo {year} {2005})},\ \Eprint
  {http://arxiv.org/abs/astro-ph/0504112} {arXiv:astro-ph/0504112} \BibitemShut
  {NoStop}%
\bibitem [{\citenamefont {Profumo}\ \emph {et~al.}(2006)\citenamefont
  {Profumo}, \citenamefont {Sigurdson},\ and\ \citenamefont
  {Kamionkowski}}]{Profumo:2006bv}%
  \BibitemOpen
  \bibfield  {author} {\bibinfo {author} {\bibfnamefont {S.}~\bibnamefont
  {Profumo}}, \bibinfo {author} {\bibfnamefont {K.}~\bibnamefont {Sigurdson}},
  \ and\ \bibinfo {author} {\bibfnamefont {M.}~\bibnamefont {Kamionkowski}},\
  }\href {\doibase 10.1103/PhysRevLett.97.031301} {\bibfield  {journal}
  {\bibinfo  {journal} {Phys. Rev. Lett.}\ }\textbf {\bibinfo {volume} {97}},\
  \bibinfo {pages} {031301} (\bibinfo {year} {2006})},\ \Eprint
  {http://arxiv.org/abs/astro-ph/0603373} {arXiv:astro-ph/0603373} \BibitemShut
  {NoStop}%
\bibitem [{\citenamefont {Bertschinger}(2006)}]{bertschinger2006effects}%
  \BibitemOpen
  \bibfield  {author} {\bibinfo {author} {\bibfnamefont {E.}~\bibnamefont
  {Bertschinger}},\ }\href {\doibase 10.1103/physrevd.74.063509} {\bibfield
  {journal} {\bibinfo  {journal} {Phys. Rev. D}\ }\textbf {\bibinfo {volume}
  {74}},\ \bibinfo {pages} {063509} (\bibinfo {year} {2006})},\ \Eprint
  {http://arxiv.org/abs/astro-ph/0607319} {arXiv:astro-ph/0607319} \BibitemShut
  {NoStop}%
\bibitem [{\citenamefont {Bringmann}\ and\ \citenamefont
  {Hofmann}(2007)}]{bringmann2007thermal}%
  \BibitemOpen
  \bibfield  {author} {\bibinfo {author} {\bibfnamefont {T.}~\bibnamefont
  {Bringmann}}\ and\ \bibinfo {author} {\bibfnamefont {S.}~\bibnamefont
  {Hofmann}},\ }\href {\doibase 10.1088/1475-7516/2007/04/016} {\bibfield
  {journal} {\bibinfo  {journal} {J. Cosmol. Astropart. Phys.}\ }\textbf
  {\bibinfo {volume} {04}},\ \bibinfo {pages} {016} (\bibinfo {year} {2007})},\
  \Eprint {http://arxiv.org/abs/hep-ph/0612238} {arXiv:hep-ph/0612238}
  \BibitemShut {NoStop}%
\bibitem [{\citenamefont {Bringmann}\ and\ \citenamefont
  {Hofmann}(2016)}]{bringmann2016erratum}%
  \BibitemOpen
  \bibfield  {author} {\bibinfo {author} {\bibfnamefont {T.}~\bibnamefont
  {Bringmann}}\ and\ \bibinfo {author} {\bibfnamefont {S.}~\bibnamefont
  {Hofmann}},\ }\href {\doibase 10.1088/1475-7516/2016/03/E02} {\ \textbf
  {\bibinfo {volume} {03}},\ \bibinfo {pages} {E02} (\bibinfo {year}
  {2016})}\BibitemShut {NoStop}%
\bibitem [{\citenamefont {Waldstein}\ \emph {et~al.}(2017)\citenamefont
  {Waldstein}, \citenamefont {Erickcek},\ and\ \citenamefont
  {Ilie}}]{Waldstein:2016blt}%
  \BibitemOpen
  \bibfield  {author} {\bibinfo {author} {\bibfnamefont {I.~R.}\ \bibnamefont
  {Waldstein}}, \bibinfo {author} {\bibfnamefont {A.~L.}\ \bibnamefont
  {Erickcek}}, \ and\ \bibinfo {author} {\bibfnamefont {C.}~\bibnamefont
  {Ilie}},\ }\href {\doibase 10.1103/PhysRevD.95.123531} {\bibfield  {journal}
  {\bibinfo  {journal} {Phys. Rev. D}\ }\textbf {\bibinfo {volume} {95}},\
  \bibinfo {pages} {123531} (\bibinfo {year} {2017})},\ \Eprint
  {http://arxiv.org/abs/1609.05927} {arXiv:1609.05927} \BibitemShut {NoStop}%
\bibitem [{\citenamefont {Gelmini}\ and\ \citenamefont
  {Gondolo}(2008)}]{Gelmini:2008sh}%
  \BibitemOpen
  \bibfield  {author} {\bibinfo {author} {\bibfnamefont {G.~B.}\ \bibnamefont
  {Gelmini}}\ and\ \bibinfo {author} {\bibfnamefont {P.}~\bibnamefont
  {Gondolo}},\ }\href {\doibase 10.1088/1475-7516/2008/10/002} {\bibfield
  {journal} {\bibinfo  {journal} {J. Cosmol. Astropart. Phys.}\ }\textbf
  {\bibinfo {volume} {10}},\ \bibinfo {pages} {002} (\bibinfo {year} {2008})},\
  \Eprint {http://arxiv.org/abs/0803.2349} {arXiv:0803.2349} \BibitemShut
  {NoStop}%
\bibitem [{\citenamefont {Eisenstein}\ and\ \citenamefont
  {Hu}(1998)}]{Eisenstein:1997ik}%
  \BibitemOpen
  \bibfield  {author} {\bibinfo {author} {\bibfnamefont {D.~J.}\ \bibnamefont
  {Eisenstein}}\ and\ \bibinfo {author} {\bibfnamefont {W.}~\bibnamefont
  {Hu}},\ }\href {\doibase 10.1086/305424} {\bibfield  {journal} {\bibinfo
  {journal} {Astrophys. J.}\ }\textbf {\bibinfo {volume} {496}},\ \bibinfo
  {pages} {605} (\bibinfo {year} {1998})},\ \Eprint
  {http://arxiv.org/abs/astro-ph/9709112} {arXiv:astro-ph/9709112} \BibitemShut
  {NoStop}%
\bibitem [{\citenamefont {Ishiyama}\ \emph {et~al.}(2010)\citenamefont
  {Ishiyama}, \citenamefont {Makino},\ and\ \citenamefont
  {Ebisuzaki}}]{ishiyama2010gamma}%
  \BibitemOpen
  \bibfield  {author} {\bibinfo {author} {\bibfnamefont {T.}~\bibnamefont
  {Ishiyama}}, \bibinfo {author} {\bibfnamefont {J.}~\bibnamefont {Makino}}, \
  and\ \bibinfo {author} {\bibfnamefont {T.}~\bibnamefont {Ebisuzaki}},\ }\href
  {\doibase 10.1088/2041-8205/723/2/l195} {\bibfield  {journal} {\bibinfo
  {journal} {Astrophys. J. Lett.}\ }\textbf {\bibinfo {volume} {723}},\
  \bibinfo {pages} {L195} (\bibinfo {year} {2010})},\ \Eprint
  {http://arxiv.org/abs/1006.3392} {arXiv:1006.3392} \BibitemShut {NoStop}%
\bibitem [{\citenamefont {Anderhalden}\ and\ \citenamefont
  {Diemand}(2013{\natexlab{a}})}]{anderhalden2013density}%
  \BibitemOpen
  \bibfield  {author} {\bibinfo {author} {\bibfnamefont {D.}~\bibnamefont
  {Anderhalden}}\ and\ \bibinfo {author} {\bibfnamefont {J.}~\bibnamefont
  {Diemand}},\ }\href {\doibase 10.1088/1475-7516/2013/04/009} {\bibfield
  {journal} {\bibinfo  {journal} {J. Cosmol. Astropart. Phys.}\ }\textbf
  {\bibinfo {volume} {04}},\ \bibinfo {pages} {009} (\bibinfo {year}
  {2013}{\natexlab{a}})},\ \Eprint {http://arxiv.org/abs/1302.0003}
  {arXiv:1302.0003} \BibitemShut {NoStop}%
\bibitem [{\citenamefont {Anderhalden}\ and\ \citenamefont
  {Diemand}(2013{\natexlab{b}})}]{anderhalden2013erratum}%
  \BibitemOpen
  \bibfield  {author} {\bibinfo {author} {\bibfnamefont {D.}~\bibnamefont
  {Anderhalden}}\ and\ \bibinfo {author} {\bibfnamefont {J.}~\bibnamefont
  {Diemand}},\ }\href {\doibase 10.1088/1475-7516/2013/08/e02} {\ \textbf
  {\bibinfo {volume} {08}},\ \bibinfo {pages} {E02} (\bibinfo {year}
  {2013}{\natexlab{b}})}\BibitemShut {NoStop}%
\bibitem [{\citenamefont {Ishiyama}(2014)}]{ishiyama2014hierarchical}%
  \BibitemOpen
  \bibfield  {author} {\bibinfo {author} {\bibfnamefont {T.}~\bibnamefont
  {Ishiyama}},\ }\href {\doibase 10.1088/0004-637x/788/1/27} {\bibfield
  {journal} {\bibinfo  {journal} {Astrophys. J.}\ }\textbf {\bibinfo {volume}
  {788}},\ \bibinfo {pages} {27} (\bibinfo {year} {2014})},\ \Eprint
  {http://arxiv.org/abs/1404.1650} {arXiv:1404.1650} \BibitemShut {NoStop}%
\bibitem [{\citenamefont {Polisensky}\ and\ \citenamefont
  {Ricotti}(2015)}]{polisensky2015fingerprints}%
  \BibitemOpen
  \bibfield  {author} {\bibinfo {author} {\bibfnamefont {E.}~\bibnamefont
  {Polisensky}}\ and\ \bibinfo {author} {\bibfnamefont {M.}~\bibnamefont
  {Ricotti}},\ }\href {\doibase 10.1093/mnras/stv736} {\bibfield  {journal}
  {\bibinfo  {journal} {Mon. Not. R. Astron. Soc.}\ }\textbf {\bibinfo {volume}
  {450}},\ \bibinfo {pages} {2172} (\bibinfo {year} {2015})},\ \Eprint
  {http://arxiv.org/abs/1504.02126} {arXiv:1504.02126} \BibitemShut {NoStop}%
\bibitem [{\citenamefont {Ogiya}\ and\ \citenamefont
  {Hahn}(2018)}]{ogiya2017sets}%
  \BibitemOpen
  \bibfield  {author} {\bibinfo {author} {\bibfnamefont {G.}~\bibnamefont
  {Ogiya}}\ and\ \bibinfo {author} {\bibfnamefont {O.}~\bibnamefont {Hahn}},\
  }\href {\doibase 10.1093/mnras/stx2639} {\bibfield  {journal} {\bibinfo
  {journal} {Mon. Not. R. Astron. Soc.}\ }\textbf {\bibinfo {volume} {473}},\
  \bibinfo {pages} {4339} (\bibinfo {year} {2018})},\ \Eprint
  {http://arxiv.org/abs/1707.07693} {arXiv:1707.07693} \BibitemShut {NoStop}%
\bibitem [{\citenamefont {Delos}\ \emph
  {et~al.}(2018{\natexlab{a}})\citenamefont {Delos}, \citenamefont {Erickcek},
  \citenamefont {Bailey},\ and\ \citenamefont
  {Alvarez}}]{delos2018ultracompact}%
  \BibitemOpen
  \bibfield  {author} {\bibinfo {author} {\bibfnamefont {M.~S.}\ \bibnamefont
  {Delos}}, \bibinfo {author} {\bibfnamefont {A.~L.}\ \bibnamefont {Erickcek}},
  \bibinfo {author} {\bibfnamefont {A.~P.}\ \bibnamefont {Bailey}}, \ and\
  \bibinfo {author} {\bibfnamefont {M.~A.}\ \bibnamefont {Alvarez}},\ }\href
  {\doibase 10.1103/physrevd.97.041303} {\bibfield  {journal} {\bibinfo
  {journal} {Phys. Rev. D}\ }\textbf {\bibinfo {volume} {97}},\ \bibinfo
  {pages} {041303(R)} (\bibinfo {year} {2018}{\natexlab{a}})},\ \Eprint
  {http://arxiv.org/abs/1712.05421} {arXiv:1712.05421} \BibitemShut {NoStop}%
\bibitem [{\citenamefont {Delos}\ \emph
  {et~al.}(2018{\natexlab{b}})\citenamefont {Delos}, \citenamefont {Erickcek},
  \citenamefont {Bailey},\ and\ \citenamefont {Alvarez}}]{delos2018density}%
  \BibitemOpen
  \bibfield  {author} {\bibinfo {author} {\bibfnamefont {M.~S.}\ \bibnamefont
  {Delos}}, \bibinfo {author} {\bibfnamefont {A.~L.}\ \bibnamefont {Erickcek}},
  \bibinfo {author} {\bibfnamefont {A.~P.}\ \bibnamefont {Bailey}}, \ and\
  \bibinfo {author} {\bibfnamefont {M.~A.}\ \bibnamefont {Alvarez}},\ }\href
  {\doibase 10.1103/physrevd.98.063527} {\bibfield  {journal} {\bibinfo
  {journal} {Phys. Rev. D}\ }\textbf {\bibinfo {volume} {98}},\ \bibinfo
  {pages} {063527} (\bibinfo {year} {2018}{\natexlab{b}})},\ \Eprint
  {http://arxiv.org/abs/1806.07389} {arXiv:1806.07389} \BibitemShut {NoStop}%
\bibitem [{\citenamefont {Angulo}\ \emph {et~al.}(2017)\citenamefont {Angulo},
  \citenamefont {Hahn}, \citenamefont {Ludlow},\ and\ \citenamefont
  {Bonoli}}]{angulo2017earth}%
  \BibitemOpen
  \bibfield  {author} {\bibinfo {author} {\bibfnamefont {R.~E.}\ \bibnamefont
  {Angulo}}, \bibinfo {author} {\bibfnamefont {O.}~\bibnamefont {Hahn}},
  \bibinfo {author} {\bibfnamefont {A.~D.}\ \bibnamefont {Ludlow}}, \ and\
  \bibinfo {author} {\bibfnamefont {S.}~\bibnamefont {Bonoli}},\ }\href
  {\doibase 10.1093/mnras/stx1658} {\bibfield  {journal} {\bibinfo  {journal}
  {Mon. Not. R. Astron. Soc.}\ }\textbf {\bibinfo {volume} {471}},\ \bibinfo
  {pages} {4687} (\bibinfo {year} {2017})},\ \Eprint
  {http://arxiv.org/abs/1604.03131} {arXiv:1604.03131} \BibitemShut {NoStop}%
\bibitem [{\citenamefont {Ishiyama}\ and\ \citenamefont
  {Ando}()}]{ishiyama2019abundance}%
  \BibitemOpen
  \bibfield  {author} {\bibinfo {author} {\bibfnamefont {T.}~\bibnamefont
  {Ishiyama}}\ and\ \bibinfo {author} {\bibfnamefont {S.}~\bibnamefont
  {Ando}},\ }\href@noop {} {\ }\Eprint {http://arxiv.org/abs/1907.03642}
  {arXiv:1907.03642} \BibitemShut {NoStop}%
\bibitem [{\citenamefont {Ogiya}\ \emph {et~al.}(2016)\citenamefont {Ogiya},
  \citenamefont {Nagai},\ and\ \citenamefont {Ishiyama}}]{ogiya2016dynamical}%
  \BibitemOpen
  \bibfield  {author} {\bibinfo {author} {\bibfnamefont {G.}~\bibnamefont
  {Ogiya}}, \bibinfo {author} {\bibfnamefont {D.}~\bibnamefont {Nagai}}, \ and\
  \bibinfo {author} {\bibfnamefont {T.}~\bibnamefont {Ishiyama}},\ }\href
  {\doibase 10.1093/mnras/stw1551} {\bibfield  {journal} {\bibinfo  {journal}
  {Mon. Not. R. Astron. Soc.}\ }\textbf {\bibinfo {volume} {461}},\ \bibinfo
  {pages} {3385} (\bibinfo {year} {2016})},\ \Eprint
  {http://arxiv.org/abs/1604.02866} {arXiv:1604.02866} \BibitemShut {NoStop}%
\bibitem [{\citenamefont {Gosenca}\ \emph {et~al.}(2017)\citenamefont
  {Gosenca}, \citenamefont {Adamek}, \citenamefont {Byrnes},\ and\
  \citenamefont {Hotchkiss}}]{gosenca20173d}%
  \BibitemOpen
  \bibfield  {author} {\bibinfo {author} {\bibfnamefont {M.}~\bibnamefont
  {Gosenca}}, \bibinfo {author} {\bibfnamefont {J.}~\bibnamefont {Adamek}},
  \bibinfo {author} {\bibfnamefont {C.~T.}\ \bibnamefont {Byrnes}}, \ and\
  \bibinfo {author} {\bibfnamefont {S.}~\bibnamefont {Hotchkiss}},\ }\href
  {\doibase 10.1103/physrevd.96.123519} {\bibfield  {journal} {\bibinfo
  {journal} {Phys. Rev. D}\ }\textbf {\bibinfo {volume} {96}},\ \bibinfo
  {pages} {123519} (\bibinfo {year} {2017})},\ \Eprint
  {http://arxiv.org/abs/1710.02055} {arXiv:1710.02055} \BibitemShut {NoStop}%
\bibitem [{\citenamefont {Navarro}\ \emph {et~al.}(1996)\citenamefont
  {Navarro}, \citenamefont {Frenk},\ and\ \citenamefont
  {White}}]{navarro1996structure}%
  \BibitemOpen
  \bibfield  {author} {\bibinfo {author} {\bibfnamefont {J.~F.}\ \bibnamefont
  {Navarro}}, \bibinfo {author} {\bibfnamefont {C.~S.}\ \bibnamefont {Frenk}},
  \ and\ \bibinfo {author} {\bibfnamefont {S.~D.~M.}\ \bibnamefont {White}},\
  }\href {\doibase 10.1086/177173} {\bibfield  {journal} {\bibinfo  {journal}
  {Astrophys. J.}\ }\textbf {\bibinfo {volume} {462}},\ \bibinfo {pages} {563}
  (\bibinfo {year} {1996})},\ \Eprint {http://arxiv.org/abs/astro-ph/9508025}
  {arXiv:astro-ph/9508025} \BibitemShut {NoStop}%
\bibitem [{\citenamefont {Navarro}\ \emph {et~al.}(1997)\citenamefont
  {Navarro}, \citenamefont {Frenk},\ and\ \citenamefont
  {White}}]{navarro1997universal}%
  \BibitemOpen
  \bibfield  {author} {\bibinfo {author} {\bibfnamefont {J.~F.}\ \bibnamefont
  {Navarro}}, \bibinfo {author} {\bibfnamefont {C.~S.}\ \bibnamefont {Frenk}},
  \ and\ \bibinfo {author} {\bibfnamefont {S.~D.~M.}\ \bibnamefont {White}},\
  }\href {\doibase 10.1086/304888} {\bibfield  {journal} {\bibinfo  {journal}
  {Astrophys. J.}\ }\textbf {\bibinfo {volume} {490}},\ \bibinfo {pages} {493}
  (\bibinfo {year} {1997})},\ \Eprint {http://arxiv.org/abs/astro-ph/9611107}
  {arXiv:astro-ph/9611107} \BibitemShut {NoStop}%
\bibitem [{\citenamefont {Lewis}\ and\ \citenamefont
  {Challinor}(2007)}]{lewis200721}%
  \BibitemOpen
  \bibfield  {author} {\bibinfo {author} {\bibfnamefont {A.}~\bibnamefont
  {Lewis}}\ and\ \bibinfo {author} {\bibfnamefont {A.}~\bibnamefont
  {Challinor}},\ }\href {\doibase 10.1103/physrevd.76.083005} {\bibfield
  {journal} {\bibinfo  {journal} {Phys. Rev. D}\ }\textbf {\bibinfo {volume}
  {76}},\ \bibinfo {pages} {083005} (\bibinfo {year} {2007})},\ \Eprint
  {http://arxiv.org/abs/astro-ph/0702600} {arXiv:astro-ph/0702600} \BibitemShut
  {NoStop}%
\bibitem [{\citenamefont {Challinor}\ and\ \citenamefont
  {Lewis}(2011)}]{challinor2011linear}%
  \BibitemOpen
  \bibfield  {author} {\bibinfo {author} {\bibfnamefont {A.}~\bibnamefont
  {Challinor}}\ and\ \bibinfo {author} {\bibfnamefont {A.}~\bibnamefont
  {Lewis}},\ }\href {\doibase 10.1103/physrevd.84.043516} {\bibfield  {journal}
  {\bibinfo  {journal} {Phys. Rev. D}\ }\textbf {\bibinfo {volume} {84}},\
  \bibinfo {pages} {043516} (\bibinfo {year} {2011})},\ \Eprint
  {http://arxiv.org/abs/1105.5292} {arXiv:1105.5292} \BibitemShut {NoStop}%
\bibitem [{\citenamefont {Aghanim}\ \emph {et~al.}()\citenamefont {Aghanim}
  \emph {et~al.}}]{aghanim2018planck}%
  \BibitemOpen
  \bibfield  {author} {\bibinfo {author} {\bibfnamefont {N.}~\bibnamefont
  {Aghanim}} \emph {et~al.} (\bibinfo {collaboration} {Planck Collaboration}),\
  }\href@noop {} {\ }\Eprint {http://arxiv.org/abs/1807.06209}
  {arXiv:1807.06209} \BibitemShut {NoStop}%
\bibitem [{\citenamefont {Hu}\ and\ \citenamefont
  {Sugiyama}(1996)}]{hu1996small}%
  \BibitemOpen
  \bibfield  {author} {\bibinfo {author} {\bibfnamefont {W.}~\bibnamefont
  {Hu}}\ and\ \bibinfo {author} {\bibfnamefont {N.}~\bibnamefont {Sugiyama}},\
  }\href {\doibase 10.1086/177989} {\bibfield  {journal} {\bibinfo  {journal}
  {Astrophys. J.}\ }\textbf {\bibinfo {volume} {471}},\ \bibinfo {pages} {542}
  (\bibinfo {year} {1996})},\ \Eprint {http://arxiv.org/abs/astro-ph/9510117}
  {arXiv:astro-ph/9510117} \BibitemShut {NoStop}%
\bibitem [{\citenamefont {Sheth}\ \emph {et~al.}(2001)\citenamefont {Sheth},
  \citenamefont {Mo},\ and\ \citenamefont {Tormen}}]{sheth2001ellipsoidal}%
  \BibitemOpen
  \bibfield  {author} {\bibinfo {author} {\bibfnamefont {R.~K.}\ \bibnamefont
  {Sheth}}, \bibinfo {author} {\bibfnamefont {H.}~\bibnamefont {Mo}}, \ and\
  \bibinfo {author} {\bibfnamefont {G.}~\bibnamefont {Tormen}},\ }\href
  {\doibase 10.1046/j.1365-8711.2001.04006.x} {\bibfield  {journal} {\bibinfo
  {journal} {Mon. Not. R. Astron. Soc.}\ }\textbf {\bibinfo {volume} {323}},\
  \bibinfo {pages} {1} (\bibinfo {year} {2001})},\ \Eprint
  {http://arxiv.org/abs/astro-ph/9907024} {arXiv:astro-ph/9907024} \BibitemShut
  {NoStop}%
\bibitem [{\citenamefont {Bardeen}\ \emph {et~al.}(1986)\citenamefont
  {Bardeen}, \citenamefont {Bond}, \citenamefont {Kaiser},\ and\ \citenamefont
  {Szalay}}]{bardeen1986statistics}%
  \BibitemOpen
  \bibfield  {author} {\bibinfo {author} {\bibfnamefont {J.~M.}\ \bibnamefont
  {Bardeen}}, \bibinfo {author} {\bibfnamefont {J.~R.}\ \bibnamefont {Bond}},
  \bibinfo {author} {\bibfnamefont {N.}~\bibnamefont {Kaiser}}, \ and\ \bibinfo
  {author} {\bibfnamefont {A.~S.}\ \bibnamefont {Szalay}},\ }\href {\doibase
  10.1086/164143} {\bibfield  {journal} {\bibinfo  {journal} {Astrophys. J.}\
  }\textbf {\bibinfo {volume} {304}},\ \bibinfo {pages} {15} (\bibinfo {year}
  {1986})}\BibitemShut {NoStop}%
\bibitem [{\citenamefont {Moore}\ \emph {et~al.}(1999)\citenamefont {Moore},
  \citenamefont {Quinn}, \citenamefont {Governato}, \citenamefont {Stadel},\
  and\ \citenamefont {Lake}}]{moore1999cold}%
  \BibitemOpen
  \bibfield  {author} {\bibinfo {author} {\bibfnamefont {B.}~\bibnamefont
  {Moore}}, \bibinfo {author} {\bibfnamefont {T.}~\bibnamefont {Quinn}},
  \bibinfo {author} {\bibfnamefont {F.}~\bibnamefont {Governato}}, \bibinfo
  {author} {\bibfnamefont {J.}~\bibnamefont {Stadel}}, \ and\ \bibinfo {author}
  {\bibfnamefont {G.}~\bibnamefont {Lake}},\ }\href {\doibase
  10.1046/j.1365-8711.1999.03039.x} {\bibfield  {journal} {\bibinfo  {journal}
  {Mon. Not. R. Astron. Soc.}\ }\textbf {\bibinfo {volume} {310}},\ \bibinfo
  {pages} {1147} (\bibinfo {year} {1999})},\ \Eprint
  {http://arxiv.org/abs/astro-ph/9903164} {arXiv:astro-ph/9903164} \BibitemShut
  {NoStop}%
\bibitem [{\citenamefont {Drakos}\ \emph {et~al.}(2019)\citenamefont {Drakos},
  \citenamefont {Taylor}, \citenamefont {Berrouet}, \citenamefont {Robotham},\
  and\ \citenamefont {Power}}]{drakos2019major}%
  \BibitemOpen
  \bibfield  {author} {\bibinfo {author} {\bibfnamefont {N.~E.}\ \bibnamefont
  {Drakos}}, \bibinfo {author} {\bibfnamefont {J.~E.}\ \bibnamefont {Taylor}},
  \bibinfo {author} {\bibfnamefont {A.}~\bibnamefont {Berrouet}}, \bibinfo
  {author} {\bibfnamefont {A.~S.}\ \bibnamefont {Robotham}}, \ and\ \bibinfo
  {author} {\bibfnamefont {C.}~\bibnamefont {Power}},\ }\href {\doibase
  10.1093/mnras/stz1307} {\bibfield  {journal} {\bibinfo  {journal} {Mon. Not.
  R. Astron. Soc.}\ }\textbf {\bibinfo {volume} {487}},\ \bibinfo {pages}
  {1008} (\bibinfo {year} {2019})},\ \Eprint {http://arxiv.org/abs/1811.12844}
  {arXiv:1811.12844} \BibitemShut {NoStop}%
\bibitem [{\citenamefont {Press}\ and\ \citenamefont
  {Schechter}(1974)}]{press1974formation}%
  \BibitemOpen
  \bibfield  {author} {\bibinfo {author} {\bibfnamefont {W.~H.}\ \bibnamefont
  {Press}}\ and\ \bibinfo {author} {\bibfnamefont {P.}~\bibnamefont
  {Schechter}},\ }\href {\doibase 10.1086/152650} {\bibfield  {journal}
  {\bibinfo  {journal} {Astrophys. J.}\ }\textbf {\bibinfo {volume} {187}},\
  \bibinfo {pages} {425} (\bibinfo {year} {1974})}\BibitemShut {NoStop}%
\bibitem [{\citenamefont {Ackermann}\ \emph
  {et~al.}(2015{\natexlab{b}})\citenamefont {Ackermann} \emph
  {et~al.}}]{ackermann2015spectrum}%
  \BibitemOpen
  \bibfield  {author} {\bibinfo {author} {\bibfnamefont {M.}~\bibnamefont
  {Ackermann}} \emph {et~al.} (\bibinfo {collaboration} {Fermi-LAT
  Collaboration}),\ }\href {\doibase 10.1088/0004-637x/799/1/86} {\bibfield
  {journal} {\bibinfo  {journal} {Astrophys. J.}\ }\textbf {\bibinfo {volume}
  {799}},\ \bibinfo {pages} {86} (\bibinfo {year} {2015}{\natexlab{b}})},\
  \Eprint {http://arxiv.org/abs/1410.3696} {arXiv:1410.3696} \BibitemShut
  {NoStop}%
\bibitem [{\citenamefont {Cirelli}\ \emph {et~al.}(2010)\citenamefont
  {Cirelli}, \citenamefont {Panci},\ and\ \citenamefont
  {Serpico}}]{Cirelli:2009dv}%
  \BibitemOpen
  \bibfield  {author} {\bibinfo {author} {\bibfnamefont {M.}~\bibnamefont
  {Cirelli}}, \bibinfo {author} {\bibfnamefont {P.}~\bibnamefont {Panci}}, \
  and\ \bibinfo {author} {\bibfnamefont {P.~D.}\ \bibnamefont {Serpico}},\
  }\href {\doibase 10.1016/j.nuclphysb.2010.07.010} {\bibfield  {journal}
  {\bibinfo  {journal} {Nucl. Phys. B}\ }\textbf {\bibinfo {volume} {840}},\
  \bibinfo {pages} {284} (\bibinfo {year} {2010})},\ \Eprint
  {http://arxiv.org/abs/0912.0663} {arXiv:0912.0663} \BibitemShut {NoStop}%
\bibitem [{\citenamefont {Diemer}\ and\ \citenamefont
  {Joyce}(2019)}]{diemer2019accurate}%
  \BibitemOpen
  \bibfield  {author} {\bibinfo {author} {\bibfnamefont {B.}~\bibnamefont
  {Diemer}}\ and\ \bibinfo {author} {\bibfnamefont {M.}~\bibnamefont {Joyce}},\
  }\href {\doibase 10.3847/1538-4357/aafad6} {\bibfield  {journal} {\bibinfo
  {journal} {Astrophys. J.}\ }\textbf {\bibinfo {volume} {871}},\ \bibinfo
  {pages} {168} (\bibinfo {year} {2019})},\ \Eprint
  {http://arxiv.org/abs/1809.07326} {arXiv:1809.07326} \BibitemShut {NoStop}%
\bibitem [{\citenamefont {Strigari}\ \emph {et~al.}(2007)\citenamefont
  {Strigari}, \citenamefont {Koushiappas}, \citenamefont {Bullock},\ and\
  \citenamefont {Kaplinghat}}]{strigari2007precise}%
  \BibitemOpen
  \bibfield  {author} {\bibinfo {author} {\bibfnamefont {L.~E.}\ \bibnamefont
  {Strigari}}, \bibinfo {author} {\bibfnamefont {S.~M.}\ \bibnamefont
  {Koushiappas}}, \bibinfo {author} {\bibfnamefont {J.~S.}\ \bibnamefont
  {Bullock}}, \ and\ \bibinfo {author} {\bibfnamefont {M.}~\bibnamefont
  {Kaplinghat}},\ }\href {\doibase 10.1103/physrevd.75.083526} {\bibfield
  {journal} {\bibinfo  {journal} {Phys. Rev. D}\ }\textbf {\bibinfo {volume}
  {75}},\ \bibinfo {pages} {083526} (\bibinfo {year} {2007})},\ \Eprint
  {http://arxiv.org/abs/astro-ph/0611925} {arXiv:astro-ph/0611925} \BibitemShut
  {NoStop}%
\bibitem [{\citenamefont {Baring}\ \emph {et~al.}(2016)\citenamefont {Baring},
  \citenamefont {Ghosh}, \citenamefont {Queiroz},\ and\ \citenamefont
  {Sinha}}]{baring2016new}%
  \BibitemOpen
  \bibfield  {author} {\bibinfo {author} {\bibfnamefont {M.~G.}\ \bibnamefont
  {Baring}}, \bibinfo {author} {\bibfnamefont {T.}~\bibnamefont {Ghosh}},
  \bibinfo {author} {\bibfnamefont {F.~S.}\ \bibnamefont {Queiroz}}, \ and\
  \bibinfo {author} {\bibfnamefont {K.}~\bibnamefont {Sinha}},\ }\href
  {\doibase 10.1103/physrevd.93.103009} {\bibfield  {journal} {\bibinfo
  {journal} {Phys. Rev. D}\ }\textbf {\bibinfo {volume} {93}},\ \bibinfo
  {pages} {103009} (\bibinfo {year} {2016})},\ \Eprint
  {http://arxiv.org/abs/1510.00389} {arXiv:1510.00389} \BibitemShut {NoStop}%
\bibitem [{\citenamefont {Martinez}(2015)}]{martinez2015robust}%
  \BibitemOpen
  \bibfield  {author} {\bibinfo {author} {\bibfnamefont {G.~D.}\ \bibnamefont
  {Martinez}},\ }\href {\doibase 10.1093/mnras/stv942} {\bibfield  {journal}
  {\bibinfo  {journal} {Mon. Not. R. Astron. Soc.}\ }\textbf {\bibinfo {volume}
  {451}},\ \bibinfo {pages} {2524} (\bibinfo {year} {2015})},\ \Eprint
  {http://arxiv.org/abs/1309.2641} {arXiv:1309.2641} \BibitemShut {NoStop}%
\bibitem [{\citenamefont {Read}\ \emph {et~al.}(2018)\citenamefont {Read},
  \citenamefont {Walker},\ and\ \citenamefont {Steger}}]{read2018case}%
  \BibitemOpen
  \bibfield  {author} {\bibinfo {author} {\bibfnamefont {J.~I.}\ \bibnamefont
  {Read}}, \bibinfo {author} {\bibfnamefont {M.~G.}\ \bibnamefont {Walker}}, \
  and\ \bibinfo {author} {\bibfnamefont {P.}~\bibnamefont {Steger}},\ }\href
  {\doibase 10.1093/mnras/sty2286} {\bibfield  {journal} {\bibinfo  {journal}
  {Mon. Not. R. Astron. Soc.}\ }\textbf {\bibinfo {volume} {481}},\ \bibinfo
  {pages} {860} (\bibinfo {year} {2018})},\ \Eprint
  {http://arxiv.org/abs/1805.06934} {arXiv:1805.06934} \BibitemShut {NoStop}%
\bibitem [{\citenamefont {Moore}(1994)}]{moore1994evidence}%
  \BibitemOpen
  \bibfield  {author} {\bibinfo {author} {\bibfnamefont {B.}~\bibnamefont
  {Moore}},\ }\href {\doibase 10.1038/370629a0} {\bibfield  {journal} {\bibinfo
   {journal} {Nature (London)}\ }\textbf {\bibinfo {volume} {370}},\ \bibinfo
  {pages} {629} (\bibinfo {year} {1994})}\BibitemShut {NoStop}%
\bibitem [{\citenamefont {Martin}\ \emph {et~al.}(2008)\citenamefont {Martin},
  \citenamefont {de~Jong},\ and\ \citenamefont
  {Rix}}]{martin2008comprehensive}%
  \BibitemOpen
  \bibfield  {author} {\bibinfo {author} {\bibfnamefont {N.~F.}\ \bibnamefont
  {Martin}}, \bibinfo {author} {\bibfnamefont {J.~T.}\ \bibnamefont {de~Jong}},
  \ and\ \bibinfo {author} {\bibfnamefont {H.-W.}\ \bibnamefont {Rix}},\ }\href
  {\doibase 10.1086/590336} {\bibfield  {journal} {\bibinfo  {journal}
  {Astrophys. J.}\ }\textbf {\bibinfo {volume} {684}},\ \bibinfo {pages} {1075}
  (\bibinfo {year} {2008})},\ \Eprint {http://arxiv.org/abs/0805.2945}
  {arXiv:0805.2945} \BibitemShut {NoStop}%
\bibitem [{\citenamefont {Woo}\ \emph {et~al.}(2008)\citenamefont {Woo},
  \citenamefont {Courteau},\ and\ \citenamefont {Dekel}}]{woo2008scaling}%
  \BibitemOpen
  \bibfield  {author} {\bibinfo {author} {\bibfnamefont {J.}~\bibnamefont
  {Woo}}, \bibinfo {author} {\bibfnamefont {S.}~\bibnamefont {Courteau}}, \
  and\ \bibinfo {author} {\bibfnamefont {A.}~\bibnamefont {Dekel}},\ }\href
  {\doibase 10.1111/j.1365-2966.2008.13770.x} {\bibfield  {journal} {\bibinfo
  {journal} {Mon. Not. R. Astron. Soc.}\ }\textbf {\bibinfo {volume} {390}},\
  \bibinfo {pages} {1453} (\bibinfo {year} {2008})},\ \Eprint
  {http://arxiv.org/abs/0807.1331} {arXiv:0807.1331} \BibitemShut {NoStop}%
\bibitem [{\citenamefont {Plummer}(1911)}]{plummer1911problem}%
  \BibitemOpen
  \bibfield  {author} {\bibinfo {author} {\bibfnamefont {H.~C.}\ \bibnamefont
  {Plummer}},\ }\href {\doibase 10.1093/mnras/71.5.460} {\bibfield  {journal}
  {\bibinfo  {journal} {Mon. Not. R. Astron. Soc.}\ }\textbf {\bibinfo {volume}
  {71}},\ \bibinfo {pages} {460} (\bibinfo {year} {1911})}\BibitemShut
  {NoStop}%
\bibitem [{\citenamefont {Kroupa}(2002)}]{kroupa2002initial}%
  \BibitemOpen
  \bibfield  {author} {\bibinfo {author} {\bibfnamefont {P.}~\bibnamefont
  {Kroupa}},\ }\href {\doibase 10.1126/science.1067524} {\bibfield  {journal}
  {\bibinfo  {journal} {Science}\ }\textbf {\bibinfo {volume} {295}},\ \bibinfo
  {pages} {82} (\bibinfo {year} {2002})},\ \Eprint
  {http://arxiv.org/abs/astro-ph/0201098} {arXiv:astro-ph/0201098} \BibitemShut
  {NoStop}%
\bibitem [{\citenamefont {Aparicio}\ \emph {et~al.}(2001)\citenamefont
  {Aparicio}, \citenamefont {Carrera},\ and\ \citenamefont
  {Mart{\'\i}nez-Delgado}}]{aparicio2001star}%
  \BibitemOpen
  \bibfield  {author} {\bibinfo {author} {\bibfnamefont {A.}~\bibnamefont
  {Aparicio}}, \bibinfo {author} {\bibfnamefont {R.}~\bibnamefont {Carrera}}, \
  and\ \bibinfo {author} {\bibfnamefont {D.}~\bibnamefont
  {Mart{\'\i}nez-Delgado}},\ }\href {\doibase 10.1086/323535} {\bibfield
  {journal} {\bibinfo  {journal} {Astron. J.}\ }\textbf {\bibinfo {volume}
  {122}},\ \bibinfo {pages} {2524} (\bibinfo {year} {2001})},\ \Eprint
  {http://arxiv.org/abs/astro-ph/0108159} {arXiv:astro-ph/0108159} \BibitemShut
  {NoStop}%
\bibitem [{\citenamefont {Widrow}(2000)}]{widrow2000distribution}%
  \BibitemOpen
  \bibfield  {author} {\bibinfo {author} {\bibfnamefont {L.~M.}\ \bibnamefont
  {Widrow}},\ }\href {\doibase 10.1086/317367} {\bibfield  {journal} {\bibinfo
  {journal} {Astrophys. J. Suppl. Ser.}\ }\textbf {\bibinfo {volume} {131}},\
  \bibinfo {pages} {39} (\bibinfo {year} {2000})}\BibitemShut {NoStop}%
\bibitem [{\citenamefont {Foreman-Mackey}\ \emph {et~al.}(2013)\citenamefont
  {Foreman-Mackey}, \citenamefont {Hogg}, \citenamefont {Lang},\ and\
  \citenamefont {Goodman}}]{Foreman_Mackey_2013}%
  \BibitemOpen
  \bibfield  {author} {\bibinfo {author} {\bibfnamefont {D.}~\bibnamefont
  {Foreman-Mackey}}, \bibinfo {author} {\bibfnamefont {D.~W.}\ \bibnamefont
  {Hogg}}, \bibinfo {author} {\bibfnamefont {D.}~\bibnamefont {Lang}}, \ and\
  \bibinfo {author} {\bibfnamefont {J.}~\bibnamefont {Goodman}},\ }\href
  {\doibase 10.1086/670067} {\bibfield  {journal} {\bibinfo  {journal} {Publ.
  Astron. Soc. Pac.}\ }\textbf {\bibinfo {volume} {125}},\ \bibinfo {pages}
  {306} (\bibinfo {year} {2013})},\ \Eprint {http://arxiv.org/abs/1202.3665}
  {arXiv:1202.3665} \BibitemShut {NoStop}%
\bibitem [{\citenamefont {Moore}(1993)}]{moore1993upper}%
  \BibitemOpen
  \bibfield  {author} {\bibinfo {author} {\bibfnamefont {B.}~\bibnamefont
  {Moore}},\ }\href {\doibase 10.1086/186967} {\bibfield  {journal} {\bibinfo
  {journal} {Astrophys. J.}\ }\textbf {\bibinfo {volume} {413}},\ \bibinfo
  {pages} {L93} (\bibinfo {year} {1993})}\BibitemShut {NoStop}%
\bibitem [{\citenamefont {van~den Bosch}\ \emph {et~al.}(2018)\citenamefont
  {van~den Bosch}, \citenamefont {Ogiya}, \citenamefont {Hahn},\ and\
  \citenamefont {Burkert}}]{van2017disruption}%
  \BibitemOpen
  \bibfield  {author} {\bibinfo {author} {\bibfnamefont {F.~C.}\ \bibnamefont
  {van~den Bosch}}, \bibinfo {author} {\bibfnamefont {G.}~\bibnamefont
  {Ogiya}}, \bibinfo {author} {\bibfnamefont {O.}~\bibnamefont {Hahn}}, \ and\
  \bibinfo {author} {\bibfnamefont {A.}~\bibnamefont {Burkert}},\ }\href
  {\doibase 10.1093/mnras/stx2956} {\bibfield  {journal} {\bibinfo  {journal}
  {Mon. Not. R. Astron. Soc.}\ }\textbf {\bibinfo {volume} {474}},\ \bibinfo
  {pages} {3043} (\bibinfo {year} {2018})},\ \Eprint
  {http://arxiv.org/abs/1711.05276} {arXiv:1711.05276} \BibitemShut {NoStop}%
\bibitem [{\citenamefont {Geringer-Sameth}\ \emph {et~al.}(2015)\citenamefont
  {Geringer-Sameth}, \citenamefont {Koushiappas},\ and\ \citenamefont
  {Walker}}]{geringer2015dwarf}%
  \BibitemOpen
  \bibfield  {author} {\bibinfo {author} {\bibfnamefont {A.}~\bibnamefont
  {Geringer-Sameth}}, \bibinfo {author} {\bibfnamefont {S.~M.}\ \bibnamefont
  {Koushiappas}}, \ and\ \bibinfo {author} {\bibfnamefont {M.}~\bibnamefont
  {Walker}},\ }\href {\doibase 10.1088/0004-637x/801/2/74} {\bibfield
  {journal} {\bibinfo  {journal} {Astrophys. J.}\ }\textbf {\bibinfo {volume}
  {801}},\ \bibinfo {pages} {74} (\bibinfo {year} {2015})},\ \Eprint
  {http://arxiv.org/abs/1408.0002} {arXiv:1408.0002} \BibitemShut {NoStop}%
\bibitem [{\citenamefont {Hayashi}\ \emph {et~al.}(2003)\citenamefont
  {Hayashi}, \citenamefont {Navarro}, \citenamefont {Taylor}, \citenamefont
  {Stadel},\ and\ \citenamefont {Quinn}}]{hayashi2003structural}%
  \BibitemOpen
  \bibfield  {author} {\bibinfo {author} {\bibfnamefont {E.}~\bibnamefont
  {Hayashi}}, \bibinfo {author} {\bibfnamefont {J.~F.}\ \bibnamefont
  {Navarro}}, \bibinfo {author} {\bibfnamefont {J.~E.}\ \bibnamefont {Taylor}},
  \bibinfo {author} {\bibfnamefont {J.}~\bibnamefont {Stadel}}, \ and\ \bibinfo
  {author} {\bibfnamefont {T.}~\bibnamefont {Quinn}},\ }\href {\doibase
  10.1086/345788} {\bibfield  {journal} {\bibinfo  {journal} {Astrophys. J.}\
  }\textbf {\bibinfo {volume} {584}},\ \bibinfo {pages} {541} (\bibinfo {year}
  {2003})},\ \Eprint {http://arxiv.org/abs/astro-ph/0203004}
  {arXiv:astro-ph/0203004} \BibitemShut {NoStop}%
\bibitem [{\citenamefont {Penarrubia}\ \emph {et~al.}(2010)\citenamefont
  {Penarrubia}, \citenamefont {Benson}, \citenamefont {Walker}, \citenamefont
  {Gilmore}, \citenamefont {McConnachie},\ and\ \citenamefont
  {Mayer}}]{penarrubia2010impact}%
  \BibitemOpen
  \bibfield  {author} {\bibinfo {author} {\bibfnamefont {J.}~\bibnamefont
  {Penarrubia}}, \bibinfo {author} {\bibfnamefont {A.~J.}\ \bibnamefont
  {Benson}}, \bibinfo {author} {\bibfnamefont {M.~G.}\ \bibnamefont {Walker}},
  \bibinfo {author} {\bibfnamefont {G.}~\bibnamefont {Gilmore}}, \bibinfo
  {author} {\bibfnamefont {A.~W.}\ \bibnamefont {McConnachie}}, \ and\ \bibinfo
  {author} {\bibfnamefont {L.}~\bibnamefont {Mayer}},\ }\href {\doibase
  10.1111/j.1365-2966.2010.16762.x} {\bibfield  {journal} {\bibinfo  {journal}
  {Mon. Not. R. Astron. Soc.}\ }\textbf {\bibinfo {volume} {406}},\ \bibinfo
  {pages} {1290} (\bibinfo {year} {2010})},\ \Eprint
  {http://arxiv.org/abs/1002.3376} {arXiv:1002.3376} \BibitemShut {NoStop}%
\bibitem [{\citenamefont {Green}\ and\ \citenamefont {van~den
  Bosch}(2019)}]{green2019tidal}%
  \BibitemOpen
  \bibfield  {author} {\bibinfo {author} {\bibfnamefont {S.~B.}\ \bibnamefont
  {Green}}\ and\ \bibinfo {author} {\bibfnamefont {F.~C.}\ \bibnamefont
  {van~den Bosch}},\ }\href {\doibase 10.1093/mnras/stz2767} {\bibfield
  {journal} {\bibinfo  {journal} {Mon. Not. R. Astron. Soc.}\ }\textbf
  {\bibinfo {volume} {490}},\ \bibinfo {pages} {2091} (\bibinfo {year}
  {2019})},\ \Eprint {http://arxiv.org/abs/1908.08537} {arXiv:1908.08537}
  \BibitemShut {NoStop}%
\bibitem [{\citenamefont {Ogiya}\ \emph {et~al.}(2019)\citenamefont {Ogiya},
  \citenamefont {Van~den Bosch}, \citenamefont {Hahn}, \citenamefont {Green},
  \citenamefont {Miller},\ and\ \citenamefont {Burkert}}]{ogiya2019dash}%
  \BibitemOpen
  \bibfield  {author} {\bibinfo {author} {\bibfnamefont {G.}~\bibnamefont
  {Ogiya}}, \bibinfo {author} {\bibfnamefont {F.~C.}\ \bibnamefont {Van~den
  Bosch}}, \bibinfo {author} {\bibfnamefont {O.}~\bibnamefont {Hahn}}, \bibinfo
  {author} {\bibfnamefont {S.~B.}\ \bibnamefont {Green}}, \bibinfo {author}
  {\bibfnamefont {T.~B.}\ \bibnamefont {Miller}}, \ and\ \bibinfo {author}
  {\bibfnamefont {A.}~\bibnamefont {Burkert}},\ }\href {\doibase
  10.1093/mnras/stz375} {\bibfield  {journal} {\bibinfo  {journal} {Mon. Not.
  R. Astron. Soc.}\ }\textbf {\bibinfo {volume} {485}},\ \bibinfo {pages} {189}
  (\bibinfo {year} {2019})},\ \Eprint {http://arxiv.org/abs/1901.08601}
  {arXiv:1901.08601} \BibitemShut {NoStop}%
\bibitem [{\citenamefont {Bonanos}\ \emph {et~al.}(2004)\citenamefont
  {Bonanos}, \citenamefont {Stanek}, \citenamefont {Szentgyorgyi},
  \citenamefont {Sasselov},\ and\ \citenamefont {Bakos}}]{bonanos2004rr}%
  \BibitemOpen
  \bibfield  {author} {\bibinfo {author} {\bibfnamefont {A.}~\bibnamefont
  {Bonanos}}, \bibinfo {author} {\bibfnamefont {K.}~\bibnamefont {Stanek}},
  \bibinfo {author} {\bibfnamefont {A.}~\bibnamefont {Szentgyorgyi}}, \bibinfo
  {author} {\bibfnamefont {D.}~\bibnamefont {Sasselov}}, \ and\ \bibinfo
  {author} {\bibfnamefont {G.}~\bibnamefont {Bakos}},\ }\href {\doibase
  10.1086/381073} {\bibfield  {journal} {\bibinfo  {journal} {Astron. J.}\
  }\textbf {\bibinfo {volume} {127}},\ \bibinfo {pages} {861} (\bibinfo {year}
  {2004})},\ \Eprint {http://arxiv.org/abs/astro-ph/0310477}
  {arXiv:astro-ph/0310477} \BibitemShut {NoStop}%
\bibitem [{bon(2007)}]{bonanos2007erratum}%
  \BibitemOpen
  \href {\doibase 10.1086/510311} {\ \textbf {\bibinfo {volume} {133}},\
  \bibinfo {pages} {756} (\bibinfo {year} {2007})}\BibitemShut {NoStop}%
\bibitem [{bon(2008)}]{bonanos2008erratum}%
  \BibitemOpen
  \href {\doibase 10.1088/0004-6256/136/2/896} {\ \textbf {\bibinfo {volume}
  {136}},\ \bibinfo {pages} {896} (\bibinfo {year} {2008})}\BibitemShut
  {NoStop}%
\bibitem [{\citenamefont {Ackermann}\ \emph {et~al.}(2014)\citenamefont
  {Ackermann} \emph {et~al.}}]{ackermann2014dark}%
  \BibitemOpen
  \bibfield  {author} {\bibinfo {author} {\bibfnamefont {M.}~\bibnamefont
  {Ackermann}} \emph {et~al.} (\bibinfo {collaboration} {Fermi-LAT
  Collaboration}),\ }\href {\doibase 10.1103/physrevd.89.042001} {\bibfield
  {journal} {\bibinfo  {journal} {Phys. Rev. D}\ }\textbf {\bibinfo {volume}
  {89}},\ \bibinfo {pages} {042001} (\bibinfo {year} {2014})},\ \Eprint
  {http://arxiv.org/abs/1310.0828} {arXiv:1310.0828} \BibitemShut {NoStop}%
\bibitem [{\citenamefont {Hooper}\ and\ \citenamefont
  {Linden}(2015)}]{hooper2015gamma}%
  \BibitemOpen
  \bibfield  {author} {\bibinfo {author} {\bibfnamefont {D.}~\bibnamefont
  {Hooper}}\ and\ \bibinfo {author} {\bibfnamefont {T.}~\bibnamefont
  {Linden}},\ }\href {\doibase 10.1088/1475-7516/2015/09/016} {\bibfield
  {journal} {\bibinfo  {journal} {J. Cosmol. Astropart. Phys.}\ }\textbf
  {\bibinfo {volume} {09}},\ \bibinfo {pages} {016} (\bibinfo {year} {2015})},\
  \Eprint {http://arxiv.org/abs/1503.06209} {arXiv:1503.06209} \BibitemShut
  {NoStop}%
\bibitem [{\citenamefont {{Fermi-LAT Collaboration}}()}]{Fermi-LAT:2019yla}%
  \BibitemOpen
  \bibfield  {author} {\bibinfo {author} {\bibnamefont {{Fermi-LAT
  Collaboration}}},\ }\href@noop {} {\ }\Eprint
  {http://arxiv.org/abs/1902.10045} {arXiv:1902.10045} \BibitemShut {NoStop}%
\bibitem [{\citenamefont {Calore}\ \emph {et~al.}(2018)\citenamefont {Calore},
  \citenamefont {Serpico},\ and\ \citenamefont {Zaldivar}}]{Calore:2018sdx}%
  \BibitemOpen
  \bibfield  {author} {\bibinfo {author} {\bibfnamefont {F.}~\bibnamefont
  {Calore}}, \bibinfo {author} {\bibfnamefont {P.~D.}\ \bibnamefont {Serpico}},
  \ and\ \bibinfo {author} {\bibfnamefont {B.}~\bibnamefont {Zaldivar}},\
  }\href {\doibase 10.1088/1475-7516/2018/10/029} {\bibfield  {journal}
  {\bibinfo  {journal} {J. Cosmol. Astropart. Phys.}\ }\textbf {\bibinfo
  {volume} {10}},\ \bibinfo {pages} {029} (\bibinfo {year} {2018})},\ \Eprint
  {http://arxiv.org/abs/1803.05508} {arXiv:1803.05508} \BibitemShut {NoStop}%
\bibitem [{\citenamefont {Hoof}\ \emph {et~al.}()\citenamefont {Hoof},
  \citenamefont {Geringer-Sameth},\ and\ \citenamefont
  {Trotta}}]{Hoof:2018hyn}%
  \BibitemOpen
  \bibfield  {author} {\bibinfo {author} {\bibfnamefont {S.}~\bibnamefont
  {Hoof}}, \bibinfo {author} {\bibfnamefont {A.}~\bibnamefont
  {Geringer-Sameth}}, \ and\ \bibinfo {author} {\bibfnamefont {R.}~\bibnamefont
  {Trotta}},\ }\href@noop {} {\ }\Eprint {http://arxiv.org/abs/1812.06986}
  {arXiv:1812.06986} \BibitemShut {NoStop}%
\bibitem [{\citenamefont {Linden}()}]{Linden:2019soa}%
  \BibitemOpen
  \bibfield  {author} {\bibinfo {author} {\bibfnamefont {T.}~\bibnamefont
  {Linden}},\ }\href@noop {} {\ }\Eprint {http://arxiv.org/abs/1905.11992}
  {arXiv:1905.11992} \BibitemShut {NoStop}%
\bibitem [{\citenamefont {Daylan}\ \emph {et~al.}(2016)\citenamefont {Daylan},
  \citenamefont {Finkbeiner}, \citenamefont {Hooper}, \citenamefont {Linden},
  \citenamefont {Portillo}, \citenamefont {Rodd},\ and\ \citenamefont
  {Slatyer}}]{Daylan:2014rsa}%
  \BibitemOpen
  \bibfield  {author} {\bibinfo {author} {\bibfnamefont {T.}~\bibnamefont
  {Daylan}}, \bibinfo {author} {\bibfnamefont {D.~P.}\ \bibnamefont
  {Finkbeiner}}, \bibinfo {author} {\bibfnamefont {D.}~\bibnamefont {Hooper}},
  \bibinfo {author} {\bibfnamefont {T.}~\bibnamefont {Linden}}, \bibinfo
  {author} {\bibfnamefont {S.~K.~N.}\ \bibnamefont {Portillo}}, \bibinfo
  {author} {\bibfnamefont {N.~L.}\ \bibnamefont {Rodd}}, \ and\ \bibinfo
  {author} {\bibfnamefont {T.~R.}\ \bibnamefont {Slatyer}},\ }\href {\doibase
  10.1016/j.dark.2015.12.005} {\bibfield  {journal} {\bibinfo  {journal} {Phys.
  Dark Univ.}\ }\textbf {\bibinfo {volume} {12}},\ \bibinfo {pages} {1}
  (\bibinfo {year} {2016})},\ \Eprint {http://arxiv.org/abs/1402.6703}
  {arXiv:1402.6703} \BibitemShut {NoStop}%
\bibitem [{\citenamefont {Dror}\ \emph {et~al.}(2019)\citenamefont {Dror},
  \citenamefont {Ramani}, \citenamefont {Trickle},\ and\ \citenamefont
  {Zurek}}]{Dror_2019}%
  \BibitemOpen
  \bibfield  {author} {\bibinfo {author} {\bibfnamefont {J.~A.}\ \bibnamefont
  {Dror}}, \bibinfo {author} {\bibfnamefont {H.}~\bibnamefont {Ramani}},
  \bibinfo {author} {\bibfnamefont {T.}~\bibnamefont {Trickle}}, \ and\
  \bibinfo {author} {\bibfnamefont {K.~M.}\ \bibnamefont {Zurek}},\ }\href
  {\doibase 10.1103/PhysRevD.100.023003} {\bibfield  {journal} {\bibinfo
  {journal} {Phys. Rev. D}\ }\textbf {\bibinfo {volume} {100}},\ \bibinfo
  {pages} {023003} (\bibinfo {year} {2019})},\ \Eprint
  {http://arxiv.org/abs/1901.04490} {arXiv:1901.04490} \BibitemShut {NoStop}%
\bibitem [{\citenamefont {Jennings}\ \emph {et~al.}()\citenamefont {Jennings},
  \citenamefont {Cordes},\ and\ \citenamefont {Chatterjee}}]{Jennings:2019qqz}%
  \BibitemOpen
  \bibfield  {author} {\bibinfo {author} {\bibfnamefont {R.~J.}\ \bibnamefont
  {Jennings}}, \bibinfo {author} {\bibfnamefont {J.~M.}\ \bibnamefont
  {Cordes}}, \ and\ \bibinfo {author} {\bibfnamefont {S.}~\bibnamefont
  {Chatterjee}},\ }\href@noop {} {\ }\Eprint {http://arxiv.org/abs/1910.08608}
  {arXiv:1910.08608} \BibitemShut {NoStop}%
\bibitem [{\citenamefont {Dai}\ and\ \citenamefont
  {Miralda-Escud{\'e}}()}]{dai2019gravitational}%
  \BibitemOpen
  \bibfield  {author} {\bibinfo {author} {\bibfnamefont {L.}~\bibnamefont
  {Dai}}\ and\ \bibinfo {author} {\bibfnamefont {J.}~\bibnamefont
  {Miralda-Escud{\'e}}},\ }\href@noop {} {\ }\Eprint
  {http://arxiv.org/abs/1908.01773} {arXiv:1908.01773} \BibitemShut {NoStop}%
\bibitem [{\citenamefont {Battaglia}\ \emph {et~al.}(2005)\citenamefont
  {Battaglia}, \citenamefont {Helmi}, \citenamefont {Morrison}, \citenamefont
  {Harding}, \citenamefont {Olszewski}, \citenamefont {Mateo}, \citenamefont
  {Freeman}, \citenamefont {Norris},\ and\ \citenamefont
  {Shectman}}]{battaglia2005radial}%
  \BibitemOpen
  \bibfield  {author} {\bibinfo {author} {\bibfnamefont {G.}~\bibnamefont
  {Battaglia}}, \bibinfo {author} {\bibfnamefont {A.}~\bibnamefont {Helmi}},
  \bibinfo {author} {\bibfnamefont {H.}~\bibnamefont {Morrison}}, \bibinfo
  {author} {\bibfnamefont {P.}~\bibnamefont {Harding}}, \bibinfo {author}
  {\bibfnamefont {E.~W.}\ \bibnamefont {Olszewski}}, \bibinfo {author}
  {\bibfnamefont {M.}~\bibnamefont {Mateo}}, \bibinfo {author} {\bibfnamefont
  {K.~C.}\ \bibnamefont {Freeman}}, \bibinfo {author} {\bibfnamefont
  {J.}~\bibnamefont {Norris}}, \ and\ \bibinfo {author} {\bibfnamefont {S.~A.}\
  \bibnamefont {Shectman}},\ }\href {\doibase 10.1111/j.1365-2966.2005.09367.x}
  {\bibfield  {journal} {\bibinfo  {journal} {Mon. Not. R. Astron. Soc.}\
  }\textbf {\bibinfo {volume} {364}},\ \bibinfo {pages} {433} (\bibinfo {year}
  {2005})},\ \Eprint {http://arxiv.org/abs/astro-ph/0506102}
  {arXiv:astro-ph/0506102} \BibitemShut {NoStop}%
\bibitem [{\citenamefont {Read}\ \emph {et~al.}(2008)\citenamefont {Read},
  \citenamefont {Lake}, \citenamefont {Agertz},\ and\ \citenamefont
  {Debattista}}]{read2008thin}%
  \BibitemOpen
  \bibfield  {author} {\bibinfo {author} {\bibfnamefont {J.}~\bibnamefont
  {Read}}, \bibinfo {author} {\bibfnamefont {G.}~\bibnamefont {Lake}}, \bibinfo
  {author} {\bibfnamefont {O.}~\bibnamefont {Agertz}}, \ and\ \bibinfo {author}
  {\bibfnamefont {V.~P.}\ \bibnamefont {Debattista}},\ }\href {\doibase
  10.1111/j.1365-2966.2008.13643.x} {\bibfield  {journal} {\bibinfo  {journal}
  {Mon. Not. R. Astron. Soc.}\ }\textbf {\bibinfo {volume} {389}},\ \bibinfo
  {pages} {1041} (\bibinfo {year} {2008})},\ \Eprint
  {http://arxiv.org/abs/0803.2714} {arXiv:0803.2714} \BibitemShut {NoStop}%
\bibitem [{\citenamefont {McMillan}(2011)}]{mcmillan2011mass}%
  \BibitemOpen
  \bibfield  {author} {\bibinfo {author} {\bibfnamefont {P.~J.}\ \bibnamefont
  {McMillan}},\ }\href {\doibase 10.1111/j.1365-2966.2011.18564.x} {\bibfield
  {journal} {\bibinfo  {journal} {Mon. Not. R. Astron. Soc.}\ }\textbf
  {\bibinfo {volume} {414}},\ \bibinfo {pages} {2446} (\bibinfo {year}
  {2011})},\ \Eprint {http://arxiv.org/abs/1102.4340} {arXiv:1102.4340}
  \BibitemShut {NoStop}%
\bibitem [{\citenamefont {Sohn}\ \emph {et~al.}(2013)\citenamefont {Sohn},
  \citenamefont {Besla}, \citenamefont {Van Der~Marel}, \citenamefont
  {Boylan-Kolchin}, \citenamefont {Majewski},\ and\ \citenamefont
  {Bullock}}]{sohn2013space}%
  \BibitemOpen
  \bibfield  {author} {\bibinfo {author} {\bibfnamefont {S.~T.}\ \bibnamefont
  {Sohn}}, \bibinfo {author} {\bibfnamefont {G.}~\bibnamefont {Besla}},
  \bibinfo {author} {\bibfnamefont {R.~P.}\ \bibnamefont {Van Der~Marel}},
  \bibinfo {author} {\bibfnamefont {M.}~\bibnamefont {Boylan-Kolchin}},
  \bibinfo {author} {\bibfnamefont {S.~R.}\ \bibnamefont {Majewski}}, \ and\
  \bibinfo {author} {\bibfnamefont {J.~S.}\ \bibnamefont {Bullock}},\ }\href
  {\doibase 10.1088/0004-637x/768/2/139} {\bibfield  {journal} {\bibinfo
  {journal} {Astrophys. J.}\ }\textbf {\bibinfo {volume} {768}},\ \bibinfo
  {pages} {139} (\bibinfo {year} {2013})},\ \Eprint
  {http://arxiv.org/abs/1210.6039} {arXiv:1210.6039} \BibitemShut {NoStop}%
\bibitem [{\citenamefont {Nesti}\ and\ \citenamefont
  {Salucci}(2013)}]{nesti2013dark}%
  \BibitemOpen
  \bibfield  {author} {\bibinfo {author} {\bibfnamefont {F.}~\bibnamefont
  {Nesti}}\ and\ \bibinfo {author} {\bibfnamefont {P.}~\bibnamefont
  {Salucci}},\ }\href {\doibase 10.1088/1475-7516/2013/07/016} {\bibfield
  {journal} {\bibinfo  {journal} {J. Cosmol. Astropart. Phys.}\ }\textbf
  {\bibinfo {volume} {07}},\ \bibinfo {pages} {016} (\bibinfo {year} {2013})},\
  \Eprint {http://arxiv.org/abs/1304.5127} {arXiv:1304.5127} \BibitemShut
  {NoStop}%
\bibitem [{\citenamefont {Helmi}\ \emph {et~al.}(2018)\citenamefont {Helmi},
  \citenamefont {Van~Leeuwen}, \citenamefont {McMillan}, \citenamefont
  {Massari}, \citenamefont {Antoja}, \citenamefont {Robin}, \citenamefont
  {Lindegren}, \citenamefont {Bastian}, \citenamefont {Arenou}, \citenamefont
  {Babusiaux} \emph {et~al.}}]{helmi2018gaia}%
  \BibitemOpen
  \bibfield  {author} {\bibinfo {author} {\bibfnamefont {A.}~\bibnamefont
  {Helmi}}, \bibinfo {author} {\bibfnamefont {F.}~\bibnamefont {Van~Leeuwen}},
  \bibinfo {author} {\bibfnamefont {P.}~\bibnamefont {McMillan}}, \bibinfo
  {author} {\bibfnamefont {D.}~\bibnamefont {Massari}}, \bibinfo {author}
  {\bibfnamefont {T.}~\bibnamefont {Antoja}}, \bibinfo {author} {\bibfnamefont
  {A.}~\bibnamefont {Robin}}, \bibinfo {author} {\bibfnamefont
  {L.}~\bibnamefont {Lindegren}}, \bibinfo {author} {\bibfnamefont
  {U.}~\bibnamefont {Bastian}}, \bibinfo {author} {\bibfnamefont
  {F.}~\bibnamefont {Arenou}}, \bibinfo {author} {\bibfnamefont
  {C.}~\bibnamefont {Babusiaux}},  \emph {et~al.} (\bibinfo {collaboration}
  {Gaia Collaboration}),\ }\href {\doibase 10.1051/0004-6361/201832698}
  {\bibfield  {journal} {\bibinfo  {journal} {Astron. Astrophys.}\ }\textbf
  {\bibinfo {volume} {616}},\ \bibinfo {pages} {A12} (\bibinfo {year}
  {2018})},\ \Eprint {http://arxiv.org/abs/1804.09381} {arXiv:1804.09381}
  \BibitemShut {NoStop}%
\bibitem [{\citenamefont {Fritz}\ \emph {et~al.}(2018)\citenamefont {Fritz},
  \citenamefont {Battaglia}, \citenamefont {Pawlowski}, \citenamefont
  {Kallivayalil}, \citenamefont {Van Der~Marel}, \citenamefont {Sohn},
  \citenamefont {Brook},\ and\ \citenamefont {Besla}}]{fritz2018gaia}%
  \BibitemOpen
  \bibfield  {author} {\bibinfo {author} {\bibfnamefont {T.}~\bibnamefont
  {Fritz}}, \bibinfo {author} {\bibfnamefont {G.}~\bibnamefont {Battaglia}},
  \bibinfo {author} {\bibfnamefont {M.}~\bibnamefont {Pawlowski}}, \bibinfo
  {author} {\bibfnamefont {N.}~\bibnamefont {Kallivayalil}}, \bibinfo {author}
  {\bibfnamefont {R.}~\bibnamefont {Van Der~Marel}}, \bibinfo {author}
  {\bibfnamefont {S.}~\bibnamefont {Sohn}}, \bibinfo {author} {\bibfnamefont
  {C.}~\bibnamefont {Brook}}, \ and\ \bibinfo {author} {\bibfnamefont
  {G.}~\bibnamefont {Besla}},\ }\href {\doibase 10.1051/0004-6361/201833343}
  {\bibfield  {journal} {\bibinfo  {journal} {Astron. Astrophys.}\ }\textbf
  {\bibinfo {volume} {619}},\ \bibinfo {pages} {A103} (\bibinfo {year}
  {2018})},\ \Eprint {http://arxiv.org/abs/1805.00908} {arXiv:1805.00908}
  \BibitemShut {NoStop}%
\end{thebibliography}%

\end{document}